\documentclass[12pt]{article} 
\usepackage{amssymb}
\usepackage{amsmath} 
\usepackage{cite}
\usepackage{axodraw}
\usepackage{hyperref}
\input xy
\xyoption{all}

\newcommand{\Z}{\mathbb{Z}}

\def\sqr#1#2{{\vcenter{\vbox{\hrule height.#2pt
            \hbox{\vrule width.#2pt height#1pt \kern#1pt
                  \vrule width.#2pt}\hrule height.#2pt}}}}

\def\square
 {\mathop{\mathchoice{\sqr{12}{15}}{\sqr{9}{12}}
{\sqr{6.3}{9}}{\sqr{4.5}{9}}}}

\def\sqra#1#2#3{{\vcenter{\vbox{\hrule height.#2pt
            \hbox{\vrule width.#2pt height#1pt \kern5pt 
#3
                  \vrule width.#2pt}\hrule height.#2pt}}}}

\numberwithin{equation}{section}

\numberwithin{table}{section}

\begin{document} 

\begin{center}

{\large\bf Quantum symmetries in orbifolds and decomposition}

\vspace*{0.2in}

Daniel G. Robbins$^1$, Eric Sharpe$^2$,
Thomas Vandermeulen$^1$

\begin{tabular}{cc}
{\begin{tabular}{l}
$^1$ Department of Physics\\
University at Albany\\
Albany, NY 12222 \end{tabular}} &
{\begin{tabular}{l}
$^2$ Department of Physics MC 0435\\
850 West Campus Drive\\
Virginia Tech\\
Blacksburg, VA  24061 \end{tabular}}
\end{tabular}

{\tt dgrobbins@albany.edu},
{\tt ersharpe@vt.edu},
{\tt tvandermeulen@albany.edu}

\end{center}

In this paper, we introduce a new set of modular-invariant phase factors
for orbifolds with trivially-acting subgroups, analogous to discrete torsion
and generalizing quantum symmetries.
After describing their basic properties,
we generalize decomposition to include orbifolds
with these new phase factors,
making a precise proposal for how such orbifolds are equivalent to 
disjoint unions of other
orbifolds without trivially-acting subgroups or one-form symmetries, 
which we check in numerous examples.

\begin{flushleft}
July 2021
\end{flushleft}

\newpage

\tableofcontents

\newpage

\section{Introduction}

This paper is motivated by renewed recent interest in orbifolds
and their anomalies, see e.g. \cite{Wang:2017loc,Bhardwaj:2017xup,Tachikawa:2017gyf,Chang:2018iay,Robbins:2019zdb,Robbins:2019ayj,yujitasi2019,Robbins:2021lry,Robbins:2021ylj}.
Specifically, the need for this work arose while studying proposals
for anomaly-resolution in \cite{Wang:2017loc,Tachikawa:2017gyf}.  As outlined
in e.g. \cite{Robbins:2021lry,Tachikawa:2017gyf}, and as will be explained in greater
detail in our upcoming work \cite{rsv}, these proposals 
implicitly require
that the `resolved' orbifolds use new degrees of freedom.

These new degrees of freedom generalize quantum symmetries of orbifolds
\cite{Ginsparg:1988ui,Vafa:1989ih},
for which reason we use the same nomenclature, and are specific
to orbifolds in which a subgroup of the orbifold group acts trivially
on the original space.  Unlike ordinary quantum symmetries, the quantum
symmetries we shall be focused on do not always arise from discrete
torsion, and so define new modular-invariant phases -- but in which the
modular invariance is achieved in a novel fashion.  If we denote the
full orbifold group by $\Gamma$ and the trivially-acting subgroup $K
\subset \Gamma$, then in the cases of most interest for us,
modular invariance is broken in $\Gamma$, but restored when the orbifold
is viewed as a $G$ orbifold, as we shall explain in detail.

The purpose of this paper is both to explore these new degrees of freedom,
which are of independent interest, and also to set the stage for
their application to anomaly resolution, which will be explored in
detail in \cite{rsv}.

Before continuing, we should mention that there exist different
generalizations of quantum symmetries in the literature.
For example, \cite{Bhardwaj:2017xup} utilizes fusion categories to build
a generalization with the property that for nonabelian orbifolds,
orbifolding by the (fusion category) quantum symmetry returns the
original space.  Our generalization, by contrast, does not involve
fusion categories (and when applied to nonabelian orbifolds need not
return the original space), but will instead provide new degrees of freedom
in orbifolds with trivially-acting subgroups.

Orbifolds with trivially-acting subgroups, and more generally
gauge theories with trivially-acting subgroups,
such as abelian gauge
theories with nonminimal charges, were explored in
\cite{Pantev:2005rh,Pantev:2005wj,Pantev:2005zs} as part of a program
of developing string compactifications on certain generalized spaces
known as stacks and gerbes.  In modern language, a gerbe is a fiber bundle
in which the fibers are `groups' of one-form symmetries, hence a sigma model
on a gerbe naturally admits a one-form symmetry corresponding to translation
along the fibers.

One of the most important outcomes of that work was the discovery of
decomposition,
first described in
\cite{Hellerman:2006zs}, an equivalence between two-dimensional
theories with one-form symmetries (and various generalizations) and disjoint
unions of other two-dimensional theories.
Decomposition has since been applied to a number of areas,
including Gromov-Witten theory \cite{ajt1,ajt2,ajt3,t1,gt1,xt1},
gauged linear sigma models \cite{Caldararu:2007tc,Hori:2011pd,Addington:2012zv,Sharpe:2012ji,Hori:2013gga,Hori:2016txh,Knapp:2019cih},
mirror symmetry \cite{Hellerman:2006zs,Gu:2018fpm,Chen:2018wep,Gu:2019zkw,Gu:2020ivl}
and heterotic string compactifications \cite{Anderson:2013sia}.
See e.g.
\cite{Sharpe:2019ddn,Tanizaki:2019rbk,Cherman:2020cvw,Eager:2020rra,Komargodski:2020mxz,Robbins:2021ylj} 
for more recent work, on topics ranging from elliptic genera to
higher-dimensional analogues.

Naturally, after introducing more general quantum symmetries in
orbifolds with trivially-acting subgroups, in this paper we will also
describe decomposition in such theories.  This is both for some semblance
of completeness, as well as because decomposition will play a crucial
role in the application to anomalies, as we will discuss in
\cite{rsv}.

We begin in section~\ref{sect:quantumsymm} by describing these new more
general quantum symmetries and their basic properties, such as their
relation to older notions of quantum symmetries, as well as the modular 
invariance of these new phases.  In section~\ref{sect:conj} we then
conjecture the form that decomposition takes in these theories, the
modification required by the presence of a quantum symmetry.

In section~\ref{sect:examples} 
we explore a number of examples, both to demonstrate
quantum symmetries explicitly, as well as to explore the effect of
decomposition.

In a series of appendices, we collect some related information.
In appendix~\ref{app:oldrev}, we review conventional quantum symmetries
in this framework, and how orbifolds-of-orbifolds returning the original theory
(or a variant thereof) is a version of decomposition.

Appendix~\ref{app:triv-pullback} is a technical demonstration that the
pullback of the quantity $d_2 B$ (which plays a role in our discussion)
is trivial in cohomology.  This will play an essential role in the
application to anomalies discussed in \cite{rsv}, and also seems to
be important for describing D-branes and open strings in these theories.

In appendix~\ref{app:open} we outline open strings and D-branes in
orbifolds with quantum symmetries.  To define a D-brane here requires
a subtle generalization.  For example, in an ordinary orbifold,
to define a D-brane, one must give a group action on the D-brane,
and if one adds discrete torsion, then that group action is
projectivized.  Here, quantum symmetries often require that
associativity of the group action is (weakly) broken.  
In this appendix we outline basics.

In appendix~\ref{app:noncent} we outline some aspects of
non-central extensions.  In the rest of this paper, we assume
that the full orbifold group $\Gamma$ is a central extension of
the effectively-acting orbifold $G$, meaning that the trivially-acting
subgroup lies within the center of $\Gamma$.  At least much of the
structure we describe appears to generalize to non-central extensions,
and we outline some details here.

Finally, a remark on nomenclature.  Across our several papers on
these matters, we have sometimes mixed
additive and multiplicative notations.  For example, a trivial
quantum symmetry $B$ is sometimes written as $B=0$
and other times is written as $B(g) = 1$ for all $g \in G$.

\section{Quantum symmetries of the $\Gamma$ orbifold}
\label{sect:quantumsymm}

\subsection{Review of ordinary quantum symmetries}
\label{sect:rev}

Let us begin with a short review of quantum symmetries in orbifolds.
Given a $G$ orbifold, the quantum symmetry 
(see e.g. \cite[section 8.5]{Ginsparg:1988ui}, \cite{Vafa:1989ih})
multiplies twist fields by phases,
so as to leave correlation functions invariant.  If $G$ is abelian,
the quantum symmetry $\hat{G} \cong G$; if $G$ is nonabelian,
the quantum symmetry $\hat{G}$ is the abelianization $G/[G,G]$.

A famous consequence of quantum symmetries is that they can undo the
orbifold:  for $G$ abelian, orbifolding the original orbifold by the
quantum symmetry returns the original theory.
(For $G$ nonabelian, to recover the original theory one must work with
fusion categories, see e.g. \cite{Bhardwaj:2017xup}; 
we shall take a different direction in
this paper.)

Those two subsequent orbifolds, the orbifold of $[X/G]$ by the
quantum symmetry $\hat{G}$, can equivalently be described as a single orbifold
$[X/ (G \times \hat{G})]$ with discrete torsion, which provides the phases
produced by the quantum symmetry.  The group $\hat{G}$ acts trivially
on $X$; it only acts via $G$-twisted sector phases.
For example, if $G = {\mathbb Z}_n$,
then orbifolding by the quantum symmetry is equivalent to
an orbifold by ${\mathbb Z}_n \times {\mathbb Z}_n$ with discrete torsion
in $H^2({\mathbb Z}_n \times {\mathbb Z}_n, U(1)) \cong {\mathbb Z}_n$,
where one ${\mathbb Z}_n$ factor acts trivially on $X$ (though its contributions
to the path integral are weighted by phases).
(See for example \cite[section 8.5]{Ginsparg:1988ui} for an explicit discussion
for the case $n=2$.)

The fact that such an orbifold is equivalent to the original space
is a special case of the decomposition for orbifolds with discrete torsion
discussed in \cite{Robbins:2021ylj}.  For example, the case $n=2$ is discussed
in detail in \cite[section 5.1]{Robbins:2021ylj}, where it is checked that,
as a consequence of decomposition,
\begin{equation}
{\rm QFT}\left(
[X / {\mathbb Z}_2 \times {\mathbb Z}_2]_{\rm d.t.} 
\right) \: = \:
{\rm QFT}\left( X \right).
\end{equation}
In appendix~\ref{app:oldrev}, we review the analogous result for general
orbifold groups $G$, demonstrating that for a quantum symmetry group
$\hat{G} = G/[G,G]$, the orbifold $[X/\Gamma]$ for 
$\Gamma = G \times \hat{G}$ with appropriate discrete torsion is
equivalent to $[X/[G,G]]$.

In this paper, we generalize quantum symmetries to orbifolds
$[X/\Gamma]$, where $\Gamma$ is no longer a product but rather a central
extension, and more importantly,
to phases which need not be determined by discrete torsion.
Despite the fact that the phases do not come from discrete torsion,
we will find that they nevertheless are modular invariant, giving us
new modular-invariant phases to add to orbifolds, at least orbifolds
in which a subgroup of the orbifold group acts trivially.
We will also find other subtleties in these new phases -- for example,
appendix~\ref{app:open} discusses how D-branes in such orbifolds
appear to be defined by non-associative analogues of equivariant
structures.

In the remainder of this section we will discuss these new phases
generalizing quantum symmetries, and discuss basic properties such
as their modular invariance.

\subsection{Basics of general construction}
\label{sect:basics}

Next, we will consider more general quantum symmetries arising
in an orbifold
$\Gamma$ that is a central extension of an effectively-acting group $G$
by a trivially-acting group $K$.  
(In appendix~\ref{app:noncent} we outline
some of the basics needed to generalize to non-central extensions,
which otherwise are left for future work.)

As in the ordinary notion, the idea behind a quantum symmetry is that
$K$ acts trivially on $X$, but nontrivially on $G$-twisted sectors.
However, since in general $\Gamma$ can be a nontrivial extension,
the $G$-twisted sectors
are not well-defined, so that description is imprecise.
A better description is that the set of $\Gamma$-twisted sectors
is acted upon\footnote{
Technically, the set of genus-one twisted sectors in the
$\Gamma$ orbifold forms a torsor under $K$.
} by $K$, with phases determined by 
$B \in H^1(G,H^1(K,U(1)))$, as follows.
Since $K$ is central, an element $B$ is equivalent to a map
\begin{equation}
B: \: G \times K \: \longrightarrow \: U(1),
\end{equation}
which we use to give relations of the form\footnote{
Note that the relations are almost but not quite symmetric under an
S transformation.  This can be derived from modular transformations.
}
\begin{equation}  \label{eq:quantsymm}
{\scriptstyle g z} \square_h \: = \: B(\pi(h), z) \left(
{\scriptstyle g} \square_h \right),
\: \: \:
{\scriptstyle g} \square_{hz} \: = \: B( \pi(g), z)^{-1} \left(
{\scriptstyle g} \square_h \right),
\end{equation}
for $z \in K$ and $g, h \in \Gamma$ a commuting pair.
(We assume that $B(g,1) = 1$ for all $g \in G$, and $B(1,k) = 1$ for all
$k \in K$.)
This is the precise meaning of the intuition that $K$ acts trivially
on $X$, but nontrivially on $G$-twisted sectors.
The resulting phases are classified by elements of
\begin{equation}
H^1(G, H^1(K,U(1))).
\end{equation}

In this paper, we will see in examples that with this choice,
the resulting $\Gamma$ orbifold is well-defined, and typically has the
form of (copies) of an orbifold by a subgroup of $G$.
We will make a precise general prediction for the form of any such
$\Gamma$ orbifold, which we will check extensively.

Before going on, let us check that the prescription above for
quantum symmetries derived from an element of $H^1(G, H^1(K,U(1)))$
is invariant under modular transformations.
Begin with the relation
\begin{equation} \label{eq:quantsymm:modtrans1}
{\scriptstyle g z} \square_h \: = \: B(\pi(h), z) \left(
{\scriptstyle g} \square_h \right),
\end{equation}
and perform a modular transformation by
\begin{equation}
\left[ \begin{array}{cc} a & b \\ c & d \end{array} \right] \: \in \:
SL(2,{\mathbb Z}).
\end{equation}
Under this modular transformation
\begin{eqnarray}
{\scriptstyle g z} \square_h & \mapsto &
{\scriptstyle (gz)^a h^b } \square_{ (gz)^c h^d},
\\
{\scriptstyle g} \square_h & \mapsto &
{\scriptstyle g^a h^b} \square_{g^c h^d},
\end{eqnarray}
and from the relation~(\ref{eq:quantsymm}),
\begin{eqnarray}
{\scriptstyle (gz)^a h^b } \square_{ (gz)^c h^d}
& = &
B( \pi( g^c h^d), z^a ) 
\left( {\scriptstyle g^a h^b} \square_{ (gz)^c h^d } \right),
\\
& = &
\frac{ B( \pi( g^c h^d), z^a ) }{
B( \pi( g^a h^b ), z^c ) }
\left( {\scriptstyle g^a h^b} \square_{g^c h^d} \right).
\end{eqnarray}
Since $B \in H^1(G, H^1(K,U(1)))$, and $K$ is assumed central, we know that
\begin{eqnarray}
B(g, z_1 z_2) & = & B(g, z_1) B(g, z_2),   \label{eq:bhom1}
\\
B(g_1 g_2, z) & = & B(g_1, z) B(g_2, z),   \label{eq:bhom2}
\end{eqnarray}
hence
\begin{eqnarray}
\frac{ B( \pi( g^c h^d), z^a ) }{
B( \pi( g^a h^b ), z^c ) }
& = &
\frac{ B( \pi( g^c h^d ), z)^a }{
B( \pi( g^a h^b, z)^c },
\\
& = &
\frac{ B( \pi( g^{ca} h^{da} ), z) }{
B( \pi(g^{ac} h^{bc} ), z) },
\\
& = &
B( \pi( h^{ad-bc} ), z) \: = \:
B( \pi(h), z).
\end{eqnarray}
Thus, we see that
\begin{equation}
{\scriptstyle (gz)^a h^b } \square_{ (gz)^c h^d}
\: = \:
B( \pi(h), z) \left(  {\scriptstyle g^a h^b} \square_{g^c h^d} \right),
\end{equation}
precisely consistent with a modular transformation of
the identity~(\ref{eq:quantsymm:modtrans1}).

In section~\ref{sect:rev} we reviewed ordinary quantum symmetries,
described by trivial extensions $\Gamma = G \times K$, and which were
produced by discrete torsion.  The relationship can be made precise using
the exact sequence
\begin{equation}  \label{eq:shortexact1}
\left( {\rm Ker}\, i^* \subset H^2(\Gamma,U(1))
\right) \: \stackrel{\beta} \longrightarrow \:
H^1(G, H^1(K,U(1))) \: \stackrel{d_2}{\longrightarrow} \:
H^3(G,U(1)),
\end{equation}
where $\iota: K \rightarrow \Gamma$ is inclusion and 
in the sequence above,
\begin{equation}
\iota^*: \: H^2(\Gamma,U(1)) \: \longrightarrow \:
H^2(K,U(1)).
\end{equation}
(This is part of a seven-term exact sequence \cite{hochschild}
slightly extending the inflation-restriction sequence
(see e.g. \cite{hochserre},
\cite[example 6.8.3]{weibel}, \cite[section I.6]{neukirch},
\cite[section 3.3]{gille-szamuely}), that is discussed in greater
detail 
in our previous work \cite{Robbins:2021ylj}.)
For later use, the maps $\beta$ and $d_2$ are given as
\begin{equation}
\beta(\omega)(\pi(g),z) \: = \: \frac{
\omega(g,z) }{ \omega(z,g) },
\end{equation}
(for $g \in \Gamma$, $z \in K$, with $\omega$ a cocycle representing
an element of $H^2(\Gamma,U(1))$ corresponding to discrete torsion)
as we shall discuss in detail in the next section, as well as
\cite[appendix C]{Robbins:2021ylj}, and
\begin{equation}
(d_2 B)( \overline{g}_1, \overline{g}_2, \overline{g}_3) \: = \:
B( \overline{g}_1, s_2 s_3 s_{23}^{-1} ),
\end{equation}
where $\overline{g}_i \in G$, $s_i = s(\overline{g}_i)$ for $s:
G \rightarrow \Gamma$ a section, as discussed in \cite[section 8]{hochschild}.
(The quantity $s_2 s_3 s_{23}^{-1}$ is the extension class in
$H^2(G,K)$ of $\Gamma$, and so vanishes in cohomology if the extension splits.)

In any event, a quantum symmetry $B \in H^1(G,H^1(K,U(1)))$ arises
from discrete torsion precisely when it is in the image of $\beta$ in the
sequence~(\ref{eq:shortexact1}) above.  Furthermore, the reader should note
from the sequence~(\ref{eq:shortexact1}) that not every quantum symmetry
$B$ can be so described -- those for which $d_2 B$ is nontrivial, are not
in the image of $\beta$ and so are not given by discrete torsion.

\subsection{Special case of discrete torsion}
\label{app:Bdt}

An important set of special cases are those in which the
quantum symmetry, defined by
$B \in H^1(G, H^1(K,U(1)))$, arises from discrete torsion 
$\omega \in H^2(\Gamma,U(1))$, meaning that $B$ is in the image of
$\beta$ in the sequence~(\ref{eq:shortexact1}).
In this section we will verify that mathematical result directly
in physics.  (See also \cite[appendix C]{Robbins:2021ylj} for another
discussion of $\beta$.)

We have described quantum symmetries as elements
$B \in H^1(G,H^1(K,U(1))$ that relate genus-one twisted sectors
as
\begin{equation}  \label{eq:quantsymm:basicreln}
{\scriptstyle gz} \square_h \: = \: B(\pi(h), z) \left( 
{\scriptstyle g} \square_h \right),
\end{equation}
for $g, h$ commuting elements of $\Gamma$, and $z \in K$.
If $K$ acts trivially on the original theory (albeit not on
$G$-twisted sectors), then this implies a relation between phases
derived from discrete torsion $\omega \in H^2(\Gamma,U(1))$, specifically
\begin{equation}
\frac{ \omega(gz,h) }{ \omega(h,gz) } \: = \:
B(\pi(h),z) \frac{\omega(g,h)}{\omega(h,g)},
\end{equation}
or equivalently,
\begin{equation} \label{eq:B}
B(\pi(h),z) \: = \:
\frac{ \omega(gz,h) }{ \omega(h,gz) }
\frac{ \omega(h,g) }{ \omega(g,h) },
\end{equation}
where we have used the fact that discrete torsion assigns the phase
\begin{equation}
\epsilon_{\rm dt}(g,h) \: = \: \frac{ \omega(g,h) }{ \omega(h,g) }
\end{equation}
to the genus-one diagram 
\begin{equation}
{\scriptstyle g} \square_h
\end{equation}
defined by commuting $g$, $h$.
In deriving this expression, we have merely assumed that the
twisted sectors are well-defined, meaning that $h$ commutes with both $g$
and $gz$.

In order to be consistent, the right-hand-side of equation~(\ref{eq:B}) must
not depend on $g$, and can only depend upon $\pi(h)$ instead of all of $h$.
We will see that the first statement is always true,
and the second is true so long as the restriction of $\omega$ to
$H^2(K,U(1))$ is trivial.  (If the restriction is not trivial,
then $B = B(h,z)$ and so defines an element of $H^1(\Gamma,H^1(K,U(1)))$
instead of $H^1(G,H^1(K,U(1)))$.)

Both of these well-definedness statements are a consequence
of the fact that, as noted in e.g. \cite[equ'n (42)]{Vafa:1986wx},
for the genus-one discrete torsion phases
\begin{equation}
\epsilon_{\rm dt}(g,h) \: = \: \frac{ \omega(g,h) }{ \omega(h,g) },
\end{equation}
(for $g$, $h$ commuting) one has
\begin{equation}  \label{eq:product}
\epsilon_{\rm dt}(x,y) \, \epsilon_{\rm dt}(x,z) \: = \:
 \epsilon_{\rm dt}(x,yz).
\end{equation}

To demonstrate that $B(\pi(h),z)$ is independent of $g$, one rearranges
\begin{equation}
\epsilon_{\rm dt}(h,g) \, \epsilon_{\rm dt}(h,z) \: = \:
 \epsilon_{\rm dt}(h,gz),
\end{equation}
to quickly find that
\begin{equation}
\frac{ \omega(gz,h) }{ \omega(h,gz) }
\frac{ \omega(h,g) }{ \omega(g,h) }
\: = \:
\frac{ \omega(z,h) }{ \omega(h,z) }.
\end{equation}
The left-hand side is equation~(\ref{eq:B}) for $B$, and the right-hand side
is the restriction of $B$ to $g=1$.  Since they are equal, we see that,
as advertised, $B$ is independent of $g$, and in particular, we can write
\begin{equation}
B(\pi(h),z) \: = \: \frac{\omega(z,h) }{ \omega(h,z) } \: = \:
\epsilon_{\rm dt}(z,h).
\end{equation}

In a similar fashion, one can show that so long as the restriction of
$\omega$ to $H^2(K,U(1))$ is trivial, $B$ only depends on $\pi(h)$, not $h$.
Specifically, suppose that $h' = h k$ for $k \in K$.
Then, using the identity~(\ref{eq:product}),
we have that
\begin{equation}
\epsilon_{\rm dt}(z,h') \: = \:
\epsilon_{\rm dt}(z,h k) \: = \: 
\epsilon_{\rm dt}(z,h) \epsilon_{\rm dt}(z,k),
\end{equation}
but since $z$, $k$ are both in $K$, so long as the restriction of
$\omega$ to $H^2(K,U(1))$ is trivial, $\epsilon(z,k) = 1$, hence
\begin{equation}
\epsilon_{\rm dt}(z,h') \: = \: \epsilon_{\rm dt}(z,h),
\end{equation}
and so $B$ depends only upon $\pi(h)$, not $h$.

We can also make contact with another expression for a phase derived
from discrete torsion.
As the expression~(\ref{eq:B}) for $B$
is independent of $g$, without loss of generality
we can take
$g = h^{-1}$, then from~(\ref{eq:B}),
we find another expression for $B$:
\begin{equation}
B(\pi(h),z) \: = \:
\frac{ \omega( h^{-1} z, h) }{ \omega(h,h^{-1} z) }
\frac{ \omega(h,h^{-1}) }{ \omega(h^{-1}, h) }.
\end{equation}
Now,
\begin{equation}
\frac{ \omega(h,h^{-1}) }{ \omega(h^{-1}, h) } \: = \: 1,
\end{equation}
and
\begin{equation}
\beta(\pi(h),\omega)(z) \: = \: 
\frac{ \omega(h, h^{-1} z) }{ \omega(h^{-1} z, h) }
\end{equation}
was the phase factor defined by discrete torsion $\omega$
in \cite[appendix C]{Robbins:2021ylj}, so we see that
\begin{equation}
B(\pi(h),z) \: = \: \frac{1}{\beta(\pi(h),\omega)(z)}.
\end{equation}
In \cite{Robbins:2021ylj} this phase factor was computed from the
mathematics of the Lyndon-Hochschild-Serre spectral sequence,
whereas here we have seen a simple physical realization.

Thus, we see that so long as the restriction of discrete torsion to
$H^2(K,U(1))$ is trivial, discrete torsion defines a quantum symmetry
$B \in H^1(G,H^1(K,U(1))$, given by $B = 1/\beta(\omega)$.
that $K$ is central in $\Gamma$.

Conversely, if $d_2 B = 1$, then we can find a corresponding element of
discrete torsion, as follows.  In this case, there exists a two-cochain
$\lambda$ such that
\begin{equation}
B\left(g_1, s_2 s_3 s_{23}^{-1} \right)
\: = \:
\frac{ \lambda(g_2, g_3) }{ \lambda(g_1 g_2, g_3) }
\frac{ \lambda(g_1, g_2 g_3) }{ \lambda(g_1, g_2) }.
\end{equation}
Then, define $\omega$ by 
\begin{equation}
\omega(s_1 k_1, s_2 k_2) \: = \:
\frac{ \lambda(g_1, g_2) }{
B\left( g_1, k_2 \right) }.
\end{equation}

Now. only some elements of $H^1(G,H^1(K,U(1)))$ can be written as
$1/\beta(\omega)$ for some $\omega$ in $H^2(\Gamma,U(1))$.
In particular, in this paper we are interested in $B$ in
$H^1(G,H^1(K,U(1)))$ whose image in $H^3(G,U(1))$ is nontrivial,
and those $B$ are not in the image of $\omega$, as follows from
the exact sequence~(\ref{eq:shortexact1}),
which was reviewed in greater detail in \cite{Robbins:2021ylj}.

For reference elsewhere, 
from the sequence~(\ref{eq:shortexact1}),
for $B \in H^1(G, H^1(K,U(1)))$ such that
$d_2 B \neq 1$ (and so are not determined by discrete torsion),
although $d_2 B$ is a nontrivial element of $H^3(G,U(1))$,
we will show in appendix~\ref{app:triv-pullback} that
$\pi^* (d_2 B)$ is trivial in $H^3(\Gamma,U(1))$ -- at least,
trivial in cohomology, though not necessarily identically $1$.

In passing, note that for $B$ determined by
discrete torsion $\omega \in H^2(\Gamma, U(1))$,
as $\epsilon_{\rm dt}(g,h) = \epsilon_{\rm dt}(h,g)^{-1}$, 
to be consistent, relation~(\ref{eq:quantsymm:basicreln})
also requires
\begin{equation}
{\scriptstyle h} \square_{gz} \: = \:
B(\pi(h),z)^{-1} \left(
{\scriptstyle h} \square_g \right),
\end{equation}
exactly as we saw is required for modular invariance in 
section~\ref{sect:basics}.

\subsection{General case}

As noted above, not all $B \in H^1(G, H^1(K,U(1)))$ are determined
by discrete torsion in $H^2(\Gamma,U(1))$.  However, we would like an
analogous description, in order to e.g. make sense of D-branes in these
orbifolds.  In this section we will discuss how one can compute
cochains $\omega$ (not cocycles) whose corresponding phases
\begin{equation}
\epsilon(g,h) \: = \: \frac{ \omega(g,h) }{ \omega(h,g) }
\end{equation}
encode $B$.

To that end, first pick a section $s: G \rightarrow \Gamma$,
and note that every $g \in \Gamma$ can be uniquely written in the
form $g = s(\pi(g)) k_g$, where $k_g$ is in (the image of) $K$,
and $\pi: \Gamma \rightarrow G$.  Then, as previously discussed, any
given twisted sector
\begin{equation}
{\scriptstyle g} \square_h \: = \:
{\scriptstyle s(\pi(g)) k_g} \square_{ s(\pi(h)) k_h }
\end{equation}
is proportional to a sector without the factors of $k$:
\begin{equation}  \label{eq:epsilon-diff-sectors}
{\scriptstyle s(\pi(g)) k_g} \square_{ s(\pi(h)) k_h }
\: = \: 
\frac{ B( \pi(h), k_g ) }{ B (\pi(g), k_h ) }
\left(
{\scriptstyle s(\pi(g)) } \square_{ s(\pi(h)) }
\right).
\end{equation}
Since, aside from the quantum symmetry, $K$ otherwise acts trivially,
we could equivalently say that we multiply the
\begin{equation}
{\scriptstyle g} \square_h
\end{equation}
by a phase
\begin{equation} \label{eq:epsdef:genl}
\epsilon(g,h) \: \equiv \:
\frac{ 
B\left( \pi(h), s(\pi(g))^{-1} g \right)
}{
B\left( \pi(g), s(\pi(h))^{-1} h \right)
}.
\end{equation}
(We suspect, but have not carefully checked, that this is
equivalent to including the coupling \cite[equ'n (2.31)]{Tachikawa:2017gyf}.)

As a consistency check, it is straightforward to show that, for example,
\begin{equation}
\frac{ 
{\scriptstyle gz} \square_h 
}{
{\scriptstyle g} \square_h
}
\: = \:
\frac{ \epsilon(gz,h) }{ \epsilon(g,h) }
\: = \: B(\pi(h), z),
\end{equation}
using the fact that $B$ is a homomorphism~(\ref{eq:bhom1}), (\ref{eq:bhom2}).
Note that although a particular realization in terms of phases $\epsilon$
depends upon the choice of section $s$, the ratio of phase factors
is independent of the choice of $s$.

The idea of multiplying a given sector by a phase factor is clearly
analogous to that in discrete torsion, and obeys similar constraints.
For example,
it is also straightforward, using the fact that $B$ is a 
homomorphism~(\ref{eq:bhom1}),
to demonstrate that
\begin{enumerate}
\item $\epsilon(g,g) = 1$,
\item $\epsilon(g,h) = \epsilon(h,g)^{-1}$,
\item $\epsilon(g,h_1 h_2) = \epsilon(g,h_1) 
\epsilon(g,h_2) \,
  \left( (d_2 B)(\pi(g), \pi(h_1), \pi(h_2) \right)^{-1}$,
\end{enumerate}
for the phases $\epsilon$ defined in~(\ref{eq:epsdef:genl}),
and where \cite[section 8]{hochschild}
\begin{equation}
(d_2 B)(\overline{g}_1, \overline{g}_2, \overline{g}_3)
\: = \:
B\left( \overline{g}_1, s(\overline{g}_2) s(\overline{g}_3) s(\overline{g}_2
\overline{g}_3)^{-1} \right)
\end{equation}
for $\overline{g}_{1,2,3} \in G$, and where $d_2$ is the same map that
appeared in the short exact sequence~(\ref{eq:shortexact1}).
In the special case that $d_2 B = 1$, 
these are the conditions that were previously derived on
phases in \cite[equ'ns (42)-(44)]{Vafa:1986wx} to guarantee modular invariance,
in the sense that
\begin{equation} \label{eq:modinv}
\epsilon(g^a h^b, g^c h^d) \: = \: \epsilon(g,h)
\end{equation}
for
\begin{equation}
\left[ \begin{array}{cc} a & b \\ c & d \end{array} \right] \: \in \:
SL(2,{\mathbb Z}).
\end{equation}
In more general cases, $d_2 B$ acts as an obstruction to this identity.

Now, despite this obstruction, this theory can nevertheless still be
modular invariant.  To see this,
the reader should note that the relation between
the identity~(\ref{eq:modinv}) and modular invariance is slightly subtle when
$K$ acts trivially.  In such a case, different sectors from possibly
different $SL(2,{\mathbb Z})$ orbits can contribute to the same
effective orbifold sector. 
In particular, one can show that the partition function is modular invariant
for any $B$, although for $d_2 B \neq 1$ the phases $\epsilon$
themselves are not $SL(2,{\mathbb Z})$-invariant.

We can see this as follows.  For $\overline{g} = \pi(g), \overline{h}
= \pi(h)$ the images in $G$ of a 
a commuting pair of elements $g, h \in \Gamma$,
define
\begin{equation}
\overline{\epsilon}(\overline{g}, \overline{h}) \: = \:
\sum_{k_1, k_2 \in K} \epsilon\left( s(\overline{g}) k_1,
s(\overline{h}) k_2 \right),
\end{equation}
where we assume that $K$ is central in $\Gamma$.
In particular, $\overline{\epsilon}$ is the sum of the contributions to
an effective-$G$-orbifold genus one sector defined by the commuting pair
$\overline{g}, \overline{h}$.
Then, we claim that
for any
\begin{equation}
\left[ \begin{array}{cc}
a & b \\
c & d \end{array} \right] \: \in \: SL(2,{\mathbb Z}),
\end{equation}
we have
\begin{equation}
\overline{\epsilon}\left( \overline{g}^a \overline{h}^b,
\overline{g}^c \overline{h}^d \right)
\: = \:
\overline{\epsilon}( \overline{g}, \overline{h} ).
\end{equation}
To see this, first note that
\begin{eqnarray}
\overline{\epsilon}(\overline{g}, \overline{h})
& = &
\sum_{k_1, k_2} \epsilon\left( s(\overline{g}) k_1,
s(\overline{h}) k_2 \right),
\\
& = &
\sum_{k_1, k_2} \frac{
B\left( \overline{h}, k_1 \right)
}{
B\left( \overline{g}, k_2 \right)
},
\end{eqnarray}
then we compute
\begin{eqnarray}
\overline{\epsilon}\left( \overline{g}^a \overline{h}^b,
\overline{g}^c \overline{h}^d \right)
& = &
\sum_{k_1, k_2} \frac{
B\left( \overline{g}^c \overline{h}^d, k_1 \right)
}{
B\left( \overline{g}^a \overline{h}^b, k_2 \right)
},
\\
& = &
\sum_{k_1, k_2} \frac{
B( \overline{g}, k_1)^c \, B(\overline{h}, k_1)^d
}{
B(\overline{g}, k_2)^a \, B(\overline{h}, k_2)^b
},
\\
& = &
\sum_{k_1, k_2} \frac{
B\left( \overline{h}, k_1^d k_2^{-b} \right)
}{
B\left( \overline{g}, k_2^a k_2^{-c} \right)
}.
\end{eqnarray}
Finally, note that we can perform a change of basis, defining
\begin{equation}
\ell_1 \: = \: k_1^d k_2^{-b}, \: \: \:
\ell_2 \: = \: k_2^a k_1^{-c},
\end{equation}
which can be inverted as 
\begin{equation}
k_1 \: = \: \ell_1^a \ell_2^b,
\: \: \:
k_2 \: = \: \ell_1^a \ell_2^d.
\end{equation}
In the new basis, we can write
\begin{eqnarray}
\overline{\epsilon}\left( \overline{g}^a \overline{h}^b,
\overline{g}^c \overline{h}^d \right)
& = &
\sum_{\ell_1, \ell_2} \frac{
B\left( \overline{h}, \ell_1 \right)
}{
B\left( \overline{g}, \ell_2 \right)
},
\\
& = &
\overline{\epsilon}(\overline{g}, \overline{h} ),
\end{eqnarray}
demonstrating that the theory defined by any $B$, not just those
with $d_2 B = 1$, is modular invariant.

Thus, theories with quantum symmetries such that $d_2 B \neq 1$ are
modular-invariant, but the reasons are subtle.  Modular invariance
acts on both the $G$ twisted sectors as well as the
$\Gamma$-twisted sectors.  However, as
\begin{equation}
(\pi(g^a h^b), \pi(g^c h^d) ) \: = \:
(\pi(g)^a \pi(h)^b, \pi(g)^c, \pi(h)^d),
\end{equation}
a modular transformation in $\Gamma$ will never slip between modular
orbits in $G$; at most, it will exchange fibers over the same modular orbit
of $G$.  The effect of the quantum symmetry is to modify the
coefficients of the $G$-twisted sector contributions, and although
$\Gamma$-modular-invariance is broken if $d_2 B \neq 1$,
$G$-modular-invariance is not, as the effect of summing up the
contributions from the action of $K$ can be arranged consistently
in modular orbits of $G$.

To summarize, so far we have found that for
 these orbifolds with quantum symmetries,
the phases~(\ref{eq:epsdef:genl}) define modular-invariant phases,
and so we see that, at least in orbifolds of this form,
with a trivially-acting subgroup, discrete torsion is not the only possible
modular-invariant phase one can add.  (See also
\cite{Sharpe:2003cs}
for a discussion of momentum/winding lattice phases in orbifolds
of non-simply-connected spaces, as another analogous example.)

As another consistency check, in the special case that
$B$ is determined by discrete torsion $\omega$, that
$B = 1/\beta(\omega)$, from~(\ref{eq:epsilon-diff-sectors}), one expects
\begin{equation}
\epsilon(g,h) \: = \: 
\frac{ \epsilon_{\rm dt}(g,h) }{ \epsilon_{\rm dt}( s(\pi(g)),
s(\pi(h)) ) },
\end{equation}
where $\epsilon_{\rm dt}$ denotes the ratio of group 2-cocycles $\omega$
defining the genus-one phase arising in discrete torsion.
Using multiplicative properties of $\epsilon_{\rm dt}$, as well as the
assumption that its restriction to $K$ is trivial, it is straightforward
to compute that
\begin{eqnarray}
\epsilon(g,h) & = &
\frac{ B(\pi(h), s(\pi(g))^{-1} g) }{ B(\pi(g), s(\pi(h))^{-1} h) }
\: = \:
\frac{\epsilon_{\rm dt}(s(\pi(g))^{-1}g, h) }{ 
\epsilon_{\rm dt}(s(\pi(h))^{-1}h, g) },
\\
& = &
\epsilon_{\rm dt}( s(\pi(g))^{-1}, h) \, \epsilon_{\rm dt}(g,h)
\, \epsilon_{\rm dt}(g, s(\pi(h))^{-1}h) 
\nonumber \\
& & \hspace*{1in} \cdot
\epsilon_{\rm dt}( s(\pi(g))^{-1}g, s(\pi(h))^{-1} h)^{-1},
\\
& = &
\epsilon_{\rm dt}(g,h) \, \epsilon_{\rm dt}(s(\pi(g))^{-1}, h)
\, \epsilon_{\rm dt}(s(\pi(g)), s(\pi(h))^{-1} h),
\\
& = &
\epsilon_{\rm dt}(g,h) 
\frac{ \epsilon_{\rm dt}( s(\pi(g)), s(\pi(h))^{-1} h) }{
\epsilon_{\rm dt}(s(\pi(g)), h) },
\\
& = &
\frac{ \epsilon_{\rm dt}(g,h) }{ 
\epsilon_{\rm dt}(s(\pi(g)), s(\pi(h)) )
},
\end{eqnarray}
as expected.

So far, given a general quantum symmetry, we have derived 
genus-one phase factors $\epsilon(g,h)$ that encode the quantum symmetry
on twisted sectors.  To make sense of D-branes, we need a bit more.
Specifically, we need an analogue of group cocycles.  Now, in general,
these phase factors cannot be described by group cocycles, as that
description only reproduces those quantum symmetries arising from discrete
torsion.  However, we can certainly construct group co{\it chains}
representing the phases $\epsilon(g,h)$.

Specifically, recall a group cochain $\omega$ is a map
$\Gamma \times \Gamma \rightarrow U(1)$, not necessarily obeying
a group cocycle condition.  We can construct a set of group
cochains $\omega(g,h)$ such that
\begin{equation}
\epsilon(g,h) \: = \: \frac{ \omega(g,h) }{ \omega(h,g) }
\end{equation}
as follows.  First, pick an ordering on the elements of the
group $\Gamma$.  (Different orderings will result in different cochains,
but the same phase $\epsilon$.)  Then, define
\begin{equation}
\omega(g,h) \: = \: \left\{ \begin{array}{cl}
\epsilon(g,h) & g \leq h, \\
1 & {\rm else}.
\end{array} \right.
\end{equation}
This construction is certainly not unique, but does demonstrate the
existence of cochains $\omega$ that reproduce the phases $\epsilon$.

Another, potentially more useful, description of the cochain is as 
follows.  Pick a section $s: G \rightarrow \Gamma$, with respect
to which any $g \in \Gamma$ can be uniquely written
$g = k s(\pi(g))$, for $k \in K$, then define
\begin{equation}  \label{eq:good-cochain-defn}
\omega(k_1 s(\pi(g_1)), k_2 s(\pi(g_2)) ) \: \equiv \:
\frac{1}{B(\pi(g_1), k_2)}.
\end{equation}
Then,
\begin{eqnarray}
\epsilon(z,h) & = &
\frac{ \omega(z,h) }{ \omega(h,z) }
\: = \:
\frac{ B(\pi(h),z) }{ B(1, s(\pi(h))^{-1} h) },
\\
& = & B(\pi(h),z),
\end{eqnarray}
using the fact that $B(1,s(\pi(h))^{-1}h) = 1$.

Let us check when the cochain $\omega$ defined above is a cocycle.
Let $g_{1,2,3}$ be three elements of $\gamma$, each written as
$g_i = k_i s(\pi(g_i))$.  Then,
\begin{eqnarray}
(d \omega)(g_1, g_2, g_3) & = &
\frac{ \omega(g_2, g_3) \omega(g_1, g_2 g_3) }{
\omega(g_1 g_2, g_3) \omega(g_1, g_2) },
\\
& = &
\frac{ B(\pi(g_1 g_2), k_3) \, B(\pi(g_1), k_2) }{
B(\pi(g_2), k_3) \, B(\pi(g_1), k_2 k_3 s(\pi(g_2)) s(\pi(g_3))
s(\pi(g_2 g_3))^{-1} ) },
\\
& = &
B\left(\pi(g_1), s(\pi(g_2)) s(\pi(g_3)) s(\pi(g_2 g_3))^{-1} \right)^{-1}.
\label{eq:pullback-b-triv}
\end{eqnarray}
Now, from \cite[section 8]{hochschild},
\begin{equation}
\pi^* (d_2 B)(g_1, g_2, g_3) \: = \:
B\left(\pi(g_1), s(\pi(g_2)) s(\pi(g_3)) s(\pi(g_2 g_3))^{-1} \right),
\end{equation}
so we see that the cochain $\omega$ defined 
in~(\ref{eq:good-cochain-defn}) 
is a cocycle if and only if the pullback of $d_2 B$ is trivial
(not just in cohomology),
and hence $B$ is in the image of $\beta$, using the exact
sequence~(\ref{eq:shortexact1}).

In passing, it is straightforward to check that when $K$ is central,
$\pi^*(d_2 B)$ is a group 3-cocycle:
\begin{eqnarray}
(d(\pi^* d_2 B)) (g_1, g_2, g_3, g_4)
& = &
\frac{
(\pi^* d_2B)(g_2, g_3, g_4) }{ (\pi^* d_2 B)(g_1 g_2, g_3, g_4) }
\frac{ (\pi^* d_2B)(g_1, g_2 g_3, g_4) }{ (\pi^* d_2B)(g_1, g_2, g_3 g_4) }
(\pi^* d_2B)(g_1, g_2, g_3),
\nonumber \\
& = &
\frac{ B(\pi(g_2), s_3 s_4 s_{34}^{-1} ) }{ B( \pi(g_1 g_2), s_3 s_4 s_{34}^{-1} ) }
\frac{ B(\pi(g_1) s_{23} s_4 s_{234}^{-1}) }{
B(\pi(g_1), s_2 s_{34} s_{234}^{-1} ) }
B(\pi(g_1), s_2 s_3 s_{23}^{-1} ),
\nonumber \\
& = & B( \pi(g_1), 1) \: = \: 1,
\end{eqnarray}
where for example $s_a = s(\pi(g_a))$.

Furthermore, from equation~(\ref{eq:pullback-b-triv}) above, we see that
$\pi^* d_2 B$ is always trivial in cohomology in $H^3(\Gamma,U(1))$, even if
$d_2 B$ is not trivial in $H^3(G,U(1))$, at least for $K$ central.
In appendix~\ref{app:triv-pullback} 
we will demonstrate that even for non-central
extensions, the pullback $\pi^* d_2 B$ is trivial in cohomology.

\section{General conjecture for decomposition}
\label{sect:conj}

In this section we will relate orbifolds with quantum symmetries to
simpler orbifolds, via a generalization of
decomposition 
\cite{Hellerman:2006zs,Sharpe:2019ddn,Tanizaki:2019rbk,Robbins:2021ylj}.
(Related observations concerning module categories
have also been made in the mathematics literature on
fusion categories, see e.g. \cite[examples 7.4.10, 9.7.2]{egno}.)

Previously, in \cite{Hellerman:2006zs,Robbins:2021ylj}, it was
possible to give a derivation of the form of decomposition by
studying the structure of D-branes.  In \cite{Hellerman:2006zs},
this meant studying group actions on honest representations,
and in \cite{Robbins:2021ylj}, this meant studying group actions on
projective representations.  Unfortunately, we do not yet
have a complete understanding of D-branes in orbifolds with
quantum symmetries -- see instead appendix~\ref{app:open} for partial results.
As a result, we cannot give a derivation of the same form.
Nevertheless, previous results do strongly constrain possible answers,
so we are able to give a fairly full accounting of possibilities,
which we check in numerous examples later in this paper.
In our upcoming work \cite{rsv} we will apply these ideas to resolutions
of anomalies.

Let $\Gamma$ be a central extension of $G$ by (abelian) $K$,
and suppose one has a quantum symmetry defined by
$B \in H^1(G,H^1(K,U(1)))$.
We begin with the case that the orbifold can be described solely
in terms of a quantum symmetry $B$.  (Many choices
of discrete torsion can be folded into $B$, as we have discussed,
but not all -- we will discuss more general cases shortly.)
This case is tightly constrained by our
previous results  \cite{Robbins:2021ylj}.  Specifically,
there we discussed discrete torsion $\omega$ whose restriction to $K$ was
trivial.  We argued there that decomposition in 
the $\Gamma$ orbifold with discrete torsion was of the form
\begin{equation}
{\rm QFT}\left( [X/\Gamma]_{\omega} \right) \: = \:
{\rm QFT}\left(
\left[ \frac{ X \times \widehat{ {\rm Coker}\, \beta(\omega) } }{
{\rm Ker} \, \beta(\omega) }
\right]_{\hat{\omega}} \right)
\end{equation}
for suitable choices of discrete torsion described in \cite{Robbins:2021ylj}. 
We can identify $\beta(\omega)^{-1}$ with a choice of
quantum symmetry, and although not every quantum symmetry can be
so described, this result does provide a large set of special cases.

In order to reproduce
that result,
we conjecture in the present case that the $\Gamma$ orbifold,
which we denote $[X/\Gamma]_B$, decomposition takes the form
\begin{equation}  \label{eq:predict}
{\rm QFT}\left( [X/\Gamma]_B \right) \: = \:
{\rm QFT}\left( \left[ \frac{X \times \widehat{{\rm Coker}\, B} }{
{\rm Ker}\, B} \right]_{\hat{\omega}} \right),
\end{equation}
with discrete torsion $\hat{\omega}$ determined just as in
decomposition 
\cite{Hellerman:2006zs,Sharpe:2019ddn,Tanizaki:2019rbk,Robbins:2021ylj},
meaning that for any irreducible representation $\rho$ of
Coker $B$, we take $\hat{\omega}(\rho)$ to be the restriction to
Ker $B$ of the image of
the extension class of $\Gamma$, an element of
$H^2(G,K)$, in $H^2(G,U(1))$.
(We are assuming, as elsewhere, that $K$ is central.)

As a special case, when $B$ is trivial,
Ker $B = G$ and
Coker $B = K$, so the prediction reduces to
\begin{equation}
{\rm QFT}\left( [X/\Gamma] \right) \: = \:
{\rm QFT}\left( \left[ \frac{X \times \hat{K} }{ G } \right]_{\hat{\omega}}
\right),
\end{equation}
which matches the prediction of decomposition 
\cite{Hellerman:2006zs,Sharpe:2019ddn,Tanizaki:2019rbk,Robbins:2021ylj}
in this case.

We will explicitly verify the
conjecture above in examples, by comparing the predictions to genus-one
partition functions, in sections~\ref{sect:ex:cyclic},
\ref{sect:ex:z2z2:z2z4:wodt}, \ref{sect:ex:z2z2:d4:wodt},
\ref{sect:ex:z2z4:z4z4:wodt}, \ref{sect:ex:d4:z4z4:wodt},
\ref{sect:ex:d4z2z2:z2z2:wodt}.

Next, let us generalize this story slightly, by explicitly allowing for
discrete torsion $\omega \in H^2(\Gamma,U(1))$ in the
$\Gamma$ orbifold, whose restriction to $K$ we will assume trivial,
in addition to the quantum symmetry $B$.  
Following the pattern of
\cite{Robbins:2021ylj}, and mindful of special cases in which
$B = 1/\beta(\omega')$ for some $\omega'$, 
we break the analysis up into cases
using the maps $\iota$, $\pi$ in
\begin{equation}
1 \: \longrightarrow \: K \: \stackrel{\iota}{\longrightarrow} \:
\Gamma \: \stackrel{\pi}{\longrightarrow} \: G \: \longrightarrow \: 1,
\end{equation}
as follows:
\begin{enumerate}
\item Suppose that $\iota^* \omega \neq 0$ as an element of
$H^2(K,U(1))$.  We do not have a conjecture for this case,
though we will compute an example in 
section~\ref{sect:ex:d4z2z2:z2z2:dta}.
\item Suppose that $\iota^* \omega = 0$ and
$\beta(\omega) \neq 0$, where $\beta(\omega) \in H^1(G,H^1(K,U(1)))$.
Then,
\begin{equation}
{\rm QFT}\left( [X/\Gamma]_{B,\omega} \right) \: = \:
{\rm QFT}\left( \left[ \frac{X \times \widehat{{\rm Coker}\, (B/ \beta(\omega))} }{
{\rm Ker}\, (B/ \beta(\omega))} \right]_{\hat{\omega}} \right),
\end{equation}
with discrete torsion $\hat{\omega}$ defined as in
\cite{Robbins:2021ylj} and restricted to Ker $B / \beta(\omega)$.

This result is more or less uniquely determined by previous results,
for reasons already described.
For example, suppose that $B$ itself is determined by
some discrete torsion $\omega'$.  (This can happen, though will not
always be the case.)  Then the quantity
\begin{equation}
B/\beta(\omega) \: = \: 1 / ( \beta(\omega') \beta(\omega) )
\: = \:
1 / \beta(\omega + \omega'),
\end{equation}
and this prescription reduces to a special case of
\cite{Robbins:2021ylj}.

We will check this prediction explicitly in examples by comparing to
genus-one partition functions, in 
sections~\ref{sect:ex:z2z2:d4:dt}, \ref{sect:ex:z2z4:z4z4:dt},
\ref{sect:ex:d4z2z2:z2z2:dtb}.

\item Suppose that $\iota^* \omega = 0$ and 
$\omega = \pi^* \overline{\omega}$ for $\overline{\omega} \in
H^2(G,U(1))$.  Then,
\begin{equation}
{\rm QFT}\left( [X/\Gamma]_{B,\omega} \right) \: = \:
{\rm QFT}\left( \left[ \frac{X \times \widehat{{\rm Coker}\, B} }{
{\rm Ker}\, B} \right]_{\overline{\omega} + \hat{\omega}_0} \right),
\end{equation}
where $\hat{\omega}_0$ is the discrete torsion predicted in the case
that $\omega$ is trivial. 

This case is also more or less uniquely determined by consistency with
previous results.  
We will check this prediction explicitly in examples by comparing to
genus-one partition functions, in sections~\ref{sect:ex:z2z2:z2z4:dt},
\ref{sect:ex:d4:z4z4:dt}.
\end{enumerate}

We note that just as in our earlier work \cite{Robbins:2021ylj},
to improve readability we unfortunately found it useful to mix
additive and multiplicative notation.

Finally, in passing we note that a relation to theories realized
as boundaries of higher-dimensional theories is implicit in
\cite[prop. 4.243]{Muller:2020phm}.

\section{Examples}
\label{sect:examples}

\subsection{Extension of cyclic groups to larger cyclic groups}
\label{sect:ex:cyclic}

\subsubsection{Extension of ${\mathbb Z}_2$ to ${\mathbb Z}_4$}

In this section we will describe an example from \cite{Robbins:2019zdb} in
the current language.  In this example, one starts with a
$G = {\mathbb Z}_2$ orbifold, 
and extend
$G$ by $K = {\mathbb Z}_2$ (with trivial action on the space, but
nontrivial action on $G$ twist fields) to 
$\Gamma = {\mathbb Z}_4$:
\begin{equation}
1 \: \longrightarrow \: {\mathbb Z}_2 \: \longrightarrow \:
{\mathbb Z}_4 \: \longrightarrow \: {\mathbb Z}_2 \:
\longrightarrow \: 1.
\end{equation}
(This example was also considered from a different perspective in
\cite[section 5.1.1]{Robbins:2021lry}.)

In this case, $H^2(\Gamma,U(1)) = 0$, so we see that
\begin{equation}
H^1(G, H^1(K,U(1))) \: = \: {\rm Hom}( {\mathbb Z}_2,
H^1( {\mathbb Z}_2, U(1)) ) \: = \: {\rm Hom}(
{\mathbb Z}_2, {\mathbb Z}_2 )
\: = \: {\mathbb Z}_2
\end{equation}
injects into $H^3(G,U(1))$, from~(\ref{eq:shortexact1}).

In any event, there are two choices of quantum symmetry
$B \in H^1(G,H^1(K,U(1)))$, and we describe both cases below.
First, consider the trivial case that $B = 0$.
In this case, Ker $B = G = {\mathbb Z}_2$ and Coker $B = K = {\mathbb Z}_2$,
so from equation~(\ref{eq:predict}) we predict that
\begin{equation}
{\rm QFT}\left( [X / {\mathbb Z}_4] \right) \: = \: 
{\rm QFT}\left( \coprod_2 [X/{\mathbb Z}_2] \right).
\end{equation}

Next, consider the case that $B$ is the nontrivial element of
$H^1(G,H^1(K,U(1))) = {\mathbb Z}_2$.
In this case, we have trivially Ker $B = {\rm Coker}\, B = 0$, 
so from equation~(\ref{eq:predict}) we predict that
\begin{equation}
{\rm QFT}\left( [X / {\mathbb Z}_4]_B \right) \: = \: 
{\rm QFT}(X).
\end{equation}

Now, let us compare to physics.  The case that $B$ is trivial
corresponds to ordinary decomposition, described in
\cite{Hellerman:2006zs}, and so will not be reviewed here.
Let us turn to the case that $B$ is nontrivial.
First, from the form of $B
\in H^1(G, H^1(K,U(1)))$ above,
there is a quantum symmetry under which
$x^2$ (the image of the generator of $K$) acts on the twist field associated
with $x$ by $(-1)$, so that the genus-one
${\mathbb Z}_4$ orbifold partition functions obey
\begin{equation}  \label{eq:ex:z4:quantsymm}
{\scriptstyle x^2} \square_{ x^n } \: = \: (-)^n \,
\left( {\scriptstyle 1} \square_{x^n} \right) ,
\end{equation}
and more generally
\begin{equation}
Z_{i,j} \: = \:(-)^i Z_{i,j-2} \: = \: 
(-)^j Z_{i-2,j},
\end{equation}
where $i, j \in \{0, \cdots, 3\}$.
In particular,
\begin{equation}
Z_{i,2} \: = \: (-)^i Z_{i,0}
\end{equation}
is equivalent to~(\ref{eq:ex:z4:quantsymm}).

Taking into account these relationships and computing the entire
genus-one partition function, we find
\begin{eqnarray}
Z\left( [X/{\mathbb Z}_4]_B \right) 
& = &
\frac{1}{| {\mathbb Z}_4|} \sum_{gh = hg} Z_{g,h},
\\
& = & \frac{1}{4} \left( Z_{0,0} + Z_{0,2} + Z_{2,0} + Z_{2,2} \right),
\\
& = & Z_{0,0} \: = \: Z(X),
\end{eqnarray}
matching the prediction.

\subsubsection{Extension of ${\mathbb Z}_2$ by ${\mathbb Z}_{k}$}

This example is a variation on the previous one.  Instead of extending
the effectively-acting ${\mathbb Z}_2$ by another ${\mathbb Z}_2$,
consider extending it by ${\mathbb Z}_{k}$ (that acts trivially on $X$)
for $k \geq 2$:
\begin{equation}
1 \: \longrightarrow \: {\mathbb Z}_{k} \: \longrightarrow \:
{\mathbb Z}_{2k} \: \longrightarrow \: {\mathbb Z}_2 \:
\longrightarrow \: 1.
\end{equation}
The possible values of the quantum symmetry $B$ are classified by
\begin{equation}
H^1(G,H^1(K,U(1))) \: = \: {\rm Hom}({\mathbb Z}_2,{\mathbb Z}_k) \: = \:
\left\{ \begin{array}{cl}
1 & k \mbox{ odd}, \\
{\mathbb Z}_2 & k \mbox{ even}.
\end{array} \right.
\end{equation}

In the case that the quantum symmetry is trivial,
Ker $B = {\mathbb Z}_2$, Coker $B = K = {\mathbb Z}_k$, and from ordinary
decomposition \cite{Hellerman:2006zs},
\begin{equation}
{\rm QFT}\left( [X/{\mathbb Z}_{2k}] \right)
\: = \: 
{\rm QFT}\left( \coprod_{k} [X/{\mathbb Z}_2] \right).
\end{equation}

Next, consider the case that $B$ is nontrivial -- for which we also assume
that $k$ is even.
In this case, 
Ker $B = 0$,
and Coker $B = {\mathbb Z}_{k/2}$.  
Thus, from equation~(\ref{eq:predict}), we predict that
\begin{equation} \label{eq:z2k:predict}
{\rm QFT}\left( [X/{\mathbb Z}_{2k} ]_B \right) \: = \:
{\rm QFT}\left( \coprod_{k/2} X \right).
\end{equation}

Now, let us check this prediction in physics.
We only consider the case that the quantum symmetry
$B$ is nontrivial (and also that $k$ is even),
as the case $B$ trivial corresponds to ordinary decomposition.
If we let $x$ denote the generator of ${\mathbb Z}_{2k}$,
so that $x^2$ generates the image of $K$,
then the quantum symmetry acts in the form
\begin{equation}
{\scriptstyle x^2} \square_x \: = \: - \left(
{\scriptstyle 1} \square_x \right),
\end{equation}
or more generally,
\begin{equation}
Z_{i,j} \: = \:(-)^i Z_{i,j-2} \: = \: 
(-)^j Z_{i-2,j},
\end{equation}
where $i, j \in \{0, \cdots, 2k-1\}$.
In this example, this is the concrete
meaning of $B$.

Taking into account these relationships and computing the entire
genus-one partition function, we find
\begin{eqnarray}
Z\left( [X/{\mathbb Z}_{2k}]_B \right) 
& = &
\frac{1}{| {\mathbb Z}_{2k} |} \sum_{gh = hg} Z_{g,h},
\\
& = &
\frac{1}{2k} \sum_{i,j = 0}^{k-1} Z_{2i,2j},
\\
& = & \frac{1}{2k} (k)^2 Z_{0,0} \: = \: \frac{k}{2} Z\left( X \right),
\\
& = & Z\left( \coprod_{k/2} X \right),
\end{eqnarray}
matching the prediction~(\ref{eq:z2k:predict}).

Another way to understand this problem is via decomposition 
\cite{Hellerman:2006zs,Sharpe:2019ddn,Tanizaki:2019rbk,Robbins:2021ylj}.
For $k$ even, as we have
assumed, the orbifold group ${\mathbb Z}_{2k}$ can be described as an
extension of ${\mathbb Z}_4$ by ${\mathbb Z}_{k/2}$:
\begin{equation}
1 \: \longrightarrow \: {\mathbb Z}_{k/2} \: \longrightarrow \:
{\mathbb Z}_{2k} \: \longrightarrow \: {\mathbb Z}_4 \: 
\longrightarrow \: 1,
\end{equation}
The ${\mathbb Z}_{k/2}$ acts trivially on both the space $X$ as
well as the twisted sectors -- the former because it is a subgroup of
the ${\mathbb Z}_k$ that acted trivially on $X$, the latter because the
action on twisted sectors was encoded in ${\mathbb Z}_2$ subgroup which
${\mathbb Z}_{k/2}$ has quotiented out.

Then, applying decomposition 
\cite{Hellerman:2006zs,Sharpe:2019ddn,Tanizaki:2019rbk,Robbins:2021ylj}, 
we have immediately that
\begin{equation}
{\rm QFT}\left( [X/{\mathbb Z}_{2k} ]_B \right) \: = \:
{\rm QFT}\left( \left[ \frac{X \times \hat{\mathbb Z}_{k/2} }{ {\mathbb Z}_4 }
\right]_B \right) \: = \:
{\rm QFT} \left( \coprod_{k/2} [ X / {\mathbb Z}_{4} ]_B \right),
\end{equation}
and as we know that
\begin{equation}
{\rm QFT}\left( [X/{\mathbb Z}_4]_B \right) \: = \: X,
\end{equation}
decomposition therefore tells us that
\begin{equation}
{\rm QFT}\left( [X/{\mathbb Z}_{2k}]_B \right) \: = \:
{\rm QFT}\left( \coprod_{k/2} X \right),
\end{equation}
in agreement with the prediction~(\ref{eq:z2k:predict}) and also with 
physics results in this case.

\subsubsection{Extension of ${\mathbb Z}_3$ to ${\mathbb Z}_9$}

In this section we start with a $G = {\mathbb Z}_3$ orbifold,
extend $G$ by $K = {\mathbb Z}_3$ to
$\Gamma = {\mathbb Z}_9$:
\begin{equation}
1 \: \longrightarrow \: {\mathbb Z}_3 \: \longrightarrow \:
{\mathbb Z}_9 \: \longrightarrow \: {\mathbb Z}_3 \: \longrightarrow \: 1.
\end{equation}

In this case,
\begin{equation}
B \: \in \: {\rm Hom}(G, H^1( K, U(1) ) \: = \: {\mathbb Z}_3.
\end{equation}
Our prediction (section~\ref{sect:conj}) falls into the following cases:
\begin{enumerate}
\item In the case $B$ is trivial, this reduces to ordinary decomposition,
for which
\begin{equation}
{\rm QFT}\left( [X/\Gamma] \right) \: = \:
{\rm QFT}\left( \left[ \frac{ X \times \hat{K} }{G} \right] \right)
\: = \:
{\rm QFT}\left( \coprod_3 [X/{\mathbb Z}_3] \right).
\end{equation}
\item In both of the cases that $B$ is nontrivial,
Ker $B = 0$ and Coker $B = 0$, so we predict (section~\ref{sect:conj}) that
\begin{equation}
{\rm QFT}\left( [X/\Gamma]_B \right) \: = \:
{\rm QFT}(X).
\end{equation}
\end{enumerate}

We can confirm these predictions explicitly at the level of partition 
functions.  The analysis for the case $B$ is trivial is standard,
and will not be repeated here.  Suppose $B$ is nontrivial.
\begin{itemize}
\item Consider genus-one $\Gamma$-twisted sectors which project
to 
\begin{equation}
{\scriptstyle 1} \square_1
\end{equation}
in the $G$ orbifold.  Phases are various powers of $B(1, z)$ for
$z \in K$, but for all $z$, $B(1,z) = 1$.  Thus, the sector above
appears with multiplicity $|K|^2 = 9$.
\item Consider, for example, genus-one $\Gamma$-twisted sectors which
project to the $G$-twisted sector
\begin{equation}
{\scriptstyle 1} \square_g
\end{equation}
for $g \neq 1$ an element of $G = {\mathbb Z}_3$.
The different $\Gamma$-twisted sectors which project to this sector
are weighted by different roots of unity, and the sum over such sectors
collapses to copies of sums over roots of unity, which vanishes.
Therefore, there are no net contributions to the
$\Gamma$-orbifold partition function from these sectors, and similarly
one can show that there are no net contributions from any sectors
that do not project to the trivial sector.
\end{itemize}
Summarizing, the genus-one partition function is given by
\begin{eqnarray}
Z\left( [X/\Gamma]_B \right) & = &
\frac{1}{\Gamma} \sum_{g, h} 
{\scriptstyle g} \square_h,
\\
& = & 
\frac{ |K|^2 }{ |\Gamma| } \left( 
{\scriptstyle 1} \square_1 \right)
\: = \: {\scriptstyle 1} \square_1,
\\
& = & Z(X),
\end{eqnarray}
confirming the prediction.

\subsection{Extension of ${\mathbb Z}_2 \times {\mathbb Z}_2$ to
${\mathbb Z}_2 \times {\mathbb Z}_4$}

In this section we begin with a
$G = {\mathbb Z}_2 \times {\mathbb Z}_2$ orbifold,
extended by $K = {\mathbb Z}_2$ to
$\Gamma = {\mathbb Z}_2 \times {\mathbb Z}_4$.
We write $G = \langle \overline{a}, \overline{b} \rangle$,
where $\overline{a}^2 = 1 = \overline{b}^2$, 
and $\Gamma = \langle a, b \rangle$, where $\overline{a} = a K$ and
$\overline{b} = b K$, with $K = \langle b^2 \rangle$.

Now, the quantum symmetry is defined by an element 
$B \in H^1(G, H^1(K,U(1)))$, which is defined by its action on 
the generators $\overline{a}$, $\overline{b}$.  We will systematically
study the orbifold $[X/\Gamma]$ for all choices of $B$, initially without
and later with discrete torsion, and compare with the
predictions of section~\ref{sect:conj}.

First, the possible values for $B$ are as follows\footnote{Here, and in many instances below, we are identifying $H^1(\Z_2,U(1))\cong\operatorname{Hom}(\Z_2,U(1))\cong\Z_2=\{+1,-1\}$.  So for the case at hand, with $K=\{1,b^2\}$, if we write $B(\overline{a})=-1$, we mean that
\begin{equation}
B(\overline{a})(1)=1,\qquad B(\overline{a})(b^2)=-1.
\end{equation}}:
\begin{enumerate}
\item $B(\overline{a}) = +1$, $B(\overline{b}) = +1$ (the trivial case),
\item $B(\overline{a}) = -1$, $B(\overline{b}) = +1$,
\item $B(\overline{a}) = +1$, $B(\overline{b}) = -1$,
\item $B(\overline{a}) = -1$, $B(\overline{b}) = -1$.
\end{enumerate}

We will first analyze orbifolds without additional discrete torsion
$\omega \in H^2(\Gamma,U(1))$, then separately study the effect
of adding $\omega \in H^2(\Gamma,U(1))$.

\subsubsection{Without discrete torsion}
\label{sect:ex:z2z2:z2z4:wodt}

For each possible $B$, we make a prediction for the structure of
the orbifold $[X/\Gamma]$ using the conjecture of section~\ref{sect:conj},
as follows:
\begin{enumerate}
\item First, consider the case that $B$ is trivial:
$B(\overline{a}) = +1$ and $B(\overline{b}) = +1$.
In this case, the $[X/\Gamma]$ orbifold can be understood via
decomposition 
\cite{Hellerman:2006zs,Sharpe:2019ddn,Tanizaki:2019rbk,Robbins:2021ylj}, which predicts
\begin{equation}
{\rm QFT}\left( [X/\Gamma] \right) \: = \:
{\rm QFT}\left( \coprod_2 [X/{\mathbb Z}_2 \times {\mathbb Z}_2] \right),
\end{equation}
a disjoint union of two copies of $[X/{\mathbb Z}_2 \times {\mathbb Z}_2]$.
For essentially the same reasons as \cite[section 6.1]{Robbins:2021ylj}, 
since the extension
is a pullback from
\begin{equation}
1 \: \longrightarrow \: {\mathbb Z}_2 \: \longrightarrow \:
{\mathbb Z}_4 \: \longrightarrow \: {\mathbb Z}_2 \: \longrightarrow \: 1,
\end{equation}
neither copy has discrete torsion.
\item Next, consider the case that
$B(\overline{a}) = -1$ and $B(\overline{b}) = +1$.
Here, Coker $B = 0$ and Ker $B = {\mathbb Z}_2 = \langle \overline{b} \rangle$,
so from section~\ref{sect:conj} we predict that
\begin{equation}
{\rm QFT}\left( [X/\Gamma]_B \right) \: = \:
{\rm QFT}\left( [X/{\mathbb Z}_2 = \langle \overline{b} \rangle] \right).
\end{equation}
\item Next, consider the case that
$B(\overline{a}) = +1$ and $B(\overline{b}) = -1$.
Here, Coker $B = 0$ and Ker $B = {\mathbb Z}_2 = \langle \overline{a} \rangle$,
so we predict that
\begin{equation}
{\rm QFT}\left(  [X/\Gamma]_B \right) \: = \:
{\rm QFT}\left( [X/{\mathbb Z}_2 = \langle \overline{a} \rangle ] \right).
\end{equation}
\item Finally, consider the case that
$B(\overline{a}) = -1$ and $B(\overline{b}) = -1$.
Here, Coker $B = 0$ and Ker $B = \langle \overline{a} \overline{b} \rangle$,
so we predict that
\begin{equation}
{\rm QFT}\left(  [X/\Gamma]_B \right) \: = \:
{\rm QFT}\left( [X/{\mathbb Z}_2 = \langle \overline{a} \overline{b} \rangle]
\right).
\end{equation}
\end{enumerate}

Finally, we check the predictions in each case by computing
genus-one partition functions.
\begin{enumerate}
\item First, consider the case that $B$ is trivial:
$B(\overline{a}) = +1$ and $B(\overline{b}) = +1$.
Essentially this case was discussed in
\cite[section 6.1]{Robbins:2021ylj}, where it was argued that the
genus-one partition function
\begin{equation}
Z\left( [X/\Gamma] \right) \: = \:
Z\left( \coprod_2 [X/ {\mathbb Z}_2 \times {\mathbb Z}_2]
 \right).
\end{equation}
\item Next, consider the case
$B(\overline{a}) = -1$ and $B(\overline{b}) = +1$.
The center $K$ acts nontrivially on twisted sectors twisted by $a$,
for example,
\begin{equation}
{\scriptstyle b^2} \square_a \: = \: - \left( 
{\scriptstyle 1} \square_a \right),
\: \: \:
{\scriptstyle b^2} \square_b \: = \: + \left(
{\scriptstyle 1} \square_b \right).
\end{equation}
Using relations of this form, it is straightforward to check that the
genus-one partition function obeys
\begin{eqnarray}
Z\left( [X/\Gamma]_B \right)
& = &
\frac{1}{|\Gamma|} \sum_{gh = hg} 
{\scriptstyle g} \square_h \, ,
\\
& = & \frac{4}{8} \left[ 
{\scriptstyle 1} \square_1 \: + \: 
{\scriptstyle 1} \square_{\overline{b}} \: + \:
{\scriptstyle \overline{b}} \square_1 \: + \:
{\scriptstyle \overline{b}} \square_{\overline{b}} \right],
\\
& = & Z \left( [X/{\mathbb Z}_2 = \langle \overline{b} \rangle ] \right),
\end{eqnarray}
matching the prediction.
\item Next, consider the case
$B(\overline{a}) = +1$ and $B(\overline{b}) = -1$.
Here, the center $K$ acts nontrivially on twisted sectors twisted by $b$,
for example,
\begin{equation}
{\scriptstyle b^2} \square_a \: = \: + \left( 
{\scriptstyle 1} \square_a \right),
\: \: \:
{\scriptstyle b^2} \square_b \: = \: - \left(
{\scriptstyle 1} \square_b \right).
\end{equation}
Using relations of this form, it is straightforward to check that the
genus-one partition function obeys
\begin{eqnarray}
Z\left( [X/\Gamma]_B \right)
& = &
\frac{1}{|\Gamma|} \sum_{gh = hg} 
{\scriptstyle g} \square_h \, ,
\\
& = & \frac{4}{8} \left[ 
{\scriptstyle 1} \square_1 \: + \:
{\scriptstyle 1} \square_{\overline{a}} \: + \:
{\scriptstyle \overline{a}} \square_1 \: + \:
{\scriptstyle \overline{a}} \square_{\overline{a}} \right],
\\
& = &
Z\left( [X/{\mathbb Z}_2 = \langle \overline{a} \rangle] \right),
\end{eqnarray}
matching the prediction.
\item Finally, consider the case
$B(\overline{a}) = -1$ and $B(\overline{b}) = -1$.
Here, the center $K$ acts nontrivially on twisted sectors twisted
by either $a$ or $b$, for example,
\begin{equation}
{\scriptstyle b^2} \square_a \: = \: - \left( 
{\scriptstyle 1} \square_a \right),
\: \: \:
{\scriptstyle b^2} \square_b \: = \: - \left(
{\scriptstyle 1} \square_b \right),
\: \: \:
{\scriptstyle b^2} \square_{ab} \: = \: + \left(
{\scriptstyle 1} \square_{ab} \right).
\end{equation}
Using relations of this form, it is straightforward to check that the
genus-one partition function obeys
\begin{eqnarray}
Z\left( [X/\Gamma]_B \right)
& = &
\frac{1}{|\Gamma|} \sum_{gh = hg} 
{\scriptstyle g} \square_h \, ,
\\
& = & \frac{4}{8} \left[ 
{\scriptstyle 1} \square_1 \: + \:
{\scriptstyle 1} \square_{\overline{a} \overline{b}} \: + \:
{\scriptstyle \overline{a} \overline{b} } \square_1 \: + \:
{\scriptstyle \overline{a} \overline{b} } \square_{
\overline{a} \overline{b} } \right],
\\
& = & Z\left( [X/{\mathbb Z}_2 = \langle \overline{a} \overline{b} \rangle]
\right),
\end{eqnarray}
matching the prediction.
\end{enumerate}

\subsubsection{With discrete torsion}
\label{sect:ex:z2z2:z2z4:dt}

So far we have considered the $[X/\Gamma]$ orbifold without discrete torsion.
Now, 
from \cite[section D.2]{Robbins:2021ylj}, we know that 
$H^2({\mathbb Z}_2 \times {\mathbb Z}_4, U(1)) = {\mathbb Z}_2$,
so one can turn on one nontrivial element of discrete torsion in the
$\Gamma = {\mathbb Z}_2 \times {\mathbb Z}_4$ orbifold.
We do so next, denoting by $\omega$ the nontrivial element of
$H^2(\Gamma,U(1))$.  Letting $\iota: K \hookrightarrow \Gamma$ denote
the inclusion and $\pi: \Gamma \rightarrow G$ the projection, 
then as observed in \cite[section 6.1]{Robbins:2021ylj},
$\iota^* \omega = 0$ and $\omega = \pi^* \overline{\omega}$,
for $\overline{\omega}$ the nontrivial element of 
$H^2({\mathbb Z}_2 \times {\mathbb Z}_2, U(1))$.  This will play an
important role in applying section~\ref{sect:conj}.

Next, we will make predictions for each of the four cases above,
using the methods of section~\ref{sect:conj}.
\begin{enumerate}
\item First, consider the case that $B$ is trivial,
$B(\overline{a}) = +1$ and $B(\overline{b}) = +1$.
This case was studied in \cite[section 6.1]{Robbins:2021ylj},
which made (and confirmed) the prediction
\begin{equation}
{\rm QFT}\left( [X/\Gamma]_{\omega} \right) \: = \:
{\rm QFT}\left( \coprod_2 [X/{\mathbb Z}_2 \times {\mathbb Z}_2]_{
\overline{\omega}} \right).
\end{equation}
\item The remaining three cases are identical to the cases studied
in the $\Gamma$ orbifold without discrete torsion.
For each $B$, Ker $B = {\mathbb Z}_2$, which does not admit discrete
torsion, so the predicted orbifold is the same.
\end{enumerate}

Finally, we check each of these predictions by computing
genus-one partition functions.
\begin{enumerate}
\item First, consider the case that $B$ is trivial,
$B(\overline{a}) = +1$ and $B(\overline{b}) = +1$.
This case was studied in \cite[section 6.1]{Robbins:2021ylj},
where it was shown that
\begin{equation}
Z\left( [X/\Gamma]_{\omega} \right) \: = \:
Z\left( \coprod_2 [X/{\mathbb Z}_2 \times {\mathbb Z}_2]_{
\overline{\omega}} \right).
\end{equation}
\item Next, consider the case that 
$B(\overline{a}) = -1$ and $B(\overline{b}) = +1$.
Here, the center $K$ acts nontrivially on twisted sectors twisted by $a$.
The genus-one partition function is given by
\begin{eqnarray}
Z\left( [X/\Gamma]_{\omega} \right) & = &
\frac{1}{|\Gamma|} \sum_{gh = hg} \epsilon(g,h)
\left( {\scriptstyle g} \square_h \right),
\\
& = & \frac{4}{8} \left[ 
{\scriptstyle 1} \square_1 \: + \: 
{\scriptstyle 1} \square_{\overline{b}} \: + \:
{\scriptstyle \overline{b}} \square_1 \: + \:
{\scriptstyle \overline{b}} \square_{\overline{b}} \right],
\\
& = & Z \left( [X/{\mathbb Z}_2 = \langle \overline{b} \rangle ] \right),
\end{eqnarray}
matching the prediction.
\item Next, consider the case that
$B(\overline{a}) = +1$ and $B(\overline{b}) = -1$.
The center $K$ acts nontrivially on twisted sectors twisted by $b$.
The genus-one partition function is given by
\begin{eqnarray}
Z\left( [X/\Gamma]_{\omega} \right)
& = &
\frac{1}{|\Gamma|} \sum_{gh = hg} \epsilon(g,h) \left(
{\scriptstyle g} \square_h \right) ,
\\
& = & \frac{4}{8} \left[ 
{\scriptstyle 1} \square_1 \: + \:
{\scriptstyle 1} \square_{\overline{a}} \: + \:
{\scriptstyle \overline{a}} \square_1 \: + \:
{\scriptstyle \overline{a}} \square_{\overline{a}} \right],
\\
& = &
Z\left( [X/{\mathbb Z}_2 = \langle \overline{a} \rangle] \right),
\end{eqnarray}
matching the prediction.
\item Finally, consider the case
$B(\overline{a}) = -1$ and $B(\overline{b}) = -1$.
Here, the center $K$ acts nontrivially on twisted sectors twisted
by either $a$ or $b$.
The genus-one partition function is given by
\begin{eqnarray}
Z\left( [X/\Gamma]_{\omega} \right)
& = &
\frac{1}{|\Gamma|} \sum_{gh = hg}  \epsilon(g,h) \left(
{\scriptstyle g} \square_h \right) ,
\\
& = & \frac{4}{8} \left[ 
{\scriptstyle 1} \square_1 \: + \:
{\scriptstyle 1} \square_{\overline{a} \overline{b}} \: + \:
{\scriptstyle \overline{a} \overline{b} } \square_1 \: + \:
{\scriptstyle \overline{a} \overline{b} } \square_{
\overline{a} \overline{b} } \right],
\\
& = & Z\left( [X/{\mathbb Z}_2 = \langle \overline{a} \overline{b} \rangle]
\right),
\end{eqnarray}
matching the prediction.
\end{enumerate}

\subsubsection{Summary}

We summarize the results of this analysis in table~\ref{table:z2z4:z2z2:summ}.

\begin{table}[h]
\begin{center}
\begin{tabular}{c|cc|cc}
Case & $B(\overline{a})$ & $B(\overline{b})$ &
Without discrete torsion & With discrete torsion \\ \hline
1 & $+1$ & $+1$ & $[X/{\mathbb Z}_2 \times {\mathbb Z}_2] \, \coprod \,
[X/{\mathbb Z}_2 \times {\mathbb Z}_2]$ &
$[X/{\mathbb Z}_2 \times {\mathbb Z}_2]_{\rm d.t.} \, \coprod \,
[X/{\mathbb Z}_2 \times {\mathbb Z}_2]_{\rm d.t.}$ \\
2 & $-1$ & $+1$ & $[X/{\mathbb Z}_2 = \langle \overline{b} \rangle]$ &
$[X/{\mathbb Z}_2 = \langle \overline{b} \rangle]$ \\
3 & $+1$ & $-1$ & $[X/{\mathbb Z}_2 = \langle \overline{a} \rangle]$ &
$[X/{\mathbb Z}_2 = \langle \overline{a} \rangle]$ \\
4 & $-1$ & $-1$ & $[X/{\mathbb Z}_2 = \langle \overline{a} \overline{b}
\rangle]$ & 
$[X/{\mathbb Z}_2 = \langle \overline{a} \overline{b}
\rangle]$
\end{tabular}
\caption{Summary of results for ${\mathbb Z}_2 \times {\mathbb Z}_4$
orbifold extending ${\mathbb Z}_2 \times {\mathbb Z}_2$
orbifold.
\label{table:z2z4:z2z2:summ}
}
\end{center}
\end{table}

\subsection{Extension of $\Z_2\times\Z_2$ to $D_4$}

In this section we consider a $G = \Z_2\times\Z_2$ theory, generated by 
$\overline{a}$ and $\overline{b}$ of order two,
extended to $\Gamma = D_4$:
\begin{equation}
1 \: \longrightarrow \:
\Z_2 \: \longrightarrow \:
 D_4 \: \longrightarrow \:
\Z_2\times\Z_2 \: \longrightarrow \:
 1, 
\end{equation}
where $\Gamma = D_4$ is the dihedral group of order 8, and $K = \Z_2$.  
We will write the elements of $D_4$ as
\begin{equation}
\{1, z, a, b, az, bz, ab, ba = abz \},
\end{equation}
where $a^2 = b^4 = 1$, $b^2 = z$, $z$ is central (generating the image of $K$),
$a$ projects to $\overline{a}$, and $b$ projects to
$\overline{b}$.
(This example was also considered from a different perspective in
\cite[section 5.2.1]{Robbins:2021lry}.)

The quantum symmetry $B$ is an element of
\begin{equation}
H^1(G,H^1(K,U(1)) \: = \:
{\rm Hom}(\Z_2 \times \Z_2, \hat{\Z}_2 ).
\end{equation}
Any nontrivial $B$ is necessarily
surjective, hence Ker $B = {\mathbb Z}_2$ and Coker $B = 0$.
From the general conjecture~(\ref{eq:predict}), we predict that
\begin{equation}
{\rm QFT}\left( [X/\Gamma]_B \right) \: = \:
{\rm QFT}\left( [X / {\mathbb Z}_2 ] \right),
\end{equation}
for any nontrivial $B$, with the choice of ${\mathbb Z}_2 \subset
{\mathbb Z}_2 \times {\mathbb Z}_2$ depending upon the choice of $B$.
If $B$ is trivial, 
then we have
\begin{equation}
{\rm QFT}\left[ [X/\Gamma] \right) \: = \:
{\rm QFT}\left( \left[ \frac{ X \times \hat{K} }{ {\mathbb Z}_2 \times
{\mathbb Z}_2 } \right]_{\hat{\omega}} \right)
\: = \:
{\rm QFT}\left( [X/ {\mathbb Z}_2 \times {\mathbb Z}_2 ] \, \coprod \,
[X/{\mathbb Z}_2 \times {\mathbb Z}_2]_{\rm d.t.} \right),
\end{equation}
as explored in detail in \cite[section 5.2]{Hellerman:2006zs}.

To make this more precise, let us consider the four possible 
values of $B \in H^1(G,H^1(K,U(1)))$, and the physical meaning of
each.  We can enumerate the possibilities as follows:
\begin{enumerate}
\item $B(\overline{a}) = +1$, $B(\overline{b}) = +1$,
\item $B(\overline{a}) = -1$, $B(\overline{b}) = +1$,
\item $B(\overline{a}) = +1$, $B(\overline{b}) = -1$,
\item $B(\overline{a}) = -1$, $B(\overline{b}) = -1$.
\end{enumerate}

We will first consider cases without discrete torsion, then add
discrete torsion $\omega \in H^2(\Gamma,U(1))$.

\subsubsection{Without discrete torsion}
\label{sect:ex:z2z2:d4:wodt}

\begin{enumerate}
\item First, consider the case in which
\begin{equation}
B(\overline{a}) \: = \:  +1 \: = \:
B(\overline{b}).
\end{equation}

This orbifold can be understood in terms of decomposition
\cite{Hellerman:2006zs,Sharpe:2019ddn,Tanizaki:2019rbk,Robbins:2021ylj},
which predicts that
\begin{equation}
{\rm QFT}\left( [X/D_4] \right) \: = \:
{\rm QFT} \left( [X/{\mathbb Z}_2 \times {\mathbb Z}_2 ] \coprod
[X/{\mathbb Z}_2 \times {\mathbb Z}_2 ]_{\rm d.t.} \right),
\end{equation}
where the second ${\mathbb Z}_2 \times {\mathbb Z}_2$ orbifold
has discrete torsion.
(See \cite[section 5.2]{Hellerman:2006zs} for further analyses of this example.)

\item Next, consider the second case, in which
$B(\overline{a}) = -1$ and
$B(\overline{b}) = +1$.  Here, Ker $B = \langle \overline{b} \rangle$,
so we predict that
\begin{equation}
{\rm QFT}\left( [X/\Gamma]_B \right) \: = \:
{\rm QFT}\left( [X/{\mathbb Z}_2 = \langle \overline{b} \rangle] \right).
\end{equation} 

Now, we check that prediction at the level of partition functions.
The center $K$ acts
trivially on the space but nontrivially on twisted sectors twisted by $a$,
as dictated by $B$.  For example:
\begin{equation}
{\scriptstyle z} \square_b \: = \: + \left( {\scriptstyle 1} \square_b \right) ,
\: \: \:
{\scriptstyle bz} \square_b \: = \: + \left( {\scriptstyle b} \square_b \right) ,
\: \: \:
{\scriptstyle bz} \square_{bz} \: = \: + \left( {\scriptstyle b} \square_{bz} 
\right) ,
\end{equation}
\begin{equation}
{\scriptstyle z} \square_a \: = \: - \left( {\scriptstyle 1} \square_a \right),
\: \: \:
{\scriptstyle z} \square_{ab} \: = \: - \left( {\scriptstyle 1} \square_{ab}
\right), \: \: \:
{\scriptstyle z} \square_{abz} \: = \: - \left( {\scriptstyle 1} \square_{abz} 
\right) ,
\end{equation}
and so forth.

For this $B$, it is straightforward to compute the genus-one
partition function
\begin{eqnarray}
Z\left( [X/D_4] \right) & = &
\frac{1}{| D_4 |} \sum_{gh = hg} {\scriptstyle g} \square_h,
\\
& = &
\frac{4}{8} \left[
{\scriptstyle 1} \square_1 \: + \:
{\scriptstyle 1} \square_{ \overline{b}} \: + \:
{\scriptstyle \overline{b}} \square_1 \: + \:
{\scriptstyle \overline{b}} \square_{\overline{b}}
\right] \: = \: Z\left( [X/{\mathbb Z}_2] \right),
\end{eqnarray}
a ${\mathbb Z}_2 = \langle \overline{b} \rangle$ orbifold, agreeing
with the prediction.

\item Next, consider the third case, in which $B(\overline{a}) = +1$
and $B(\overline{b}) = -1$.
Here, Ker $B = \langle \overline{a} \rangle$, so we predict that
\begin{equation}
{\rm QFT}\left( [X/\Gamma]_B \right) \: = \:
{\rm QFT}\left( [X/{\mathbb Z}_2 = \langle \overline{a} \rangle] \right).
\end{equation}

Now, we check that prediction at the level of partition functions.
The center $K$ acts trivially on the space, but nontrivially
on twisted sectors twisted by $b$, as dictated by $B$,
for example:
\begin{equation}
{\scriptstyle z} \square_b \: = \: - \left( {\scriptstyle 1} \square_b \right) ,
\: \: \:
{\scriptstyle bz} \square_b \: = \: - \left( {\scriptstyle b} \square_b \right) ,
\: \: \:
{\scriptstyle bz} \square_{bz} \: = \: - \left( {\scriptstyle b} \square_{bz} 
\right) ,
\end{equation}
\begin{equation}
{\scriptstyle z} \square_a \: = \: + \left( {\scriptstyle 1} \square_a \right),
\: \: \:
{\scriptstyle z} \square_{abz} \: = \: - \left( {\scriptstyle 1} \square_{abz} 
\right) ,
\end{equation}
and so forth.  

For this $B$,
it is now straightforward to compute the genus-one partition function:
\begin{eqnarray}
Z\left( [X/D_4] \right) & = &
\frac{1}{| D_4 |} \sum_{gh = hg} {\scriptstyle g} \square_h,
\\
& = & \frac{1}{8} \left[
{\scriptstyle 1} \square_1 \: + \:
{\scriptstyle z} \square_1 \: + \:
{\scriptstyle 1} \square_z \: + \:
{\scriptstyle z} \square_z \: + \:
{\scriptstyle 1} \square_a \: + \: 
\cdots  \: + \:
{\scriptstyle az} \square_{az} \right],
\\
& = &
\frac{4}{8} \left[ 
{\scriptstyle 1} \square_1 \: + \:
{\scriptstyle 1} \square_{\overline{a}} \: + \:
{\scriptstyle \overline{a}} \square_1 \: + \:
{\scriptstyle \overline{a}} \square_{ \overline{a}} \right]
\: = \: Z\left( [X/{\mathbb Z}_2 ] \right).
\end{eqnarray}
All sectors involving a ``$b$'' have cancelled out,
and the theory is equivalent to a ${\mathbb Z}_2$ orbifold by
${\mathbb Z}_2 = \langle \overline{a} \rangle$,
agreeing with the prediction.

\item Next, consider the fourth case, in which
$B(\overline{a}) = -1$ and $B(\overline{b}) = -1$.
Here, Ker $B = \langle \overline{a} \overline{b} \rangle$, so we predict that
\begin{equation}
{\rm QFT}\left( [X/\Gamma]_B \right) \: = \:
{\rm QFT}\left( [X/{\mathbb Z}_2 = \langle \overline{a} \overline{b}
\rangle] \right).
\end{equation}

Now, we check that prediction at the level of partition functions.
In this case, $B$ dictates the relations
\begin{equation}
{\scriptstyle z} \square_b \: = \: - \left( {\scriptstyle 1} \square_b \right) ,
\: \: \:
{\scriptstyle bz} \square_b \: = \: - \left( {\scriptstyle b} \square_b \right) ,
\: \: \:
{\scriptstyle bz} \square_{bz} \: = \: - \left( {\scriptstyle b} \square_{bz} 
\right) ,
\end{equation}
\begin{equation}
{\scriptstyle z} \square_a \: = \: - \left( {\scriptstyle 1} \square_a \right),
\: \: \:
{\scriptstyle z} \square_{ab} \: = \: + \left( {\scriptstyle 1} \square_{ab}
\right), \: \: \:
{\scriptstyle z} \square_{abz} \: = \: + \left( {\scriptstyle 1} \square_{abz} 
\right) ,
\end{equation}
and so forth.

For this $B$, it is straightforward to compute the genus-one
partition function:
\begin{eqnarray}
Z\left( [X/D_4] \right) & = &
\frac{1}{| D_4 |} \sum_{gh = hg} {\scriptstyle g} \square_h,
\\
& = &
\frac{4}{8} \left[
{\scriptstyle 1} \square_1 \: + \:
{\scriptstyle 1} \square_{ \overline{a} \overline{b} } \: + \:
{\scriptstyle \overline{a} \overline{b}} \square_1 \: + \:
{\scriptstyle \overline{a} \overline{b} } \square_{\overline{a} \overline{b}}
\right] \: = \: \Z\left( [X/{\mathbb Z}_2] \right).
\end{eqnarray}
This is a ${\mathbb Z}_2 = \langle \overline{a} \overline{b} \rangle$
orbifold, agreeing with the prediction.

\end{enumerate}

\subsubsection{With discrete torsion}
\label{sect:ex:z2z2:d4:dt}

Now, the $\Gamma = D_4$ orbifold admits one possible nonzero value of
discrete torsion, as 
\cite[appendix D.3]{Robbins:2021ylj}
\begin{equation}
H^2(D_4,U(1)) \: = \: {\mathbb Z}_2.
\end{equation}
Before describing the result of turning on that discrete torsion,
let us walk through its analysis formally.
A $D_4$ orbifold with trivially-acting central ${\mathbb Z}_2$ and
discrete torsion was studied in
\cite[section 5.5]{Robbins:2021ylj}, and in the notation of that
reference, $\iota^* \omega = 0$ but $\beta(\omega) \neq 0$.

Now, $\beta(\omega) \in H^1(G,H^1(K,U(1)))$ where in this case,
$G = D_4/{\mathbb Z}_2 = {\mathbb Z}_2 \times {\mathbb Z}_2$.
For a general $\omega \in H^2(\Gamma,U(1))$, one defines
\cite{Robbins:2021ylj}
\begin{equation}
\beta(\omega)(q,k) \: = \:
\frac{ \omega( k s(q), s(q)^{-1} )
}{
\omega( s(q)^{-1}, k s(q) )
},
\end{equation}
where $s: G \rightarrow \Gamma$ is a section, which here we can take to be
\begin{equation}
s(1) \: = \: 1, \: \: \:
s(\overline{a}) \: = \: a, \: \: \:
s(\overline{b}) \: = \: b, \: \: \:
s(\overline{a} \overline{b}) \: = \: ab,
\end{equation}
for which one computes the nontrivial cases 
\cite[section 5.5]{Robbins:2021ylj}
\begin{equation}
\beta(\omega)(\overline{a}, z) \: = \: -1, \: \: \:
\beta(\omega)(\overline{b}, z) \: = \: +1, \: \: \:
\beta(\omega)(\overline{a}\overline{b}, z) \: = \: -1.
\end{equation}
(As noted in section~\ref{app:Bdt}, $\beta(\omega)$ is independent of the
choice of section $s$.)

The effect of $\beta(\omega)$, it should now be clear, is to modify
the map $B$, essentially by flipping its action on $a$-twisted sectors.
We shall see this explicitly next.

Next, we systematically describe the results for $[X/D_4]$ orbifolds with
discrete torsion, in all cases.
\begin{enumerate}
\item We begin with the first case, with trivial $B$.
This is an example of decomposition with discrete torsion,
described in \cite[section 5.5]{Robbins:2021ylj}. The result is 
\begin{equation}
{\rm QFT}\left( [X/D_4]_{\rm d.t.} \right) \: = \:
{\rm QFT} \left( [X/{\mathbb Z}_2] \right),
\end{equation}
where the ${\mathbb Z}_2$ is generated by $\overline{b}$.

\item Next, consider the case $B(\overline{a}) = -1$,
$B(\overline{b}) = +1$.  In this case, $B/\beta(\omega)$ is trivial,
so we predict
\begin{equation}
{\rm QFT}\left( [X/\Gamma]_{B,\omega} \right) \: = \:
{\rm QFT}\left( \coprod_2 [X/ {\mathbb Z}_2 \times {\mathbb Z}_2]_{
\hat{\omega}} \right),
\end{equation}
for some discrete torsion $\hat{\omega}$ on the components.

Next, we check the prediction at the level of partition functions.
Here, it is straightforward to compute
\begin{eqnarray}
Z\left( [X/D_4]_{\rm d.t.} \right) & = &
\frac{1}{| D_4 |} \sum_{gh = hg} \epsilon(g,h)\left(
 {\scriptstyle g} \square_h \right),
\\
& = &
\frac{4}{8} \left[
{\scriptstyle 1} \square_1 \: + \: 
{\scriptstyle 1} \square_{\overline{a}} \: + \:
{\scriptstyle 1} \square_{\overline{b}} \: + \:
{\scriptstyle 1} \square_{\overline{a} \overline{b}} \: + \:
{\scriptstyle \overline{a}} \square_1 \: + \:
{\scriptstyle \overline{a}} \square_{\overline{a}} \: + \:
{\scriptstyle \overline{b}} \square_{\overline{1}} \: + \:
{\scriptstyle \overline{b}} \square_{\overline{b}}
\right. \nonumber \\
& & \hspace*{2.5in} \left.
 \: + \:
{\scriptstyle \overline{a} \overline{b}} \square_1 \: + \:
{\scriptstyle \overline{a} \overline{b}} \square_{\overline{a} \overline{b}}
\right],
\\
& = &
[X/D_4] \: = \: [X / {\mathbb Z}_2 \times {\mathbb Z}_2] \, \coprod \,
[X / {\mathbb Z}_2 \times {\mathbb Z}_2]_{\rm d.t.}.
\end{eqnarray}
Thus, this case is equivalent to the original $D_4$ orbifold,
agreeing with our prediction.

\item Next, consider the third case,
$B(\overline{a}) = +1$, $B(\overline{b}) = -1$.
In this case, $(B/\beta(\omega))(\overline{a}) = -1$,
$(B/\beta(\omega))(\overline{b}) = -1$, so that
Ker $B/\beta(\omega) = \langle \overline{a} \overline{b} \rangle$,
with no cokernel,
hence we predict
\begin{equation}
{\rm QFT}\left( [X/\Gamma]_{B,\omega} \right) \: = \:
{\rm QFT}\left( [X/{\mathbb Z}_2 = \langle \overline{a} \overline{b} \rangle ]
\right).
\end{equation}

Now, we will check that prediction.
The genus-one partition function is given by
\begin{eqnarray}
Z\left( [X/D_4]_{\rm d.t.} \right) & = &
\frac{1}{| D_4 |} \sum_{gh = hg} \epsilon(g,h)\left(
 {\scriptstyle g} \square_h \right),
\\
& = &
\frac{4}{8} \left[
{\scriptstyle 1} \square_1 \: + \: 
{\scriptstyle 1} \square_{\overline{a} \overline{b}} \: + \:
{\scriptstyle \overline{a} \overline{b}} \square_1 \: + \:
{\scriptstyle \overline{a} \overline{b}} \square_{ \overline{a} \overline{b}}
\right] \: = \:
Z\left( [X/{\mathbb Z}_2 ] \right).
\end{eqnarray}
Thus, this case is equivalent to a
${\mathbb Z}_2 = \langle \overline{a} \overline{b} \rangle$ orbifold,
agreeing with our prediction.

\item Now, consider the fourth case.  
Here, $(B/\beta(\omega))(\overline{a}) = +1$,
$(B/\beta(\omega))(\overline{b} = -1$, hence
Ker $B/\beta(\omega) = \langle \overline{a} \rangle$, with no cokernel,
and so we predict
\begin{equation}
{\rm QFT}\left( [X/\Gamma]_{B,\omega} \right) \: = \:
{\rm QFT}\left( [X/{\mathbb Z}_2 = \langle \overline{a} \rangle] \right).
\end{equation}

Next, we check the prediction.
The genus-one partition function is given by
\begin{eqnarray}
Z\left( [X/D_4]_{\rm d.t.} \right) & = &
\frac{1}{| D_4 |} \sum_{gh = hg} \epsilon(g,h)
\left(  {\scriptstyle g} \square_h \right),
\\
& = &
\frac{4}{8} \left[
{\scriptstyle 1} \square_1 \: + \: 
{\scriptstyle 1} \square_{\overline{a}} \: + \:
{\scriptstyle \overline{a}} \square_1 \: + \:
{\scriptstyle \overline{a}} \square_{\overline{a}}
\right] \: = \: Z\left( [X/{\mathbb Z}_2 ] \right),
\end{eqnarray}
the ${\mathbb Z}_2 = \langle \overline{a} \rangle$ orbifold of $X$,
using the discrete torsion phases in \cite[table D.4]{Robbins:2021ylj}.
This agrees with our prediction, as expected.

\end{enumerate}

\subsubsection{Summary}

We summarize the results of this analysis in table~\ref{table:d4:z2z2:summ}.
As anticipated earlier, the effect of turning on discrete torsion
in the $D_4$ orbifold is to exchange the $B(\overline{a})$ values.

\begin{table}[h]
\begin{center}
\begin{tabular}{c|cc|cc}
Case & $B(\overline{a})$ & $B(\overline{b})$ &
Without discrete torsion & With discrete torsion \\ \hline
1 & $+1$ & $+1$ & $[X/{\mathbb Z}_2 \times {\mathbb Z}_2] \, \coprod \,
[X/{\mathbb Z}_2 \times {\mathbb Z}_2]_{\rm d.t.}$ &
$[X/{\mathbb Z}_2 = \langle \overline{b} \rangle]$ \\
2 & $-1$ & $+1$ & $[X/{\mathbb Z}_2=\langle \overline{b} \rangle]$ &
$[X/{\mathbb Z}_2 \times {\mathbb Z}_2] \, \coprod \,
[X/{\mathbb Z}_2 \times {\mathbb Z}_2]_{\rm d.t.}$ \\
3 & $+1$ & $-1$ & $[X/{\mathbb Z}_2 = \langle \overline{a}\rangle]$ 
 & $[X/{\mathbb Z}_2 = \langle \overline{a} \overline{b} \rangle]$ \\
4 & $-1$ & $-1$ & $[X/{\mathbb Z}_2 = \langle \overline{a} \overline{b}\rangle]$
 & $[X/{\mathbb Z}_2 = \langle \overline{a} \rangle]$
\end{tabular}
\caption{Summary of results for $D_4$ orbifold extending
${\mathbb Z}_2 \times {\mathbb Z}_2$ orbifold.
\label{table:d4:z2z2:summ}
}
\end{center}
\end{table}

\subsection{Extension of ${\mathbb Z}_2 \times {\mathbb Z}_4$ to
${\mathbb Z}_4 \rtimes {\mathbb Z}_4$}

Consider an orbifold $[X/ {\mathbb Z}_2 \times {\mathbb Z}_4]$,
where the orbifold group is 
extended by $K = {\mathbb Z}_2$ to 
$\Gamma = {\mathbb Z}_4 \rtimes {\mathbb Z}_4$, a semidirect
product of two copies of ${\mathbb Z}_4$.  We will use the
same notation as in 
\cite{Robbins:2021ylj}, and describe ${\mathbb Z}_4 \rtimes {\mathbb Z}_4$
as generated by $x$, $y$ subject to the constraints
\begin{equation}
x^4 \: = \: 1 \: = \: y^4, \: \: \:
y = xyx.
\end{equation}
In this notation, we take 
$K = \langle x^2 \rangle$, so that $K$ is central in ${\mathbb Z}_4 \rtimes
{\mathbb Z}_4$, and $G = {\mathbb Z}_2 \times {\mathbb Z}_4$ is
generated by $xK$ and $yK$.

As in the previous example, we will walk through the physical implications of
each choice of $B$, 
to get a more complete
picture of the physics here.

In this case, the possible $B \in H^1(G,H^1(K,U(1))$ are
determined by their action on the generators of $G$, as listed below:
\begin{enumerate}
\item $B(xK) = +1$, $B(yK)=+1$,
\item $B(xK) = +1$, $B(yK) = -1$,
\item $B(xK) = -1$, $B(yK) = +1$,
\item $B(xK) = -1$, $B(yK) = -1$.
\end{enumerate}

First we will analyze these cases without discrete torsion,
then we will consider the effect of adding
discrete torsion $\omega \in H^2(\Gamma,U(1))$.

\subsubsection{Without discrete torsion}
\label{sect:ex:z2z4:z4z4:wodt}

We list the predictions of our conjecture~(\ref{eq:predict}) below.
\begin{enumerate}
\item In the first case, $B$ is trivial, so this case reduces
to ordinary decomposition
\cite{Hellerman:2006zs,Sharpe:2019ddn,Tanizaki:2019rbk,Robbins:2021ylj}, 
for which
\begin{equation}
{\rm QFT}\left( [X/{\mathbb Z}_4 \rtimes {\mathbb Z}_4] \right)
\: = \:
{\rm QFT}\left( [X/{\mathbb Z}_2 \times {\mathbb Z}_4] \, \coprod \,
[X / {\mathbb Z}_2 \times {\mathbb Z}_4]_{\rm d.t.} \right).
\end{equation}
\item In the second case, where $B(xK)$ is trivial but $B(yK)$ nontrivial,
we have that Ker $B = {\mathbb Z}_2 \times {\mathbb Z}_2 = 
\langle xK, y^2 K \rangle$ and
Coker $B = 0$, hence from~(\ref{eq:predict}) we predict that
\begin{equation}
{\rm QFT}\left(  [X/{\mathbb Z}_4 \rtimes {\mathbb Z}_4] \right)
\: = \:
{\rm QFT}\left( [X / {\mathbb Z}_2 \times {\mathbb Z}_2] \right).
\end{equation}
In this case, since only the trivial irreducible representation of $K$
appears, there is no discrete torsion.
\item In the third case, where $B(xK)$ is nontrivial but $B(yK)$ trivial,
we have that Ker $B = {\mathbb Z}_4 = \langle yK \rangle$ and Coker $B = 0$,
hence from~(\ref{eq:predict}) we predict that
\begin{equation}
{\rm QFT}\left(  [X/{\mathbb Z}_4 \rtimes {\mathbb Z}_4] \right)
\: = \:
{\rm QFT}\left( [X/{\mathbb Z}_4] \right).
\end{equation}
\item In the fourth case, where both $B(xK)$ and $B(yK)$ are nontrivial,
Coker $B = 0$ again but Ker $B = ({\mathbb Z}_2 \times {\mathbb Z}_4)/
{\mathbb Z}_2 = {\mathbb Z}_4 = \langle xyK \rangle$, 
and from~(\ref{eq:predict})
we predict that
\begin{equation}
{\rm QFT}\left(  [X/{\mathbb Z}_4 \rtimes {\mathbb Z}_4] \right)
\: = \:
{\rm QFT}\left( [X/{\mathbb Z}_4] \right).
\end{equation}
\end{enumerate}

Now, we will check the prediction in each case.
\begin{enumerate}
\item First, consider the trivial case
$B(xK) = +1$ and $B(yK) = +1$.
This corresponds to an ordinary ${\mathbb Z}_4 \rtimes {\mathbb Z}_4$
orbifold, with $K = {\mathbb Z}_2$ acting trivially both on the underlying
space as well as on the twisted sectors of $G$.  This case can be
described by ordinary decomposition 
\cite{Hellerman:2006zs,Sharpe:2019ddn,Tanizaki:2019rbk,Robbins:2021ylj},
which predicts 
\begin{equation}
{\rm QFT}\left( [X/{\mathbb Z}_4 \rtimes {\mathbb Z}_4] \right)
\: = \: {\rm QFT}\left( 
[X/{\mathbb Z}_2 \times {\mathbb Z}_4] \, \coprod \,
[X/{\mathbb Z}_2 \times {\mathbb Z}_4]_{\rm d.t.} \right),
\end{equation}
a disjoint union of two copies of $[X/{\mathbb Z}_2 \times {\mathbb Z}_4]$,
with discrete torsion in one copy.  This can be seen directly in 
e.g. partition functions, as follows.  Some of the
${\mathbb Z}_2 \times {\mathbb Z}_4$ twisted sectors are missing from
the ${\mathbb Z}_4 \rtimes {\mathbb Z}_4$ orbifold partition function,
such as
\begin{equation}
{\scriptstyle xK} \square_{yK} \, ,
\end{equation}
as the two group elements do not lift to commuting pairs in
${\mathbb Z}_4 \rtimes {\mathbb Z}_4$.
These are precisely the same pairs of group elements in
${\mathbb Z}_2 \times {\mathbb Z}_4$ that can be weighted by nontrivial
discrete torsion phases, see e.g. \cite[table D.2]{Robbins:2021ylj}.
Summing the two partition functions cancels out the missing twisted
sectors, recovering the correct ${\mathbb Z}_4 \rtimes {\mathbb Z}_4$
orbifold partition function.

\item Next, we consider the case
$B(xK) = +1$ and $B(yK) = -1$.  This implies that the central $K$
acts trivially on the space, but nontrivially on twisted sectors twisted
by $y$, as illustrated below:
\begin{equation}  \label{eq:z2z4:z4z4:case2:beta}
{\scriptstyle x^2} \square_y \: = \: - \left( {\scriptstyle 1} \square_y \right),
\: \: \:
{\scriptstyle x^2} \square_{y^2} \: = \: + \left(
{\scriptstyle 1} \square_{y^2} \right), 
\: \: \:
{\scriptstyle x^2} \square_x \: = \: + \left(
{\scriptstyle 1} \square_x \right).
\end{equation}
Using these relations, it is straightforward to check that in the
genus-one partition function of the ${\mathbb Z}_4 \rtimes
{\mathbb Z}_4$ orbifold, all twisted sectors with a $y$ or $y^3$ cancel
out, and checking multiplicities, one finds
\begin{equation}
Z\left( [X/{\mathbb Z}_4 \rtimes {\mathbb Z}_4] \right) \: = \:
Z\left( [X/{\mathbb Z}_2 \times {\mathbb Z}_2] \right),
\end{equation}
where the ${\mathbb Z}_2 \times {\mathbb Z}_2 = \langle xK, y^2K \rangle$.

\item Next, we consider the case 
$B(xK) = -1$ and $B(yK) = +1$.
This implies that the central $K$ acts nontrivially on twisted sectors
twisted by $x$, as illustrated below:
\begin{equation}
{\scriptstyle x^2} \square_x \: = \: - \left( {\scriptstyle 1} \square_x \right),
\: \: \:
{\scriptstyle x^2} \square_{x^2} \: = \: + \left(
{\scriptstyle 1} \square_{x^2} \right),
\: \: \:
{\scriptstyle x^2} \square_y \: = \: + \left(
{\scriptstyle 1} \square_y \right).
\end{equation}
As a result, twisted sector contributions with an odd number of $x$'s along
either leg cancel out from the genus-one partition function.  
Taking into account multiplicities, we find that the partition function
is given by
\begin{equation}
Z\left( [X/{\mathbb Z}_4 \rtimes {\mathbb Z}_4] \right) \: = \:
Z\left( [X/{\mathbb Z}_4] \right),
\end{equation}
where the effectively-acting ${\mathbb Z}_4$ on the right is 
$\langle yK \rangle$.

\item Next, we consider the case $B(xK) = -1$ and
$B(yK) = -1$.  
This implies that the central $K$ acts nontrivially on twisted sectors
twisted by either $x$ or $y$, as illustrated below:
\begin{equation}
{\scriptstyle x^2} \square_x \: = \: - \left( {\scriptstyle 1} \square_x \right),
\: \: \:
{\scriptstyle x^2} \square_{x^2} \: = \: + \left(
{\scriptstyle 1} \square_{x^2} \right),
\: \: \:
{\scriptstyle x^2} \square_y \: = \: - \left(
{\scriptstyle 1} \square_y \right).
\end{equation}
\begin{equation}
{\scriptstyle x^2} \square_{xy} \: = \: + \left( 
{\scriptstyle 1} \square_{xy} \right),
\: \: \:
{\scriptstyle x^3} \square_{xy^3} \: = \: + \left(
{\scriptstyle x} \square_{xy^3} \right).
\end{equation}
As a result, genus-one twisted sectors such that the sum of the number of
$x$'s and $y$'s on any leg are odd will cancel out.
The partition function of this theory has the form
\begin{eqnarray}
\lefteqn{
Z\left( [X/{\mathbb Z}_4 \rtimes {\mathbb Z}_4] \right)
} \nonumber \\
& = &
\frac{1}{| {\mathbb Z}_4 \rtimes {\mathbb Z}_4|}
\sum_{gh = hg}  
{\scriptstyle g} \square_h \, ,
\\
& = &
\frac{4}{|{\mathbb Z}_4 \rtimes {\mathbb Z}_4|}
\left[
{\scriptstyle 1} \square_1 \: + \:
{\scriptstyle 1} \square_{y^2 K} \: + \:
{\scriptstyle 1} \square_{xy K} \: + \:
{\scriptstyle 1} \square_{xy^3 K} \: + \:
{\scriptstyle xy K} \square_1 \: + \:
{\scriptstyle xy K} \square_{xy K} \: + \:
{\scriptstyle xy K} \square_{y^2 K}
\right. \nonumber \\
& & \hspace*{1in} \left.
 \: + \:
{\scriptstyle xy K} \square_{xy^3 K}
 \: + \:
{\scriptstyle y^2 K} \square_1
 \: + \:
{\scriptstyle y^2 K} \square_{y^2 K} \: + \:
{\scriptstyle y^2 K} \square_{xy K} \: + \:
{\scriptstyle y^2 K} \square_{xy^3 K}
\right. \nonumber \\
& & \hspace*{1in} \left.
 \: + \:
{\scriptstyle xy^3 K} \square_1 \: + \:
{\scriptstyle xy^3 K} \square_{xy K}
 \: + \:
{\scriptstyle xy^3 K} \square_{xy^3 K} \: + \:
{\scriptstyle xy^3 K} \square_{y^2 K} \right],
 \\
& = &
Z\left( [X/{\mathbb Z}_4 ] \right),
\end{eqnarray}
where the ${\mathbb Z}_4 = \langle xy K \rangle$, using the fact that
$(xy)^2 = y^2$, $(xy)^3 = xy^3$ in ${\mathbb Z}_4 \rtimes {\mathbb Z}_4$.

\end{enumerate}

As expected, each case matches our
prediction~(\ref{eq:predict}).

\subsubsection{With discrete torsion}
\label{sect:ex:z2z4:z4z4:dt}

Now, let us turn on discrete torsion in the
${\mathbb Z}_4 \rtimes {\mathbb Z}_4$ orbifold.  Since
$H^2({\mathbb Z}_4 \rtimes {\mathbb Z}_4,U(1)) = {\mathbb Z}_2$,
there is exactly one nontrivial choice.

Before describing the physics results in this case,
let us take a moment to analyze this case formally and predict
the result.  In the notation of
\cite{Robbins:2021ylj}, this is a case in which
$\iota^* \omega = 0$ and $\beta(\omega) \neq 0$,
where $\beta(\omega)$ is the image of $\omega \in H^2(\Gamma,U(1))$
in $H^1(G,H^1(K,U(1)))$.  
As computed in \cite[section 5.6]{Robbins:2021ylj},
\begin{equation}
\beta(\omega)(xK,x^2) \: = \: +1, \: \: \:
\beta(\omega)(yK,x^2) \: = \: -1,
\end{equation}
so the effect of turning on $\omega$ should be to flip the value of
$B(yK)$.

Now, let us check this prediction, by walking through each case
explicitly.
\begin{enumerate}
\item In the first case, $B$ itself is trivial.  
Now, the $\Gamma = {\mathbb Z}_4 \rtimes {\mathbb Z}_4$ orbifold with
trivially-acting central $K = {\mathbb Z}_2 = \langle x^2 \rangle$
and discrete torsion was discussed in detail in
\cite[section 5.6]{Robbins:2021ylj}, where it was argued that
\begin{equation}
{\rm QFT}\left( [X/\Gamma]_{\rm d.t.} \right) \: = \:
{\rm QFT}\left( [X/{\mathbb Z}_2 \times {\mathbb Z}_2] \right).
\end{equation}

\item Next, we consider the case $B(xK) = +1$,
$B(yK)=-1$, with discrete torsion in the
$\Gamma = {\mathbb Z}_4 \rtimes {\mathbb Z}_4$ orbifold.
As observed in the case without discrete torsion, 
$B$ encodes an action of $K$ on the twisted sectors twisted by $y$,
as described in equation~(\ref{eq:z2z4:z4z4:case2:beta}.  In particular,
$B$ relates by a sign twisted sectors with an odd number of powers of
$y$.  The twisted sector phases of ${\mathbb Z}_4 \rtimes {\mathbb Z}_4$,
as listed in \cite[table D.6]{Robbins:2021ylj}, have the same effect:
twisted sectors with odd powers of $y$ are also related by signs.
The combined effect of both $B$ and the discrete torsion is,
therefore, to cancel out, leaving us with a
$\Gamma = {\mathbb Z}_4 \rtimes {\mathbb Z}_4$ orbifold in which
$K = \langle x^2 \rangle$ acts trivially on both the underlying
space $X$ as well as twisted sectors.  This can be described by
ordinary decomposition, and as we described earlier, and can be checked
directly in partition functions, we have
\begin{equation}
{\rm QFT}\left( [X/{\mathbb Z}_4 \rtimes {\mathbb Z}_4] \right)
\: = \: {\rm QFT}\left( 
[X/{\mathbb Z}_2 \times {\mathbb Z}_4] \, \coprod \,
[X/{\mathbb Z}_2 \times {\mathbb Z}_4]_{\rm d.t.} \right),
\end{equation}
a disjoint union of two ${\mathbb Z}_2 \times {\mathbb Z}_4$
orbifolds.

\item Next, we consider the case
$B(xK) = -1$, $B(yK) = +1$.
In this case, as before, roughly, the $B$
by itself would cancel out genus-one twisted sectors with odd powers of $x$,
and the discrete torsion would cancel out genus-one twisted sectors with
odd powers of $y$; however, sectors involving e.g. $xy$, $x y^3$ survive.
Using \cite[table D.6]{Robbins:2021ylj},
the genus-one partition function is
\begin{eqnarray}
\lefteqn{
Z\left( [X/{\mathbb Z}_4 \rtimes {\mathbb Z}_4]_{\rm d.t.} \right)
} \nonumber \\
& = &
\frac{1}{| {\mathbb Z}_4 \rtimes {\mathbb Z}_4|}
\sum_{gh = hg} \epsilon(g,h) \left( 
{\scriptstyle g} \square_h \right),
\\
& = &
\frac{4}{|{\mathbb Z}_4 \rtimes {\mathbb Z}_4|}
\left[
{\scriptstyle 1} \square_1 \: + \:
{\scriptstyle 1} \square_{y^2 K} \: + \:
{\scriptstyle 1} \square_{xy K} \: + \:
{\scriptstyle 1} \square_{xy^3 K} \: + \:
{\scriptstyle xy K} \square_1 \: + \:
{\scriptstyle xy K} \square_{xy K} \: + \:
{\scriptstyle xy K} \square_{y^2 K}
\right. \nonumber \\
& & \hspace*{1in} \left.
 \: + \:
{\scriptstyle xy K} \square_{xy^3 K}
 \: + \:
{\scriptstyle y^2 K} \square_1 
 \: + \:
{\scriptstyle y^2 K} \square_{y^2 K} \: + \:
{\scriptstyle y^2 K} \square_{xy K} \: + \:
{\scriptstyle y^2 K} \square_{xy^3 K}
\right. \nonumber \\
& & \hspace*{1in} \left.
 \: + \:
{\scriptstyle xy^3 K} \square_1 \: + \:
{\scriptstyle xy^3 K} \square_{xy K}
 \: + \:
{\scriptstyle xy^3 K} \square_{xy^3 K} \: + \:
{\scriptstyle xy^3 K} \square_{y^2 K} \right],
 \\
& = &
Z\left( [X/{\mathbb Z}_4 ] \right),
\end{eqnarray}
where the ${\mathbb Z}_4 = \langle xy K \rangle$, using the fact that
$(xy)^2 = y^2$, $(xy)^3 = xy^3$ in ${\mathbb Z}_4 \rtimes {\mathbb Z}_4$.

\item Finally, we consider the case $B(xK) = -1$ and
$B(yK) = -1$, for a ${\mathbb Z}_4 \rtimes {\mathbb Z}_4$ orbifold with
discrete torsion. Using the methods above, it is straightforward to
compute that
\begin{eqnarray}
\lefteqn{
Z\left( [X/{\mathbb Z}_4 \rtimes {\mathbb Z}_4]_{\rm d.t.} \right)
} \nonumber \\
& = & \frac{1}{| {\mathbb Z}_4 \rtimes {\mathbb Z}_4|}
\sum_{gh = hg} \epsilon(g,h) \left(
{\scriptstyle g} \square_h \right),
\\
& = &
\frac{4}{| {\mathbb Z}_4 \rtimes {\mathbb Z}_4 |}
\left[
{\scriptstyle 1} \square_1 \: + \:
{\scriptstyle 1} \square_{y K} \: + \:
{\scriptstyle 1} \square_{y^2 K} \: + \:
{\scriptstyle 1} \square_{y^3 K} \: + \:
{\scriptstyle y K} \square_1 \: + \: 
{\scriptstyle y K} \square_{y K} \: + \:
{\scriptstyle y K} \square_{y^2 K} \: + \:
{\scriptstyle y K} \square_{y^3 K}
\right. \nonumber \\
& & \hspace*{1in} \left.
 \: + \:
{\scriptstyle y^2 K} \square_{1} \: + \:
{\scriptstyle y^2 K} \square_{y K}
 \: + \:
{\scriptstyle y^2 K} \square_{y^2 K} \: + \:
{\scriptstyle y^2 K} \square_{y^3 K} \: + \:
{\scriptstyle y^3 K} \square_1 \: + \:
{\scriptstyle y^3 K} \square_{y K}
\right. \nonumber \\
& & \hspace*{3in} \left.
 \: + \:
{\scriptstyle y^3 K} \square_{y^2 K} \: + \:
{\scriptstyle y^3 K} \square_{y^3 K}
\right],
 \\
& = & 
Z\left( [X/{\mathbb Z}_4] \right),
\end{eqnarray}
where the ${\mathbb Z}_4 = \langle y K\rangle$.

\end{enumerate}

\subsubsection{Summary}

These results are summarized in table~\ref{table:z4z4:z2z2:summ}.
As anticipated, the effect of discrete torsion is to flip the value
of $\beta(yK)$.

\begin{table}[h]
\begin{center}
\begin{tabular}{c|cc|cc}
Case & $B(xK)$ & $B(yK)$ &
Without discrete torsion & With discrete torsion \\ \hline
1 & $+1$ & $+1$ & $[X/{\mathbb Z}_2 \times {\mathbb Z}_4] \, \coprod \,
[X/{\mathbb Z}_2 \times {\mathbb Z}_4]_{\rm d.t.}$  &
$[X/{\mathbb Z}_2 \times {\mathbb Z}_2]$ \\
2 & $+1$ & $-1$ & $[X/{\mathbb Z}_2 \times {\mathbb Z}_2 = 
\langle xK, y^2K \rangle]$
& $[X/{\mathbb Z}_2 \times {\mathbb Z}_4] \, \coprod \,
[X/{\mathbb Z}_2 \times {\mathbb Z}_4]_{\rm d.t.}$ \\
3 & $-1$ & $+1$ & $[X/{\mathbb Z}_4 = \langle yK \rangle]$ &
$[X / {\mathbb Z}_4 = \langle xy K \rangle]$ \\
4 & $-1$ & $-1$ &
 $[X / {\mathbb Z}_4  = \langle xy K \rangle]$
 & $[X/{\mathbb Z}_4 = \langle y K \rangle]$
\end{tabular}
\caption{Summary of results for $D_4$ orbifold extending
${\mathbb Z}_2 \times {\mathbb Z}_2$ orbifold.
\label{table:z4z4:z2z2:summ}
}
\end{center}
\end{table}

\subsection{Extension of $D_4$ to
${\mathbb Z}_4 \rtimes {\mathbb Z}_4$}

In this section, we will discuss an example of a
$D_4$ orbifold being extended by
$K = {\mathbb Z}_2$
to $\Gamma = {\mathbb Z}_4 \rtimes {\mathbb Z}_4$.  In this
example, the extension is
the central subgroup 
$K = \langle y^2 \rangle$, in the notation of \cite{Robbins:2021ylj}.

We will follow the same notation as \cite{Robbins:2021ylj},
so we denote the elements of $D_4$ by
\begin{equation}
\{1, z, a, b, az, bz, ab, ba = abz \},
\end{equation}
where 
\begin{equation}
a^2 \: = \: 1 \: = \: b^4, \: \: \:
b^2 \: = \: z,
\end{equation}
with $z$ central in $D_4$.  The projection $\pi: \Gamma \rightarrow D_4$
maps
\begin{equation}
\pi(x) \: = \: b \: = \: \pi(x y^2), \: \: \:
\pi(y) \: = \: a \: = \: \pi(y^3).
\end{equation}

The possible $B \in H^1(G,H^1(K,U(1))$
are enumerated by their valued on the generators below:
\begin{enumerate}
\item $B(a) = +1$, $B(b)=+1$, 
\item $B(a) = +1$, $B(b) = -1$,
\item $B(a) = -1$, $B(b) = +1$,
\item $B(a) = -1$, $B(b) = -1$.
\end{enumerate}

We will first consider cases without discrete torsion,
then cases with discrete torsion
$\omega \in H^2({\mathbb Z}_4 \rtimes {\mathbb Z}_4,U(1))$.

\subsubsection{Without discrete torsion}
\label{sect:ex:d4:z4z4:wodt}

From section~\ref{sect:conj}, we predict that
\begin{equation}
{\rm QFT}\left( [X/\Gamma]_B \right) \: = \:
{\rm QFT}\left( \left[ 
\frac{ X \times \widehat{{\rm Coker}\: B} }{ {\rm Ker}\: B } \right]_{
\hat{\omega}_0} \right),
\end{equation}
where $\hat{\omega}_0$ is the discrete torsion that one computes from
ordinary decomposition, without a quantum symmetry.
Here, since $K$ is central, $\hat{\omega}_0$ is the image of the
extension class under the corresponding irreducible representation.
We studied this example without a quantum symmetry in
\cite[section 6.2]{Robbins:2021ylj}, where we argued that
$\hat{\omega}_0 = 0$, so to predict each case, one merely needs to 
compute the kernel and cokernel of $B$, which we do below.
\begin{enumerate}
\item First, consider the case $B(a) = +1$, $B(b) = +1$,
meaning the case without a quantum symmetry.  Since $B$ is
trivial, the kernel is $G = D_4$, and the cokernel is $\hat{\mathbb Z}_2$,
so the result reduces to that of ordinary decomposition:
\begin{equation}
{\rm QFT}\left( [X/\Gamma]_B \right) \: = \:
{\rm QFT}\left( \coprod_2 [X/D_4] \right).
\end{equation}
\item Next, consider the case $B(a) = +1$, $B(b) = -1$.
In this case, the cokernel of $B$ is trivial, but
\begin{equation}
{\rm Ker}\: B \: = \: \langle a, b^2 \rangle \: = \:
\langle y K, x^2 K \rangle \: = \:
{\mathbb Z}_2 \times {\mathbb Z}_2,
\end{equation}
hence we predict
\begin{equation}
{\rm QFT}\left( [X/\Gamma]_B \right) \: = \:
{\rm QFT}\left( [X / {\mathbb Z}_2 \times {\mathbb Z}_2 ] \right).
\end{equation}
\item Next, consider the case $B(a) = -1$, $B(b) = +1$.
In this case, the cokernel of $B$ is trivial, but
\begin{equation}
{\rm Ker}\: B \: = \: \langle a^2, b \rangle \: = \: \langle b \rangle
\: = \: \langle x K \rangle \: = \: {\mathbb Z}_4,
\end{equation}
hence we predict
\begin{equation}
{\rm QFT}\left( [X/\Gamma]_B \right) \: = \:
{\rm QFT}\left( [X/{\mathbb Z}_4] \right).
\end{equation}
\item Next, consider the case $B(a) = -1$, $B(b) = -1$.
In this case, the cokernel of $B$ is trivial, but
\begin{equation}
{\rm Ker}\: B \: = \: \langle a^2, ab, b^2 \rangle \: = \:
\langle ab, b^2 \rangle \: = \: \langle xyK, y^2K \rangle \: = \:
{\mathbb Z}_2 \times {\mathbb Z}_2,
\end{equation}
hence we predict
\begin{equation}
{\rm QFT}\left( [X/\Gamma]_B \right) \: = \:
{\rm QFT}\left( [X/{\mathbb Z}_2 \times {\mathbb Z}_2] \right).
\end{equation}
\end{enumerate}

Next, we briefly review the physics each case for $B$, without discrete torsion.
In each case, we find that partition function computations verify the
predictions above.  (We have included the case of trivial $B$ for
completeness.)
\begin{enumerate}
\item First, consider the case that
$B(a) = +1$ and $B(b)=+1$.
This is an ordinary ${\mathbb Z}_4 \rtimes {\mathbb Z}_4$
orbifold, with $K = {\mathbb Z}_2$ acting trivially both on the underlying
space as well as on the twisted sectors of $G$. 
This case can be described by ordinary decomposition
\cite{Hellerman:2006zs,Sharpe:2019ddn,Tanizaki:2019rbk,Robbins:2021ylj},
which predicts (see \cite[section 6.2]{Robbins:2021ylj})
\begin{equation}
{\rm QFT}\left( [X/{\mathbb Z}_4 \rtimes {\mathbb Z}_4] \right)
\: = \:
{\rm QFT}\left( \coprod_2 [X/D_4] \right),
\end{equation}
or in other words, that this theory is equivalent to a disjoint
union of two $D_4$ orbifolds, neither with discrete torsion.

\item Next, we consider the case that
$B(a) = +1$ and $B(b) = -1$.
This implies that the central $K$ acts trivially on the space, but 
nontrivially on twisted sectors twisted by $x$, for example
\begin{equation}
{\scriptstyle y^2} \square_x \: = \: - \left(
{\scriptstyle 1} \square_x \right),
\: \: \:
{\scriptstyle x y^2} \square_y \: = \: + \left(
{\scriptstyle x} \square_y \right),
\: \: \:
{\scriptstyle y^2} \square_{y^2} \: = \: + \left(
{\scriptstyle 1} \square_{y^2} \right)
\: = \:  + \left( {\scriptstyle 1} \square_1 \right).
\end{equation}
Essentially as a result, in the genus-one partition function,
twisted sectors with an odd number of powers of $x$ on either leg cancel
out.  It is straightforward to compute that
the genus-one partition function is
\begin{eqnarray}
Z\left( [X / {\mathbb Z}_4 \rtimes {\mathbb Z}_4] \right) 
& = &
\frac{1}{| {\mathbb Z}_4 \rtimes {\mathbb Z}_4 |}
\sum_{gh = hg} 
{\scriptstyle g} \square_h \, ,
\\
& = & Z\left( [X/{\mathbb Z}_2 \times {\mathbb Z}_2] \right),
\end{eqnarray}
an orbifold by ${\mathbb Z}_2 \times {\mathbb Z}_2 =
\langle x^2 K, y K \rangle$.

\item  Next, we consider the case that
$B(a) = -1$ and $B(b) = +1$.  
This implies that the central $K$ acts trivially on the space but
nontrivially on twisted sectors twisted by $y$, for example
\begin{equation}
{\scriptstyle y^2} \square_y \: = \: - \left(
{\scriptstyle 1} \square_y \right),
\: \: \:
{\scriptstyle y^2} \square_x \: = \: + \left(
{\scriptstyle 1} \square_x \right).
\end{equation}
Essentially as a result, in the genus-one partition function,
twisted sectors with an odd number of powers of $y$ on either leg cancel
out.  It is straightforward to compute that the genus-one partition function
is
\begin{eqnarray}
Z\left( [X / {\mathbb Z}_4 \rtimes {\mathbb Z}_4] \right) 
& = &
\frac{1}{| {\mathbb Z}_4 \rtimes {\mathbb Z}_4 |}
\sum_{gh = hg} 
{\scriptstyle g} \square_h \, ,
\\
& = &
Z\left( [X/{\mathbb Z}_4] \right),
\end{eqnarray}
an orbifold by ${\mathbb Z}_4 = \langle x K \rangle$.

\item Finally, we consider the case that
$B(a) = -1$ and $B(b) = -1$.  This implies that the central
$K$ acts trivially on the space but nontrivially on twisted sectors
twisted by either $a$ or $b$, for example
\begin{equation}
{\scriptstyle y^2} \square_x \: = \: - \left(
{\scriptstyle 1} \square_x \right),
 \: \: \:
{\scriptstyle y^2} \square_y \: = \: - \left(
{\scriptstyle 1} \square_y \right),
\: \: \:
{\scriptstyle y^2} \square_{xy} \: = \: + \left(
{\scriptstyle 1} \square_{xy} \right).
\end{equation}
Essentially as a result, in the genus-one partition function, twisted sector
with an odd number of powers of $x$ and $y$ on either leg cancel out.
It is straightforward to compute that the genus-one partition function is
\begin{eqnarray}
Z\left( [X / {\mathbb Z}_4 \rtimes {\mathbb Z}_4] \right) 
& = &
\frac{1}{| {\mathbb Z}_4 \rtimes {\mathbb Z}_4 |}
\sum_{gh = hg} 
{\scriptstyle g} \square_h \, ,
\\
& = &
Z\left( [X/{\mathbb Z}_2 \times {\mathbb Z}_2 ] \right),
\end{eqnarray}
an orbifold by ${\mathbb Z}_2 \times {\mathbb Z}_2 =
\langle x^2 K, xy K\rangle$, without discrete torsion.

\end{enumerate}

\subsubsection{With discrete torsion}
\label{sect:ex:d4:z4z4:dt}

Now, let us turn on discrete torsion $\omega$ in the ${\mathbb Z}_4 \rtimes
{\mathbb Z}_4$ orbifold.  As discussed in
\cite[section 6.2]{Robbins:2021ylj}, 
$\iota^* \omega = 0$, $\beta(\omega) = 0$, and in fact
this discrete torsion is a pullback
from $\overline{\omega} \in H^2(D_4,U(1))$.

From section~\ref{sect:conj},
we predict
\begin{equation}
{\rm QFT}\left( [X/\Gamma]_{B,\omega} \right) \: = \:
{\rm QFT}\left( \left[ \frac{ X \times \widehat{ {\rm Coker}\: B } }{
{\rm Ker}\: B} \right]_{\overline{\omega} + \hat{\omega}_0} \right).
\end{equation}
From our discussion in the case without discrete torsion,
we know that $\hat{\omega}_0 = 0$, and we have already computed the
kernels and cokernels, so we merely need to add the
restriction of $\overline{\omega}$ to the kernel in each case to
get our prediction, as we list below.  
We include the case that $B$ is trivial for
completeness.
\begin{enumerate}
\item First, consider the case that $B$ is trivial:
$B(a) = +1$, $B(b) = +1$.  In this case, the kernel is $D_4$ and
the cokernel is $\hat{\mathbb Z}_2$, so we find
\begin{equation}
{\rm QFT}\left( [X/\Gamma]_{B,\omega} \right) \: = \:
{\rm QFT}\left( \coprod_2 [X/D_4]_{\overline{\omega}} \right),
\end{equation}
as discussed in \cite[section 6.2]{Robbins:2021ylj}.
\item Next, consider the case that 
$B(a) = +1$, $B(b) = -1$.  Here, the cokernel is trivial and the
kernel is $\langle yK, x^2K \rangle =
{\mathbb Z}_2 \times {\mathbb Z}_2$, so we predict
\begin{equation}
{\rm QFT}\left( [X/\Gamma]_{B,\omega} \right) \: = \:
{\rm QFT}\left( [X/{\mathbb Z}_2 \times {\mathbb Z}_2]_{\overline{\omega}}
\right).
\end{equation}
\item Next, consider the case that $B(a) = -1$, $B(b) = +1$.  Here,
the cokernel is trivial and the kernel is $\langle xK \rangle =
{\mathbb Z}_4$.  Since this group has no discrete torsion, the
restriction of $\overline{\omega}$ is trivial, so we predict
\begin{equation}
{\rm QFT}\left( [X/\Gamma]_{B,\omega} \right) \: = \:
{\rm QFT}\left( [X/{\mathbb Z}_4] \right).
\end{equation}
\item Next, consider the case that $B(a) = -1$, $B(b) = -1$.
Here, the cokernel is trivial and the kernel is 
$\langle xy K, y^2 K \rangle = {\mathbb Z}_2 \times
{\mathbb Z}_2$, so we predict
\begin{equation}
{\rm QFT}\left( [X/\Gamma]_{B,\omega} \right) \: = \:
{\rm QFT}\left( [X/{\mathbb Z}_2 \times {\mathbb Z}_2]_{\overline{\omega}}
\right).
\end{equation}
\end{enumerate}

Next, we compute genus-one partition functions.  In each case, we find
that the partition function computations are consistent with the
predictions above.
\begin{enumerate}

\item First, consider the case that
$B(a) = +1$ and $B(b)=+1$.
This is an ordinary ${\mathbb Z}_4 \rtimes {\mathbb Z}_4$
orbifold with discrete torsion, with trivially-acting
$K = {\mathbb Z}_2 = \langle y^2 \rangle$.  This particular
example was discussed in \cite[section 6.2]{Robbins:2021ylj},
where it was argued
\begin{equation}
{\rm QFT}\left( [X/{\mathbb Z}_4 \rtimes {\mathbb Z}_4]_{\rm d.t.} \right)
\: = \:
{\rm QFT}\left( \coprod_2 [X/D_4]_{\overline{\omega}} \right),
\end{equation}
or in other words that this theory is equivalent to a disjoint union of 
two $D_4$ orbifolds, each with discrete torsion $\overline{\omega}$.

\item Next, we consider the case that
$B(a) = +1$ and $B(b) = -1$, and turn on discrete torsion
in the ${\mathbb Z}_4 \rtimes {\mathbb Z}_4$ orbifold.
The analysis proceeds very similarly to the case we considered previously
without discrete torsion, as the cancellations due to $\beta$ involve
pairs which have the same discrete torsion phases,
using the phases in \cite[table D.6]{Robbins:2021ylj}.
It is straightforward to check that the
genus-one partition function is
\begin{eqnarray}
Z\left( [X / {\mathbb Z}_4 \rtimes {\mathbb Z}_4]_{\rm d.t.} \right) 
& = &
\frac{1}{| {\mathbb Z}_4 \rtimes {\mathbb Z}_4 |}
\sum_{gh = hg}  \epsilon(g,h) \left(
{\scriptstyle g} \square_h \right) ,
\\
& = &
Z\left( [X/{\mathbb Z}_2 \times {\mathbb Z}_2]_{\rm d.t.} \right),
\end{eqnarray}
an orbifold by ${\mathbb Z}_2 \times {\mathbb Z}_2 =
\langle x^2 K, y K \rangle$ with discrete torsion.

\item Next, we consider the case that
$B(a) = -1$ and $B(b) = +1$,
and turn on discrete torsion in the ${\mathbb Z}_4 \rtimes {\mathbb Z}_4$
orbifold.  The analysis proceeds very similarly to the case we considered
previously without discrete torsion, as the cancellations due to
$\beta$ involve pairs which have the same discrete torsion phases,
using the phases in \cite[table D.6]{Robbins:2021ylj}.  It is
straightforward to check that the genus-one partition function is
\begin{eqnarray}
Z\left( [X / {\mathbb Z}_4 \rtimes {\mathbb Z}_4]_{\rm d.t.} \right) 
& = &
\frac{1}{| {\mathbb Z}_4 \rtimes {\mathbb Z}_4 |}
\sum_{gh = hg}  \epsilon(g,h) \left(
{\scriptstyle g} \square_h \right) ,
\\
& = &
Z\left( [X/{\mathbb Z}_4] \right),
\end{eqnarray}
an orbifold by ${\mathbb Z}_4 = \langle x K \rangle$.

\item Finally, we consider the case that
$B(a) = -1$ and $B(b) = -1$,
and turn on discrete torsion in the ${\mathbb Z}_4 \rtimes {\mathbb Z}_4$
orbifold.  The analysis proceeds very similarly to the case we considered
previously without discrete torsion, as the cancellations due to
$\beta$ involve pairs which have the same discrete torsion phases,
using the phases in \cite[table D.6]{Robbins:2021ylj}.  It is
straightforward to check that the genus-one partition function is
\begin{eqnarray}
Z\left( [X / {\mathbb Z}_4 \rtimes {\mathbb Z}_4]_{\rm d.t.} \right) 
& = &
\frac{1}{| {\mathbb Z}_4 \rtimes {\mathbb Z}_4 |}
\sum_{gh = hg}  \epsilon(g,h) \left(
{\scriptstyle g} \square_h \right) ,
\\
& = &
Z\left( [X/{\mathbb Z}_2 \times {\mathbb Z}_2]_{\rm d.t.} \right),
\end{eqnarray}
an orbifold by ${\mathbb Z}_2 \times {\mathbb Z}_2 =
\langle x^2 K, xy K\rangle$, with discrete torsion.
\end{enumerate}

\subsubsection{Summary}

We summarize the results of this analysis in table~\ref{table:z4z4:d4:summ}.

\begin{table}[h]
\begin{center}
\begin{tabular}{c|cc|cc}
Case & $B(a)$ & $B(b)$ &
Without discrete torsion & With discrete torsion \\ \hline
1 & $+1$ & $+1$ &
$[X/D_4] \, \coprod \, [X/D_4]$ &
$[X/D_4]_{\overline{\omega}} \, \coprod \, [X/D_4]_{\overline{\omega}}$ \\
2 & $+1$ & $-1$ & $[X/{\mathbb Z}_2 \times {\mathbb Z}_2 = \langle x^2 K, y K \rangle]$
& $[X/{\mathbb Z}_2 \times {\mathbb Z}_2 = \langle x^2 K, y K \rangle ]_{\rm d.t.}$\\
3 & $-1$ & $+1$ & $[X/{\mathbb Z}_4 = \langle x K \rangle]$
 & $[X/{\mathbb Z}_4 = \langle x K \rangle]$ \\
4 & $-1$ & $-1$ & $[X/{\mathbb Z}_2 \times {\mathbb Z}_2 = 
\langle x^2 K, xy K \rangle]$ & 
$[X/{\mathbb Z}_2 \times {\mathbb Z}_2 =  
\langle x^2 K, xy K \rangle]_{\rm d.t.}$
\end{tabular}
\caption{Summary of results for ${\mathbb Z}_4 \rtimes {\mathbb Z}_4$
orbifold extending $D_4$ orbifold.
\label{table:z4z4:d4:summ}
}
\end{center}
\end{table}

\subsection{Extension of ${\mathbb Z}_2 \times {\mathbb Z}_2$ to
$D_4 \times {\mathbb Z}_2 \times {\mathbb Z}_2$}
\label{sect:ex:d4z2z2:z2z2}

In this section, we begin with a
$G = {\mathbb Z}_2 \times {\mathbb Z}_2 = \langle \overline{a},
\overline{b} \rangle$ orbifold,
which we centrally extend by $K = ({\mathbb Z}_2)^3$ to
$\Gamma = D_4 \times {\mathbb Z}_2 \times {\mathbb Z}_2$,
where we write $({\mathbb Z}_2)^2 = \langle x, y\rangle$.
In particular, in this example, the restriction of discrete torsion in
$\Gamma$ to $K$ can be nontrivial, so we will be able to explore
cases in which $\iota^* \omega \neq 0$.

\subsubsection{Without discrete torsion}
\label{sect:ex:d4z2z2:z2z2:wodt}

First, let us assume that the $\Gamma$ orbifold has no discrete torsion,
and work out the consequences.
We will describe details for some representative 
values of $B \in H^1(G,H^1(K,U(1)))$ 
below.   

First, if $B$ is trivial, then the $[X/\Gamma]$ orbifold is described
by decomposition 
\cite{Hellerman:2006zs,Sharpe:2019ddn,Tanizaki:2019rbk,Robbins:2021ylj}, 
which predicts
\begin{equation}
{\rm QFT}\left( [X/\Gamma] \right) \: = \:
{\rm QFT}\left( \coprod_4 [X/D_4] \right),
\end{equation}
and furthermore (see esp. \cite[section 5.2]{Hellerman:2006zs}),
\begin{equation} \label{eq:d4z2z2:d4decomp}
{\rm QFT}\left( [X/D_4] \right) \: = \:
{\rm QFT}\left( [X/{\mathbb Z}_2 \times {\mathbb Z}_2] \, \coprod \,
[X/{\mathbb Z}_2 \times {\mathbb Z}_2]_{\rm d.t.} \right),
\end{equation}
so in particular,
\begin{equation}
{\rm QFT}\left( [X/\Gamma] \right) \: = \:
{\rm QFT}\left( \coprod_4  [X/{\mathbb Z}_2 \times {\mathbb Z}_2]
\, \coprod \,
\coprod_4 [X/{\mathbb Z}_2 \times {\mathbb Z}_2]_{\rm d.t.}
\right).
\end{equation}
It is straightforward to check using by-now standard methods that closed-string
partition functions agree with the predictions above, so we move on to
less trivial cases.

Next, consider the case that $B(\overline{a}) = (+,-,+)$
but $B(\overline{b})$ is trivial.  The notation means that $z \in D_4$
and $y$ act trivially on $\overline{a}$, but $x$ acts nontrivially.
In this case, Ker $B = {\mathbb Z}_2 = \langle \overline{b} \rangle$,
and Coker $B = ({\mathbb Z}_2)^2$, so we predict
\begin{equation}
{\rm QFT}\left( [X/\Gamma]_B \right) \: = \:
{\rm QFT}\left( \left[ \frac{ X \times \widehat{ {\mathbb Z}_2 \times
{\mathbb Z}_2 } }{ {\mathbb Z}_2 } \right] \right) \: = \:
{\rm QFT}\left( \coprod_4 [X/{\mathbb Z}_2 = \langle \overline{b} \rangle
] \right).
\end{equation}
The same kernel and cokernel arise, and so one has the same prediction,
for the cases $\beta(\overline{a}) = (+,+,-)$ and $(+,-,-)$
while $\beta(\overline{b}) = (+,+,+)$.

It is straightforward to verify this prediction using
partition functions.  Returning to the case 
$\beta(\overline{a}) = (+,-,+)$, in the partition function
\begin{equation}
Z\left( [X/\Gamma]_{B} \right) \: = \:
\frac{1}{|\Gamma|} \sum_{gh=hg} 
{\scriptstyle g} \square_h \, ,
\end{equation}
it is straightforward to check that all contributions from genus-one
twisted sectors involving $a$ are cancelled out by corresponding sectors
with $x$'s.  The remaining genus-one
twisted sectors project to those of a $\langle \overline{b} \rangle =
{\mathbb Z}_2$ orbifold, with multiplicity $4^3$ (arising from
possible multiplications by $z$, $x$, and $y$.)  As a result,
\begin{eqnarray}
Z\left( [X/\Gamma]_{B} \right) & = &
\frac{ 4^3 }{|\Gamma|} \left( \langle \overline{b} \rangle 
\mbox{ twisted sectors} \right)
\: = \: 
Z\left( \coprod_4 [X/ {\mathbb Z}_2 = \langle \overline{b} \rangle] \right),
\end{eqnarray}
confirming the prediction above.  The other cases above are treated
identically.

Certainly a variety of other cases exist, but as we feel we have
sufficient examples of this form, 
for the sake of brevity we will move on.

\subsubsection{With discrete torsion in ${\mathbb Z}_2 \times {\mathbb Z}_2 \subset \Gamma$}
\label{sect:ex:d4z2z2:z2z2:dta}

Next, let us assume that there is discrete torsion in
${\mathbb Z}_2 \times {\mathbb Z}_2 \subset \Gamma$.
Note that in this case, $\iota^* \omega \neq 0$.
As a result, we do not have a prediction for decomposition here,
but we can collect results experimentally.

If $B$ is trivial, then this reduces to an example studied
in \cite[section 4.5]{Robbins:2021ylj}, where it was shown
\begin{equation}
{\rm QFT}\left( [X/\Gamma]_{\omega} \right) \: = \:
{\rm QFT}\left( [X/D_4] \right) \: = \:
{\rm QFT}\left(  [X/{\mathbb Z}_2 \times {\mathbb Z}_2] \, \coprod \,
[X/{\mathbb Z}_2 \times {\mathbb Z}_2]_{\rm d.t.} \right).
\end{equation}

Suppose next that $B(\overline{a}) = (+,-,+)$ and
$B(\overline{b})$ trivial.
We will see that the genus-one partition function
is the same as that of $[X/D_4]$.

First, consider the genus-one twisted sectors that project to
\begin{equation}
{\scriptstyle 1} \square_1 \, .
\end{equation}
Excluding contributions from $z$ (the generator of the center of $D_4$),
the contributing sectors are
\begin{eqnarray}
\lefteqn{
{\scriptstyle 1} \square_1 \: + \:
{\scriptstyle x} \square_1 \: + \:
{\scriptstyle y} \square_1 \: + \:
{\scriptstyle xy} \square_1 \: + \:
{\scriptstyle 1} \square_x \: + \:
{\scriptstyle 1} \square_y \: + \:
{\scriptstyle 1} \square_{xy} \: + \:
{\scriptstyle x} \square_x \: - \:
{\scriptstyle x} \square_y \: - \:
{\scriptstyle x} \square_{xy} \: - \:
{\scriptstyle y} \square_x 
} \nonumber \\
& & \hspace*{2in}
\: + \:
{\scriptstyle y} \square_y \: - \:
{\scriptstyle y} \square_{xy} \: - \:
{\scriptstyle xy} \square_x \: - \: 
{\scriptstyle xy} \square_y \: + \:
{\scriptstyle xy} \square_{xy} \, ,
\end{eqnarray}
where the signs are entirely due to discrete torsion.
Since none of the group elements project to elements of $G$ involving
$\overline{a}$ or products thereof, $B$ does not contribute any relative signs.
Thus, we see that, ignoring contributions from $z$'s, these genus
one sectors are equivalent to
\begin{equation}
(10-6) \left( 
{\scriptstyle 1} \square_1 \right) \: = \:
(4)  \left( 
{\scriptstyle 1} \square_1 \right).
\end{equation}

Now, consider the genus-one twisted sectors that project to 
\begin{equation}
{\scriptstyle \overline{a}} \square_1 \, .
\end{equation}
Excluding contributions from $z$ (the generator of the center of $D_4$),
the contributing sectors are
\begin{eqnarray}
\lefteqn{
{\scriptstyle a} \square_1 \: + \:
{\scriptstyle xa} \square_1 \: + \:
{\scriptstyle ya} \square_1 \: + \:
{\scriptstyle xya} \square_1 \: + \:
{\scriptstyle a} \square_x \: + \:
{\scriptstyle a} \square_y \: + \:
{\scriptstyle a} \square_{xy} \: + \:
{\scriptstyle xa} \square_x \: - \:
{\scriptstyle xa} \square_y \: - \:
{\scriptstyle xa} \square_{xy} \: - \:
{\scriptstyle ya} \square_x 
} \nonumber \\
& & \hspace*{2in}
\: + \:
{\scriptstyle ya} \square_y \: - \:
{\scriptstyle ya} \square_{xy} \: - \:
{\scriptstyle xya} \square_x \: - \: 
{\scriptstyle xya} \square_y \: + \:
{\scriptstyle xya} \square_{xy} \, ,
\end{eqnarray}
where the signs listed are entirely due to discrete torsion.
In this case, $B$ contributes nontrivial relative phases.
For example,
\begin{equation}
{\scriptstyle a} \square_x \: = \: - \left( 
{\scriptstyle a} \square_1 \right).
\end{equation}
Taking into account these signs, as well as discrete torsion,
we find that the sum of the genus-one twisted sectors above is again
\begin{equation}
(4) \left( {\scriptstyle \overline{a}} \square_1 \right).
\end{equation}
Other genus-one sectors involving factors of $a$ are similar.

Putting these contributions together, and taking into account 
contributions from $z$'s, we find
\begin{eqnarray}
Z\left( [X/\Gamma]_{B,\omega} \right)
& = &
\frac{ (4) }{ |D_4 \times ({\mathbb Z}_2)^2| }\left(
\mbox{sectors of $D_4$ orbifold} \right),
\\
& = & Z\left( [X/D_4] \right) \: = \:
Z\left( [X/{\mathbb Z}_2 \times {\mathbb Z}_2] \, \coprod \,
[X/{\mathbb Z}_2 \times {\mathbb Z}_2]_{\rm d.t.} \right).
\end{eqnarray}

\subsubsection{With discrete torsion in $D_4 \subset \Gamma$}
\label{sect:ex:d4z2z2:z2z2:dtb}

Finally, let us consider the possibility that there is discrete
torsion in $D_4 \subset \Gamma$.  
In these cases, $\iota^* \omega = 0$ but $\beta(\omega) \neq 0$.

If $B$ is trivial, then from decomposition, this reduces
to $|{\mathbb Z}_2 \times {\mathbb Z}_2| = 4$ copies of
$[X/D_4]_{\omega}$, for which it was shown in 
\cite[section 5.5]{Robbins:2021ylj} that
\begin{equation}
{\rm QFT}\left( [X/D_4]_{\omega} \right) \: = \:
{\rm QFT}\left( [X/{\mathbb Z}_2 = \langle \overline{b} \rangle]
\right),
\end{equation}
hence
\begin{equation}
{\rm QFT}\left( [X/\Gamma]_{\omega} \right) \: = \:
{\rm QFT}\left( \coprod_4 [X/D_4]_{\omega} \right) \: = \:
{\rm QFT}\left( \coprod_4 [X/{\mathbb Z}_2 = \langle \overline{b} \rangle]
\right).
\end{equation}

Suppose instead that $B(\overline{a}) = (+,-,+)$ and
$B(\overline{b})$ is trivial.
Then from section~\ref{sect:conj}, we predict that
\begin{equation}
{\rm QFT}\left( [X/\Gamma]_{B, \omega} \right) \: = \:
{\rm QFT}\left( \left[ \frac{ X \times \widehat{ {\rm Coker}\,
B/\beta(\omega) } }{ {\rm Ker} B/\beta(\omega) } \right] \right),
\end{equation}
where $\beta(\omega)$ was computed in the closely-related case of
a $D_4$ orbifold with discrete torsion and trivially-acting center
in \cite[section 5.5]{Robbins:2021ylj}.  
In the notation of this section,
\begin{equation}
\beta(\omega)(\overline{a}) \: = \: (-,+,+),
\: \: \:
\beta(\omega)(\overline{b}) \: = \: (+,+,+),
\end{equation}
so that
\begin{equation}
(B/\beta(\omega))(\overline{a}) \: = \: (-,-,+),
\: \: \:
(B/\beta(\omega))(\overline{b}) \: = \: (+,+,+).
\end{equation}
From this we compute
\begin{equation}
{\rm Ker}\, B/\beta(\omega) \: = \:
{\mathbb Z}_2 = \langle \overline{b} \rangle,
\: \: \:
{\rm Coker}\, B/\beta(\omega) \: = \: {\mathbb Z}_2 \times {\mathbb Z}_2.
\end{equation}
Hence, in this case, we predict
\begin{equation}
{\rm QFT}\left( [X/\Gamma]_{B, \omega} \right) \: = \:
{\rm QFT}\left( \coprod_4 [X/{\mathbb Z}_2 = \langle \overline{b} \rangle]
\right).
\end{equation}

It is straightforward to check this statement at the level of
genus-one partition functions.  Briefly, for $B(\overline{a}) = (+,-,+)$
and $B(\overline{b})$ trivial, with discrete torsion in $D_4$ but not
${\mathbb Z}_2 \times {\mathbb Z}_2$, it is straightforward to check
that, because of $B(\overline{a})$, all contributions from sectors
involving $\overline{a}$ cancel out, and between $z$, $x$, $y$,
the remaining sectors have multiplicities as, for example
\begin{equation}
{\scriptstyle b} \square_1 \: = \: (4^3) \left( 
{\scriptstyle \overline{b} } \square_1 \right),
\end{equation}
and the restriction of the $D_4$ discrete torsion to
$\langle \overline{b} \rangle \subset D_4$ is trivial.
The genus-one partition function then has the form
\begin{eqnarray}
Z\left( [X/\Gamma]_{B, \omega} \right)
& = &
\frac{ 4^3 }{|\Gamma|} \left( 
{\scriptstyle 1} \square_1 \: + \:
{\scriptstyle \overline{b}}\square_1 \: + \:
{\scriptstyle 1} \square_{ \overline{b}} \: + \:
{\scriptstyle \overline{b}} \square_{\overline{b}} \right),
\\
& = &
(2)  \left( 
{\scriptstyle 1} \square_1 \: + \:
{\scriptstyle \overline{b}}\square_1 \: + \:
{\scriptstyle 1} \square_{ \overline{b}} \: + \:
{\scriptstyle \overline{b}} \square_{\overline{b}} \right),
\\
& = & 
(4) Z\left( [X/{\mathbb Z}_2 = \langle \overline{b} \rangle] \right)
\: = \:
Z\left( \coprod_4 [X/{\mathbb Z}_2 = \langle \overline{b} \rangle] \right),
\end{eqnarray}
confirming the prediction.

\subsubsection{Summary}

So far we have worked out predictions for several cases.
Computing partition functions is straightforward using the methods
already described, so we omit their explicit description;
suffice it to say, in all cases, the genus-one partition functions
match predictions.  We summarize the results in
table~\ref{table:d4z2z2:z2z2:summ}.

\begin{table}[h]
\begin{center}
\begin{tabular}{c|c|c|c|c|c}
$B(\overline{a})$ & $B(\overline{b})$ & W/o d.t. & D.t. in $({\mathbb Z}_2)^2$
& D.t. in $D_4$ \\ \hline
$(+++)$ & $(+++)$ & 
$\coprod_4 [X/D_4]$
& $[X/D_4]$
& $\coprod_4 [X/{\mathbb Z}_2 = \langle \overline{b} \rangle]$
\\[1ex] \hline
$(+ \pm \mp)$ &  & 
& 
& 
\\[-1ex]
$(+--)$ & \raisebox{1.5ex}{$(+++)$} &
\raisebox{1.5ex}{$\coprod_4 [X/{\mathbb Z}_2 = \langle \overline{b} \rangle]$}
 & \raisebox{1.5ex}{$[X/D_4]$} 
 & \raisebox{1.5ex}{$\coprod_4 [X/{\mathbb Z}_2 = \langle \overline{b} \rangle]$}
\\[1ex] \hline
\end{tabular}
\caption{Summary of results for $D_4 \times ({\mathbb Z}_2)^2$ orbifold.
In terms of ${\mathbb Z}_2 \times {\mathbb Z}_2 = \langle x, y \rangle$,
results are symmetric between actions involving $x$, $y$, and $xy$.
In all cases, predictions match partition functions.
The reader should note that the orbifold $[X/D_4]$ is reducible,
and is a sum of two copies of $[X/{\mathbb Z}_2 \times {\mathbb Z}_2]$,
as indicated in equation~(\ref{eq:d4z2z2:d4decomp}).
\label{table:d4z2z2:z2z2:summ}
}
\end{center}
\end{table}

\section{Conclusions}

In this paper, we have described a new set of modular-invariant
phase factors for orbifolds with trivially-acting subgroups, generalizing
quantum symmetries.  We have also described decomposition
\cite{Hellerman:2006zs,Sharpe:2019ddn,Tanizaki:2019rbk,Robbins:2021ylj} 
for orbifolds with quantum symmetries, and detailed both the form
of the resulting theories and decomposition in a variety of examples.

In our next paper \cite{rsv}, we will apply these new phase factors
to resolve anomalies in orbifolds.

One matter left for future work is the definition of D-branes and the
open string sector in the presence of quantum symmetries.
In appendix~\ref{app:open} we outline basics, in particular the fact
that if $d_2 B \neq 0$, then associativity of the group action on the 
D-brane must be (weakly) broken.  We intend to explore this in more detail
in future work.

Another matter left for future work concerns non-central extensions.
In most of this paper, we assume that the full orbifold group
$\Gamma$ is a central extension of the effectively-acting group $G$,
meaning that the trivially-acting subgroup $K$ lies within the
center of $\Gamma$.  However, much of the structure we describe
also seems to apply to non-central extensions.  We outline some basics
in appendix~\ref{app:noncent}, and intend to explore non-central
extensions in more detail in future work.

\section*{Acknowledgements}

We would like to thank R.~Donagi, T.~Pantev, R.~Szabo, 
and Y.~Tachikawa for useful
conversations.  D.R. was partially supported by
NSF grant PHY-1820867.
E.S. was partially supported by NSF grant
PHY-2014086.

\appendix

\section{Notes on conventional quantum symmetries in orbifolds}
\label{app:oldrev}

It is a well-known result that in an orbifold $[X/G]$, there is an
abelian $G/[G,G]$ symmetry such that orbifolding by the latter returns
$[X/[G,G]]$.  Indeed, much of this paper is devoted to generalizing this
result.
In this appendix we will display this old result in modern language,
in the spirit of decomposition \cite{Robbins:2021ylj}.

Given some $G$ orbifold of a space $X$, 
we extend $G$ to $\Gamma = G \times \hat{G}$, where
$\hat{G} = G_{\rm ab} = G/[G,G]$ 
acts trivially on $X$, and nontrivially on $G$-twisted sectors.

We turn on discrete torsion defined by 
the two-cocycle $\omega$
\begin{equation}
\omega( (g_1, g_1'), (g_2, g_2')) \: = \: \alpha(g_1, g_2')
\end{equation}
where
\begin{equation}
\alpha: \: G \times \hat{G} \: \longrightarrow \: U(1)
\end{equation}
is a homomorphism in both arguments.
(It is straightforward to check that $\omega$ is a two-cocycle,
so long as $\alpha$ is a homomorphism in both arguments.)
Then the $G$ action on $H^1(\hat{G},U(1))$ is
\begin{equation}
(g \cdot \phi)(g') \: = \: \alpha(g,g') \phi(g').
\end{equation}

We want to show that for some choice of $\alpha$, this action is transitive,
meaning that all of $H^1(\hat{G},U(1))$ is in a single $G$-orbit.  Since we
can start with the trivial element of $H^1(G',U(1))$, namely $\phi(g') = 1$,
we just have to show that we can pick $\alpha$ such that
every possible $\phi(g')$ is given by $\alpha(g,g')$ for some $g$.
However, since
\begin{equation}
H^1(\hat{G},U(1)) \: \cong \: G_{\rm ab} \: = \: G/[G,G],
\end{equation}
this is always possible.  Let $f: G/[G,G] \rightarrow H^1(\hat{G},U(1))$ denote
the isomorphism, and $p: G \rightarrow G/[G,G]$ the projection, then as
$f \circ p$ is surjective, we can take
\begin{equation}
\alpha(g,g') \: = \: (f \circ p)(g)(g').
\end{equation}

Now, let us apply the generalized decomposition of
\cite{Robbins:2021ylj} to this case.
Clearly the restriction of $\omega$ to $K = \hat{G}$ is trivial,
but $\beta(\omega) \neq 0$, and in fact is given by
\cite[equ'n (C.28)]{Robbins:2021ylj}
\begin{eqnarray}
\beta(\omega)(g,z) & = &
\frac{
\omega(g,g^{-1}z) 
}{
\omega(g^{-1}z,g)
}
\: = \:
\frac{
\omega( (g,1), (g^{-1},z) )
}{
\omega( (g^{-1},z), (g,1) )
}
\: = \:
\frac{
\alpha(g,z)
}{
\alpha(g^{-1},1)
},
\\
& = & (f \circ p)(g)(z).
\end{eqnarray}
The kernel of this map is $[G,G]$, and there is no cokernel,
as it is surjective.

Then, decomposition implies
\begin{equation}
{\rm QFT} \left( [X/G \times \hat{G}]_{\omega = \alpha} \right) \: = \:
{\rm QFT}\left( \left[ \frac{ X \times {\rm Coker}\: \beta(\omega) }{
{\rm Ker}\: \beta(\omega) } \right] \right)
\: = \:
{\rm QFT}\left( [X / [G,G] \right).
\end{equation}

\section{Triviality of $\pi^* (d_2 B)$ in cohomology}
\label{app:triv-pullback}

As observed in section~\ref{sect:basics}, not every quantum symmetry
$B \in H^1(G, H^1(K,U(1)))$ is determined by discrete torsion.
Those which are not determined by discrete torsion have
nontrivial images $d_2 B \in H^3(G, U(1))$.  In this appendix we will
show that although their images can be nontrivial elements of
$H^3(G, U(1))$, their pullbacks $\pi^* (d_2 B) \in H^3(\Gamma, U(1))$
are trivial as elements of $H^3(\Gamma,U(1))$ -- trivial in cohomology,
in other words, though not necessarily identically equal to $1$.

Given $B \in H^1(G, H^1(K,U(1)))$ and a section $s: G \rightarrow \Gamma$, 
define
\begin{equation}
\lambda(g_1,g_2) \: = \: B(\pi(g_1), s_1 g_2 s_2^{-1} s_1^{-1} ),
\end{equation}
where $g_i \in \Gamma$ and $s_i = s(\pi(g_i))$. 
Then,
\begin{eqnarray}
(d \lambda)(g_1, g_2, g_3) & = &
\frac{
B(\pi(g_2), s_2 g_3 s_3^{-1} s_2^{-1}) \,
B(\pi(g_1), s_1 g_2 g_3 s_{23}^{-1} s_1^{-1})
}{
B(\pi(g_1 g_2), s_{12} g_3 s_3^{-1} s_{12}^{-1}) \,
B(\pi(g_1), s_1 g_2 s_2^{-1} s_1^{-1}
},
\\
& = &
B\left(\pi(g_1), s_1 g_2 g_3 s_{23}^{-1} s_1^{-1} \cdot
s_{12} s_3 g_3^{-1} s_{12}^{-1} \cdot
s_1 s_2 g_2^{-1} s_1^{-1} \right)
\nonumber \\
& & 
\cdot
B\left(\pi(g_2), s_2 g_3 s_3^{-1} s_2^{-1} \cdot
s_1^{-1} s_{12} s_3 g_3^{-1} s_{12}^{-1} s_1 \right),
\\
& = &
B\left(\pi(g_1), s_1 s_2 g_2^{-1} s_1^{-1} \cdot
s_1 g_2 g_3 s_{23}^{-1} s_1^{-1} \cdot
s_{12} s_3 g_3^{-1} s_{12}^{-1} \right)
\nonumber \\
& & \cdot
B\left( \pi(g_2), s_2 g_3 s_3^{-1} s_2^{-1} \cdot
s_1^{-1} s_{12} s_2^{-1} \cdot
s_2 s_3 g_3^{-1} s_2^{-1} \cdot
s_2 s_{12}^{-1} s_1 \right),
\\
& = &
B\left( \pi(g_1), s_1 s_2 s_{12}^{-1} \cdot
s_{12} g_3 s_{23}^{-1} s_1^{-1} \cdot
s_{12} s_2^{-1} s_1^{-1} \cdot
s_1 s_2 s_3 g_3^{-1} s_{12}^{-1} \right),
\\
& = &
B\left( \pi(g_1), s_1 s_2 s_3 s_{23}^{-1} s_1^{-1} \right),
\\
& = & (d_2 B)( \pi(g_1), \pi(g_2), \pi(g_3) )
\: = \: ( \pi^* d_2 B)(g_1, g_2, g_3).
\end{eqnarray} 
In the expression above, we have not assumed that $\Gamma$ is a central
extension, instead using the more general conventions outlined in
appendix~\ref{app:noncent}.

In any event, we now see that
$\pi^* (d_2 B)$ is trivial in cohomology, even when $\Gamma$ is not
a central extension of $G$.

\section{Open string sector}
\label{app:open}

In this appendix we outline some ideas regarding open string sectors
in these theories with quantum symmetries.  Briefly,
ordinarily one would describe D-branes in terms of (possibly projective)
equivariant structures on some sheaf or bundle on the covering
space; however, if $d_2 B \neq 1$, then associativity is broken.
That said, associativity is broken by $\pi^* (d_2 B)$, which as we
saw in appendix~\ref{app:triv-pullback} is trivial
in cohomology, so associativity holds `up to homotopy,'
and is only broken in a weak sense.

In this appendix, we will only outline such weakly associative
versions of (projectively) equivariant structures.
We leave a more detailed examination of the open string sector
for future work.

To define D-branes, let us attempt to proceed in the same fashion as for
discrete torsion in \cite{Douglas:1998xa,Sharpe:2000ki,Sharpe:2000wu},
in which the action of the group $\Gamma$ on the Chan-Paton factors
is twisted by a group cochain.

Let $\omega$ be a 2-cochain in $\Gamma$ such that
\begin{equation}  \label{eq:cochain-constr}
(d \omega)(g_1, g_2, g_3) \: = \:
B( \pi(g_1), s(\pi(g_2)) s(\pi(g_3)) s( \pi(g_2 g_3) )^{-1}
\: = \: \pi^* (d_2 B)(g_1, g_2, g_3)^{-1},
\end{equation}
where $s$ denotes a normalized section $s: G \rightarrow \Gamma$
(normalized meaning that $s(1) = 1$), and where we also assume
that the cochain is normalized so that
\begin{equation} \label{eq:norm}
\omega(g,1) \: = \: \omega(1,g) \: = \: 1.
\end{equation}
(Note in passing that this includes as a special case
$\omega$ that are coclosed, corresponding to elements of discrete torsion.)
Such a cochain $\omega$ exists precisely because $\pi^* (d_2 B)$ is trivial
in cohomology, even if $d_2 B$ itself is not.

Let $\phi$ be a twisted representation of $\Gamma$, twisted in the
sense that
\begin{equation}
\phi(g_1) \phi(g_2) \: = \:
\omega(g_1, g_2) \, \phi(g_1 g_2).
\end{equation}
(The normalization condition allows us to consistently identify
$\phi(1)$ with the identity.)
This is very similar to a projective representation, except that
$\omega$ is a cochain rather than a cocycle.  As a result,
multiplication is not quite associative:
\begin{eqnarray}
\phi(g_1) \left( \phi(g_2) \phi(g_3) \right)
& = &
\omega(g_1, g_2 g_3) \, \omega(g_2, g_3) \, \phi(g_1 g_2 g_3),
\\
\left( \phi(g_1) \phi(g_2) \right) \phi(g_3) 
& = &
\omega(g_1, g_2) \, \omega( g_1 g_2, g_3) \, \phi(g_1 g_2 g_3),
\end{eqnarray}
hence
\begin{eqnarray}
\phi(g_1) \left( \phi(g_2) \phi(g_3) \right)
& = & 
(d \omega)(g_1, g_2, g_3) \, \left( \phi(g_1) \phi(g_2) \right) \phi(g_3).
\end{eqnarray}

If $d \omega = 1$, as happens in the case of discrete torsion,
then we see that multiplication of the $\phi$ is associative,
and we see that the $\phi$ form a projective representation of $\Gamma$.

However, if $d \omega \neq 1$, then associativity does not hold,
and we do not have a projective representation of $\Gamma$.
Furthermore, if we implement this group action in boundary OPEs
between topological defect lines at the boundary, this means that
such boundary OPEs are
nonassociative.

This may seem rather strange,
but nonassociative boundary OPEs have been considered previously
in \cite{Cornalba:2001sm,Herbst:2001ai,Herbst:2002fk,Herbst:2003we}
in the context of D-branes in the presence of\footnote{
We would like to thank R.~Szabo for a useful discussion of this matter.
}
nontrivial $H$ flux.  In fact, the analogy can be made more precise.
Discrete torsion in orbifolds can be described in terms of
$B$ fields, see \cite{Sharpe:2000ki,Sharpe:2000wu}, and the image
of $H^1(G,H^1(K,U(1)))$ in $H^3(G,U(1))$ is an analogue of the
curvature $H$.  If the element of $H^1(G,H^1(K,U(1)))$
is in the image of discrete torsion, then we get an ordinary
associative projective equivariant structure.  If the element of
$H^1(G,H^1(K,U(1)))$ has a nonzero image in $H^3(G,U(1))$, which
morally is the analogue of a non-trivial $H$, then the equivariant
structure is nonassociative, in precise analogy with the nonassociative
boundary OPEs arising in cases with nontrivial $H$ flux in
\cite{Cornalba:2001sm,Herbst:2001ai,Herbst:2002fk,Herbst:2003we}.

Now, to be clear, no examples exist in which
$\phi$ acts as a linear transformation on a finite-dimensional vector space.
In such a case, $\phi(k)$ is simply a finite-dimensional matrix,
and matrix multiplication is always associative in finite dimensions.
However, if the vector space is infinite-dimensional, then,
matrix multiplication can be nonassociative, see for example
\cite{blp,bernkopf,ka,aelpm,cooke}.
An example from \cite{blp} is defined by the matrices
\begin{equation}
V \: = \: \left[ \begin{array}{cccc}
1 & 1 & 1 & \cdots \\
0 & 1 & 1 & \cdots \\
0 & 0 & 1 & \cdots \\
\vdots & \vdots & \vdots & \ddots \end{array} \right],
\: \: \:
U \: = \: \left[ \begin{array}{rrrrc}
1 & -1 & 0 & 0 &\cdots \\
0 & 1 & -1 & 0 & \cdots \\
0 & 0 & 1 & -1 & \cdots \\
\vdots & \vdots & \vdots & \vdots & \ddots \end{array} \right],
\: \: \:
A \: = \: \left[ \begin{array}{rrrc}
0 & 0 & 0 & \cdots \\
-1 & 0 & 0 & \cdots \\
-1 & -1 & 0 & \cdots \\
\vdots & \vdots & \vdots & \ddots \end{array} \right],
\end{equation}
where in $U$ each row has one successive $(+1,-1)$ pair with other
entries equal to zero.  It is easy to see that
$UA$ and $VU$ are both the identity, hence
\begin{equation}
V(UA) \: = \: V \: \neq \: A \: = \: (VU)A.
\end{equation}
In passing, physically this would require infinite-rank Chan-Paton factors
in the case that the quantum symmetry does not arise from discrete torsion.
Similarly, in the case that the curvature $H$ of the $B$ field is
nonzero in de Rham cohomology, any sheaf twisted by $H$ also has
infinite rank (see e.g. \cite[section 4.1]{brylinski}), 
in line with the role of $H$ in
nonassociative open string products in
\cite{jackiw,Gunaydin:1985ur,Cornalba:2001sm,Herbst:2001ai,Herbst:2002fk,Herbst:2003we,Blumenhagen:2013zpa}.

In principle, given such a nonassociative equivariant structure,
one could then construct closed-string phases in much the same fashion
described for discrete torsion in
e.g. \cite{Aspinwall:2000xv}, though we shall not try to do so here.

We have outlined one possible way to understand open string sectors
in orbifolds with $B$ such that $d_2 B \neq 1$, breaking associativity,
but there are other
approaches.  For example,
\begin{itemize}
\item Another approach is described in
\cite[section 2.3]{Willerton:2008gyk}, where groups $G$ with an anomaly
in $H^3(G,U(1))$ are described as groupoids, which have representations on
2-vector spaces.  Similar approaches are described in 
\cite{Roche:1990hs,Baez:2008hz}, \cite[sections 8-9]{Freed:1994ad},
\cite[appendix E]{Moore:1988qv}.
\item Non-associative equivariant structures in a different context are
described in
\cite{Bunk:2018qvk,Bunk:2020rju,Mickelsson:2019par,Mickelsson:2020wec}.
\end{itemize}

We leave a 
detailed examination of open string sectors for
future work.

\section{Non-central extensions}
\label{app:noncent}

In the rest of this paper, we have focused on central extensions.
However, some preliminary computations suggest that similar results should
hold for non-central extensions.  We do not claim a complete understanding
of such cases, but collect here a few pertinent results.

\subsection{Basics}

Consider an orbifold of a space $X$ by a group $\Gamma$, where
a subgroup $K \subset \Gamma$ acts trivially, with effectively-acting
coset $G = \Gamma/K$:
\begin{equation}
1 \: \longrightarrow \: K \: \longrightarrow \: \Gamma \: 
\stackrel{\pi}{\longrightarrow} \:
G \: \longrightarrow \: 1.
\end{equation}
We do not assume that $K$ is necessarily central.

We can describe the quantum symmetry as follows. 
For $g \in \Gamma$, $k \in K$, the quantum symmetry is a phase
assigned to twist fields, depending on their projection to $G$,
which we take to obey
\begin{equation}
B( \pi(g_1) \pi(g_2), k) \: = \:
B( \pi(g_1), k) \,
B( \pi(g_2), g_1^{-1} k g_1 ).
\end{equation}
As the action of the identity should be trivial, we also require
\begin{equation}
B( 1, k) \: = \: 1 \: = \: B( \overline{g}, 1)
\end{equation}
for $\overline{g} \in G$, $k \in K$.
For $g, h \in \Gamma$, $z \in K$, and $s: G \rightarrow \Gamma$
a section (meaning $\pi \circ s$ is the identity on $G$), 
the quantum symmetry is a relation
\begin{eqnarray}
{\scriptstyle gz} \square_h & = &
B\left( \pi(h), s(\pi(h)) z s(\pi(h))^{-1} \right)
\left( 
{\scriptstyle g} \square_h \right),
\label{eq:defn1} \\
{\scriptstyle g} \square_{hz} & = &
B\left( \pi(g), s(\pi(g)) z s(\pi(g))^{-1} \right)^{-1}
\left( 
{\scriptstyle g} \square_h \right),
\end{eqnarray}
where $B: G \times K \rightarrow U(1)$ is a map such that
\begin{eqnarray}
B(\overline{g}_1 \overline{g}_2, k) & = & B(\overline{g}_1,k) 
\, B(\overline{g}_2, s(\overline{g}_1)^{-1} k s(\overline{g}_1) ),
\\
B(\overline{g}, k_1 k_2) & = & B(\overline{g}, k_1) \, B(\overline{g},k_2),
\end{eqnarray}
for $\overline{g}, \overline{g}_{1,2} \in G$.

The phase above is independent of the choice
of section $s$:  over any $\overline{h} \in G$,
the value of any two choices of section will differ by
an element $k \in K$, and note
\begin{eqnarray}
B\left( \pi(h), ( s k ) z (s k)^{-1} \right)
& = &
B\left( \pi(h), (s k s^{-1}) (s z s^{-1}) (s k^{-1} s^{-1}) \right),
\\
& = &
B \left( \pi(h), s k s^{-1} \right) \,
B \left( \pi(h), s k^{-1} s^{-1} \right) \,
B \left( \pi(h), s z s^{-1} \right),
\\
& = &
B \left( \pi(h), s z s^{-1} \right).
\end{eqnarray}

Given a section, we can interpret these quantum symmetries as defining
discrete-torsion-like phases associated to sectors in the
effective $G$, specifically,
\begin{equation} \label{eq:qs-phase-defn}
{\scriptstyle g} \square_h \: = \: 
\epsilon(g,h) \left( 
{\scriptstyle \pi(g)} \square_{\pi(h)} \right),
\end{equation}
for commuting $g, h \in \Gamma$, where
\begin{equation}
\epsilon(g,h) \: = \:
\frac{
B\left( \pi(h), s_h s_g^{-1} g s_h^{-1} \right)
}{
B\left( \pi(g), s_g s_h^{-1} h s_g^{-1} \right),
},
\end{equation}
where we have abbreviated $s_g = s(\pi(g))$.
As a consistency check, note that
\begin{eqnarray}
\frac{
{\scriptstyle gz} \square\limits_h 
}{
{\scriptstyle g} \square\limits_h 
} & = &
\frac{
\epsilon(gz, h) 
}{
\epsilon(g,h)
},
\\
& = &
\frac{
B( \pi(h), s_h s_g^{-1} g z s_h^{-1} )
}{
B( \pi(h), s_h s_g^{-1} g s_h^{-1} )
},
\\
& = &
B(\pi(h), s_h s_g^{-1} g z (s_g^{-1} g)^{-1} s_h^{-1} ),
\\
& = &
B( \pi(h), s_h s_g^{-1} g s_h^{-1}) \,
B( \pi(h), s_h z s_h^{-1} ) \,
B( \pi(h), s_h g^{-1} s_g s_h^{-1} ),
\\
& = &
B(\pi(h), s_h z s_h^{-1} ),
\end{eqnarray}
matching the original definition~(\ref{eq:defn1}).

Now, let us consider the genus-one phases $\epsilon(g,h)$ arising from
a quantum symmetry in a $\Gamma$-orbifold, 
given in equation~(\ref{eq:qs-phase-defn}).
It is straightforward to show that $\epsilon(g,g) = 1$ and
$\epsilon(g,h) = \epsilon(h,g)^{-1}$.  It is also straightforward to see
that
\begin{eqnarray}
\epsilon(g, h_1 h_2) & = &
\epsilon(g, h_1) \epsilon(g, h_2)
B\left( \pi(g), s_g s_1^{-1} h_1 s_2^{-1} h_1^{-1} s_{12} s_g^{-1} \right)
\nonumber \\
& & \hspace*{0.5in} \cdot
B\left( \pi(h_1), s_{12} s_g^{-1} g s_{12}^{-1} s_1 g^{-1} s_g s_1^{-1} \right)
\nonumber \\
& & \hspace*{0.5in} \cdot
B\left( \pi(h_2), s_1^{-1} s_{12} s_g^{-1} g s_{12}^{-1} s_1 s_2 g^{-1} s_g s_2^{-1}
\right).
\end{eqnarray}

Using the fact that
\begin{eqnarray}
\lefteqn{
s_1^{-1} s_{12} s_g^{-1} g s_{12}^{-1} s_1 s_2 g^{-1} s_g s_2^{-1}
} \nonumber \\
& = &
( s_1^{-1} s_{12} s_2^{-1}) (s_2 s_g^{-1} g s_2^{-1}) ( s_2 s_{12}^{-1} s_1 )
( s_2 g^{-1} s_g s_2^{-1} ),
\end{eqnarray}
where each factor on the right lies in $K$, we can then write
\begin{eqnarray}
\lefteqn{
B\left( \pi(h_2), s_1^{-1} s_{12} s_g^{-1} g s_{12}^{-1} s_1 s_2 g^{-1} s_g s_2^{-1}
\right)
} \nonumber \\
& = &
B( \pi(h_2), s_1^{-1} s_{12} s_2^{-1} )
B( \pi(h_2), s_2 s_{12}^{-1} s_1 )
B( \pi(h_2), s_2 s_g^{-1} g s_2^{-1} )
B( \pi(h_2), s_2 g^{-1} s_g s_2^{-1} ),
\nonumber \\
& = & 1.
\end{eqnarray}

Thus, we can write
\begin{eqnarray}
\epsilon(g, h_1 h_2) & = &
\epsilon(g, h_1) \epsilon(g, h_2)
(d_2 B)(\pi(g), \pi(h_1) \pi(h_2))^{-1}
\nonumber \\
& & \hspace*{0.5in} \cdot
B\left(\pi(g), ( s_g s_1^{-1} h_1 s_2^{-1} h_1^{-1} s_{12} )
( s_{12}^{-1} s_2 s_1 s_g^{-1} ) \right)
\nonumber \\
& & \hspace*{0.5in} \cdot
B\left( \pi(h_1), s_{12} s_g^{-1} g s_{12}^{-1} s_1 g^{-1} s_g s_1^{-1} \right),
\end{eqnarray}
where $d_2$ is the map $H^1(G,H^1(K,U(1))) \rightarrow H^3(G,U(1))$
appearing in section~\ref{sect:basics}, so that explicitly
\begin{equation}
(d_2 B)( \pi(g), \pi(h_1), \pi(h_2)) \: = \: 
B\left(\pi(g), s_g s_1^{-1} s_2^{-2} s_{12} s_g^{-1} \right).
\end{equation}

Now, to settle all issues associated with quantum symmetries in
non-central extensions, there are several more things we need to
do at this point.  For example, we need to demonstrate that $B$ only depends
upon the conjugacy class of $\overline{g}$, to show that $B$ acts on
twist fields of $G$; we need to elaborate on what circumstances this is
a symmetry, when the correlation functions are invariant; and we need to
discuss why physically in this more general case the quantum symmetries
are classified by $H^1(G,H^1(K,U(1))$, defined with the crossed module
condition.  We leave all of that for future work, and focus here on
merely outlining a few other matters.  We conclude this appendix
with an example.

\subsection{Example} 

Consider the case that $K = {\mathbb Z}_4 = \langle i \rangle \subset
\Gamma = {\mathbb H}$, the group of quaternions.  We have the short exact
sequence
\begin{equation}
1 \: \longrightarrow \: {\mathbb Z}_4 \: \longrightarrow \: {\mathbb H}
\: \longrightarrow \: {\mathbb Z}_2 \: \longrightarrow \: 1.
\end{equation}
Specifically, consider the orbifold $[X/\Gamma]_B$, where the quantum
symmetry $B$ is defined to be the nontrivial element of 
$H^1(G, H^1(K, U(1)))$. 

Explicitly, if we let $\xi$ denote the generator of 
$G = {\mathbb Z}_2$, and $i$ the generator of $K = {\mathbb Z}_4$,
then we take
\begin{equation}
B(\xi, k) \: = \: \exp(\pi i/2).
\end{equation}
Let us quickly check that this has the crossed homomorphism structure.
Take the section $s: G \rightarrow \Gamma$ to be defined by
$s(\xi) = j$, so that $s(\xi)^{-1} = -j$.
Then,
\begin{eqnarray}
B(\xi, i) & = & \exp(\pi i/2),
\\
B(\xi, s(\xi)^{-1} (i) s(\xi) ) & = &
B( \xi, (-j)(i)(j) ) \: = \: B(\xi, -jk) \: = \: B(\xi, i^3),
\\
& = & \exp(3 \pi i/2),
\\
B(\xi^2,i) & = & 1,
\end{eqnarray}
and indeed one can see that
\begin{equation}
B(\xi^2, i) \: = \: B(\xi, i) \,
B(\xi, s(\xi)^{-1} (i) s(\xi) )
\end{equation}
as expected.

Using this quantum symmetry, one can compute
\begin{eqnarray}
{\scriptstyle (-)} \square_{\pm j, \pm k} & = &
B( \xi, i^2) \left( {\scriptstyle 1} \square_{\pm j, \pm k} \right)
\: = \:
(-) \left( {\scriptstyle 1} \square_{\pm j, \pm k} \right),
\\
{\scriptstyle -j} \square_{\pm j} & = &
B(\xi, i^2) \left( {\scriptstyle j} \square_{\pm j} \right)
\: = \:
(-) \left(  {\scriptstyle j} \square_{\pm j} \right),
\\
{\scriptstyle -k} \square_{\pm k} & = & (-)
\left( {\scriptstyle k} \square_{\pm k} \right),
\end{eqnarray}
hence the genus-one partition function is given by
\begin{eqnarray}
Z\left( [X/{\mathbb H}]_B \right)
& = & 
\frac{1}{| {\mathbb H} |}
\sum_{gh = hg} {\scriptstyle g} \square_h,
\\
& = & \frac{4^2}{8} \left( {\scriptstyle 1} \square_1 \right) \: = \:
2 \left( {\scriptstyle 1} \square_1 \right),
\end{eqnarray}
since all genus-one sectors that do not project to the trivial sector cancel
out.  Thus, at the level of genus-one partition functions, we find
\begin{equation}
[X/{\mathbb H}]_B \: = \: \coprod_2 X \: = \: X \coprod X.
\end{equation}


\begin{thebibliography}{199}

\addcontentsline{toc}{section}{References}

\bibitem{Robbins:2021lry}
D.~Robbins, E.~Sharpe and T.~Vandermeulen,
``Anomalies, extensions and orbifolds,''
{\tt arXiv:2106.00693}.


\bibitem{Wang:2017loc}
J.~Wang, X.~G.~Wen and E.~Witten,
``Symmetric gapped interfaces of SPT and SET states: systematic constructions,''
Phys. Rev. X \textbf{8} (2018) 031048,
{\tt arXiv:1705.06728}.

\bibitem{Bhardwaj:2017xup}
L.~Bhardwaj and Y.~Tachikawa,
``On finite symmetries and their gauging in two dimensions,''
JHEP \textbf{03} (2018) 189,
{\tt arXiv:1704.02330}.

\bibitem{Tachikawa:2017gyf}
Y.~Tachikawa,
``On gauging finite subgroups,''
SciPost Phys. \textbf{8} (2020) 015,
{\tt arXiv:1712.09542}.

\bibitem{Chang:2018iay}
C.~M.~Chang, Y.~H.~Lin, S.~H.~Shao, Y.~Wang and X.~Yin,
``Topological defect lines and renormalization group flows in two dimensions,''
JHEP \textbf{01} (2019) 026,
{\tt arXiv:1802.04445}.



\bibitem{Robbins:2019zdb}
D.~Robbins and T.~Vandermeulen,
``Orbifolds from modular orbits,''
Phys. Rev. D \textbf{101} (2020)  106021,
{\tt arXiv:1911.05172}.


\bibitem{Robbins:2019ayj}
D.~Robbins and T.~Vandermeulen,
``Modular orbits at higher genus,''
JHEP \textbf{02} (2020) 113,
{\tt arXiv:1911.06306}.

\bibitem{yujitasi2019} Y. Tachikawa, TASI 2019 lectures,
available at
{\tt https://member.ipmu.jp/yuji.tachikawa/lectures/2019-top-anom/tasi2019.pdf}

\bibitem{Robbins:2021ylj}
D.~Robbins, E.~Sharpe and T.~Vandermeulen,
``A generalization of decomposition in orbifolds,''
{\tt arXiv:2101.11619}.

\bibitem{rsv} D. Robbins, E. Sharpe, T. Vandermeulen,
``Anomaly resolution via decomposition,'' to appear.

\bibitem{Ginsparg:1988ui}
P.~H.~Ginsparg, 
``Applied conformal field theory,''
pp. 1-168 of {\it Les Houches 1988:
Fields, Strings, Critical Phenomena}, 
ed. E. Br\'ezin, J. Zinn-Justin, North-Holland, Amsterdam, 1990,
{\tt arXiv:hep-th/9108028}.

\bibitem{Vafa:1989ih}
C.~Vafa, 
``Quantum symmetries of string vacua,''
Mod. Phys. Lett. A \textbf{4} (1989) 1615-1626.

\bibitem{Pantev:2005rh}
T.~Pantev and E.~Sharpe, 
``Notes on gauging noneffective group actions,''
{\tt arXiv:hep-th/0502027}.

\bibitem{Pantev:2005wj}
T.~Pantev and E.~Sharpe,
``String compactifications on Calabi-Yau stacks,''
Nucl. Phys. B \textbf{733} (2006) 233-296,
{\tt arXiv:hep-th/0502044}.

\bibitem{Pantev:2005zs}
T.~Pantev and E.~Sharpe,
``GLSM's for gerbes (and other toric stacks),''
Adv. Theor. Math. Phys. \textbf{10} (2006)  77-121,
{\tt arXiv:hep-th/0502053}.

\bibitem{Hellerman:2006zs}
S.~Hellerman, A.~Henriques, T.~Pantev, E.~Sharpe and M.~Ando,
``Cluster decomposition, T-duality, and gerby CFT's,''
Adv. Theor. Math. Phys. \textbf{11} (2007)  751-818,
{\tt arXiv:hep-th/0606034}. 


\bibitem{ajt1} E. Andreini, Y. Jiang, H.-H. Tseng, ``On Gromov-Witten
theory of root gerbes,''
{\tt arXiv:0812.4477}.

\bibitem{ajt2} E. Andreini, Y. Jiang, H.-H. Tseng, ``Gromov-Witten theory
of product stacks,''
Comm. Anal. Geom. {\bf 24} (2016) 223-277,
{\tt arXiv:0905.2258}.

\bibitem{ajt3} E. Andreini, Y. Jiang, H.-H. Tseng, ``Gromov-Witten theory
of root gerbes I:  structure of genus $0$ moduli spaces,''
J. Diff. Geom. {\bf 99} (2015) 1-45,
{\tt arXiv:0907.2087}.

\bibitem{t1} H.-H. Tseng, ``On degree zero elliptic orbifold
Gromov-Witten invariants,''
Int. Math. Res. Notices {\bf 2011} (2011) 2444-2468,
{\tt arXiv:0912.3580}.

\bibitem{gt1} A. Gholampour, H.-H. Tseng, ``On Donaldson-Thomas invariants
of threefold stacks and gerbes,''
Proc. Amer. Math. Soc. {\bf 141} (2013) 191-203,
{\tt arXiv:1001.0435}.

\bibitem{xt1} X. Tang, H.-H. Tseng, ``Duality theorems of \'etale
gerbes on orbifolds,''
Adv. Math. {\bf 250} (2014) 496-569,
{\tt arXiv:1004.1376}.

\bibitem{Caldararu:2007tc}
A.~C\u{a}ld\u{a}raru, J.~Distler, S.~Hellerman, T.~Pantev and E.~Sharpe,
``Non-birational twisted derived equivalences in abelian GLSMs,''
Commun. Math. Phys. \textbf{294} (2010) 605-645,
{\tt arXiv:0709.3855}.

\bibitem{Hori:2011pd}
K.~Hori,
``Duality in two-dimensional (2,2) supersymmetric non-abelian gauge theories,''
JHEP \textbf{10} (2013) 121,
{\tt arXiv:1104.2853}.

\bibitem{Addington:2012zv}
N.~M.~Addington, E.~P.~Segal and E.~Sharpe,
``D-brane probes, branched double covers, and noncommutative resolutions,''
Adv. Theor. Math. Phys. \textbf{18} (2014) 1369-1436,
{\tt arXiv:1211.2446}.

\bibitem{Sharpe:2012ji}
E.~Sharpe,
``Predictions for Gromov-Witten invariants of noncommutative resolutions,''
J. Geom. Phys. \textbf{74} (2013) 256-265,
{\tt arXiv:1212.5322}.

\bibitem{Hori:2013gga}
K.~Hori and J.~Knapp,
``Linear sigma models with strongly coupled phases - one parameter models,''
JHEP \textbf{11} (2013) 070,
{\tt arXiv:1308.6265}.

\bibitem{Hori:2016txh}
K.~Hori and J.~Knapp,
``A pair of Calabi-Yau manifolds from a two parameter non-Abelian gauged linear sigma model,''
{\tt arXiv:1612.06214}.

\bibitem{Knapp:2019cih}
J.~Knapp and E.~Sharpe,
``GLSMs, joins, and nonperturbatively-realized geometries,''
JHEP \textbf{12} (2019) 096,
{\tt arXiv:1907.04350}.


\bibitem{Gu:2018fpm}
W.~Gu and E.~Sharpe,
``A proposal for nonabelian mirrors,''
{\tt arXiv:1806.04678}.

\bibitem{Chen:2018wep}
Z.~Chen, W.~Gu, H.~Parsian and E.~Sharpe,
``Two-dimensional supersymmetric gauge theories with exceptional gauge groups,''
Adv. Theor. Math. Phys. \textbf{24} (2020)  67-123,
{\tt arXiv:1808.04070}.

\bibitem{Gu:2019zkw}
W.~Gu, H.~Parsian and E.~Sharpe,
``More non-Abelian mirrors and some two-dimensional dualities,''
Int. J. Mod. Phys. A \textbf{34} (2019) 1950181,
{\tt arXiv:1907.06647}.

\bibitem{Gu:2020ivl}
W.~Gu, E.~Sharpe and H.~Zou,
``Notes on two-dimensional pure supersymmetric gauge theories,''
JHEP \textbf{04} (2021) 261,
{\tt arXiv:2005.10845}.




\bibitem{Anderson:2013sia}
L.~B.~Anderson, B.~Jia, R.~Manion, B.~Ovrut and E.~Sharpe,
``General aspects of heterotic string compactifications on stacks and gerbes,''
Adv. Theor. Math. Phys. \textbf{19} (2015) 531-611,
{\tt arXiv:1307.2269}.

\bibitem{Sharpe:2019ddn}
E.~Sharpe,
``Undoing decomposition,''
Int. J. Mod. Phys. A \textbf{34} (2020) 1950233,
{\tt arXiv:1911.05080}.

\bibitem{Tanizaki:2019rbk}
Y.~Tanizaki and M.~\"Unsal,
``Modified instanton sum in QCD and higher-groups,''
JHEP \textbf{03} (2020) 123,
{\tt arXiv:1912.01033}.

\bibitem{Cherman:2020cvw}
A.~Cherman and T.~Jacobson,
``Lifetimes of near eternal false vacua,''
Phys. Rev. D \textbf{103} (2021) 105012,
{\tt arXiv:2012.10555}.

\bibitem{Eager:2020rra}
R.~Eager and E.~Sharpe,
``Elliptic genera of pure gauge theories in two dimensions with semisimple non-simply-connected gauge groups,''
{\tt arXiv:2009.03907}.

\bibitem{Komargodski:2020mxz}
Z.~Komargodski, K.~Ohmori, K.~Roumpedakis and S.~Seifnashri,
``Symmetries and strings of adjoint QCD$_{2}$,''
JHEP \textbf{03} (2021) 103,
{\tt arXiv:2008.07567}.

\bibitem{hochschild} G. Hochschild, ``Basic constructions in group
extension theory,'' pp. 183-201 in
{\it Contributions to algebra:  a collection of papers dedicated to
Ellis Kolchin}, ed. by H. Bass, P. J. Cassidy, J. Kovacic,
Academic Press, New York, 1977.


\bibitem{hochserre} G. Hochschild, J.-P. Serre,
``Cohomology of group extensions,''
Trans. Amer. Math. Soc. {\bf 74} (1953) 110-134.



\bibitem{weibel} C. Weibel, {\it An introduction to homological algebra},
Cambridge University Press, Cambridge, UK, 1994.

\bibitem{neukirch} J. Neukirch, A. Schmidt, K. Wingberg,
{\it Cohomology of number fields}, second edition, 
Springer-Verlag, Berlin, 2008.


\bibitem{gille-szamuely} P. Gille, T. Szamuely,
{\it Central simple algebras and Galois cohomology},
Cambridge studies in adv. math. {\bf 101}, Cambridge University Press,
Cambridge, 2006.

\bibitem{Vafa:1986wx}
C.~Vafa,
``Modular invariance and discrete torsion on orbifolds,''
Nucl. Phys. B \textbf{273} (1986) 592-606.


\bibitem{egno} P. Etingof, S. Gelaki, D. Nikshych, V. Ostrik,
{\it Tensor categories}, 
Math. surveys and monographs 205,
American Mathematical Society, Providence, Rhode Island, 2015.


\bibitem{Muller:2020phm}
L.~M\"uller,
``Extended functorial field theories and anomalies in quantum field theories,''
{\tt arXiv:2003.08217}.


\bibitem{Sharpe:2003cs}
E.~Sharpe,
``Discrete torsion and shift orbifolds,'' 
Nucl. Phys. B \textbf{664} (2003) 21-44,
{\tt arXiv:hep-th/0302152}.

\bibitem{Douglas:1998xa}
M.~R.~Douglas,
``D-branes and discrete torsion,''
{\tt arXiv:hep-th/9807235}.

\bibitem{Sharpe:2000ki}
E.~R.~Sharpe,
``Discrete torsion,''
Phys. Rev. D \textbf{68} (2003) 126003,
{\tt arXiv:hep-th/0008154}.

\bibitem{Sharpe:2000wu}
E.~R.~Sharpe,
``Recent developments in discrete torsion,''
Phys. Lett. B \textbf{498} (2001) 104-110,
{\tt arXiv:hep-th/0008191}.

\bibitem{Cornalba:2001sm}
L.~Cornalba and R.~Schiappa,
``Nonassociative star product deformations for D-brane world volumes in curved backgrounds,''
Commun. Math. Phys. \textbf{225} (2002) 33-66,
{\tt arXiv:hep-th/0101219}.



\bibitem{Herbst:2001ai}
M.~Herbst, A.~Kling and M.~Kreuzer,
``Star products from open strings in curved backgrounds,''
JHEP \textbf{09} (2001) 014,
{\tt arXiv:hep-th/0106159}.


\bibitem{Herbst:2002fk}
M.~Herbst, A.~Kling and M.~Kreuzer,
``Noncommutative tachyon action and D-brane geometry,''
JHEP \textbf{08} (2002) 010,
{\tt arXiv:hep-th/0203077}.

\bibitem{Herbst:2003we}
M.~Herbst, A.~Kling and M.~Kreuzer,
``Cyclicity of nonassociative products on D-branes,''
JHEP \textbf{03} (2004) 003,
{\tt arXiv:hep-th/0312043}.

\bibitem{blp} D. Bossaller, S. R. L\'opez-Permouth,
``On the associativity of infinite matrix multiplication,''
{\tt arXiv:1803.09779}.

\bibitem{bernkopf} M Bernkopf, ``A history of infinite matrices,''
Arch. Hist. Exact Sci. {\bf 4} (1968) 308-358.

\bibitem{ka} K Keremedis, A Abian,
  ``On the associativity and commutativity of multiplication of
    infinite matrices,''
Int. J. Math. Educ. Sci. Technol. {\bf 19} (1988) 175-182.

\bibitem{aelpm} L. M. Al-Essa, S. R. L\'opez-Permouth, N. M. Muthana,
``Modules over infinite-dimensional algebras,''
Linear and Multilinear Algebra {\bf 66} (2018) 488-496.

\bibitem{cooke} R. Cooke, {\it Infinite matrices and sequence spaces},
Dover, New York, 1955.

\bibitem{brylinski} J.-L. Brylinski,
{\it Loop spaces, characteristic classes, and geometric quantization},
Birkh\"auser, Boston, 1993.


\bibitem{Blumenhagen:2013zpa}
R.~Blumenhagen, M.~Fuchs, F.~Ha\ss{}ler, D.~L\"ust and R.~Sun,
``Non-associative deformations of geometry in double field theory,''
JHEP \textbf{04} (2014) 141,
{\tt arXiv:1312.0719}.

\bibitem{jackiw} R. Jackiw, ``Three-cocycle in mathematics and physics,''
Phys. Rev. Lett. {\bf 54} (1985) 159-162.

\bibitem{Gunaydin:1985ur}
M.~Gunaydin and B.~Zumino,
``Magnetic charge and non-associative algebras,''
pp. 43-53 in 
{\it Old and new problems in fundamental physics:  meeting in honour
of G. C. Wick}, Quaderni, Pisa: Scuola Normale Superiore, 1986.


\bibitem{Aspinwall:2000xv}
P.~S.~Aspinwall,
``A note on the equivalence of Vafa's and Douglas's picture of discrete torsion,''
JHEP \textbf{12} (2000) 029,
{\tt arXiv:hep-th/0009045}.




\bibitem{Willerton:2008gyk} 
S.~Willerton,
``The twisted Drinfeld double of a finite group via gerbes and finite groupoids,''
Algebr. Geom. Topol. \textbf{8} (2008)  1419-1457,
{\tt arXiv:math/0503266}.

\bibitem{Roche:1990hs}
P.~Roche, V.~Pasquier and R.~Dijkgraaf,
``QuasiHopf algebras, group cohomology and orbifold models,''
Nucl. Phys. B Proc. Suppl. \textbf{18} (1990) 60-72.

\bibitem{Baez:2008hz}
J.~C.~Baez, A.~Baratin, L.~Freidel and D.~K.~Wise,
``Infinite-dimensional representations of 2-groups,''
Mem. Am. Math. Soc. \textbf{1032} (2012) 1-112,
{\tt arXiv:0812.4969}.


\bibitem{Freed:1994ad}
D.~S.~Freed,
``Higher algebraic structures and quantization,''
Commun. Math. Phys. \textbf{159} (1994) 343-398,
{\tt arXiv:hep-th/9212115}.

\bibitem{Moore:1988qv}
G.~W.~Moore and N.~Seiberg,
``Classical and quantum conformal field theory,''
Commun. Math. Phys. \textbf{123} (1989) 177-254.



\bibitem{Bunk:2018qvk}
S.~Bunk, L.~M\"uller and R.~J.~Szabo,
``Geometry and 2-Hilbert space for nonassociative magnetic translations,''
Lett. Math. Phys. \textbf{109} (2019)  1827-1866,
{\tt arXiv:1804.08953}.

\bibitem{Bunk:2020rju}
S.~Bunk, L.~M\"uller and R.~J.~Szabo,
``Smooth 2-group extensions and symmetries of bundle gerbes,''
{\tt arXiv:2004.13395}.

\bibitem{Mickelsson:2019par}
J.~Mickelsson,
``Non-associative magnetic translations: A QFT construction,''
{\tt arXiv:1905.01944}.

\bibitem{Mickelsson:2020wec}
J.~Mickelsson and M.~Murray,
``Non associative magnetic translations from parallel transport in projective Hilbert bundles,''
J. Geom. Phys. \textbf{163} (2021) 104152,
{\tt arXiv:2011.11431}.



\end{thebibliography}
\end{document}